\documentclass[twocolumn,tighten]{aastex63} 
\hypersetup{linkcolor=red,citecolor=blue,filecolor=cyan,urlcolor=magenta}
\usepackage{amsmath}
\usepackage{natbib}
\usepackage{enumitem}

\newcommand{\hdtwo}{\object[HD 222925]{HD~222925}}

\newcommand{\loggf}{\mbox{$\log(gf)$}}
\newcommand{\kmsec}{\mbox{km~s$^{\rm -1}$}}
\newcommand{\logg}{\mbox{log~{\it g}}}

\newcommand{\teff}{\mbox{$T_{\rm eff}$}}
\newcommand{\vt}{\mbox{$v_{\rm t}$}}
\newcommand{\rpro}{\mbox{{\it r}-process}}
\newcommand{\spro}{\mbox{{\it s}-process}}

\newcommand{\logeps}[1]{$\log\varepsilon$(#1)}
\newcommand{\rs}{\textit{r}$+$\textit{s}}

\shorttitle{$R$-process Abundances in HD 222925}
\shortauthors{Roederer et al.}
\accepted{for publication in the Astrophysical Journal Supplement Series}

\begin{document}

\title{%
The $R$-Process Alliance:\
A Nearly Complete $R$-Process Abundance Template
Derived from Ultraviolet Spectroscopy of
the $R$-Process-Enhanced Metal-Poor Star HD~222925\footnote{%
Based on observations made with the NASA/ESA 
Hubble Space Telescope,
obtained at the Space Telescope Science Institute (STScI), which is 
operated by the Association of Universities for 
Research in Astronomy, Inc.\ (AURA) under NASA contract NAS~5-26555.
These observations are primarily associated with program GO-15657.
Data from programs 
GO-7348,
GO-8197,
GO-9804,
GO-12554,
GO-14161, and
GO-14765 are also used.
This paper also includes data gathered with the 6.5~meter 
Magellan Telescopes located at Las Campanas Observatory, Chile.
}
}

\author{Ian U.\ Roederer}
\affiliation{%
Department of Astronomy, University of Michigan,
1085 S.\ University Ave., Ann Arbor, MI 48109, USA}
\affiliation{%
Joint Institute for Nuclear Astrophysics -- Center for the
Evolution of the Elements (JINA-CEE), USA}
\email{Email:\ iur@umich.edu}

\author{James E.\ Lawler}
\affiliation{%
Department of Physics, University of Wisconsin-Madison,
Madison, WI 53706, USA}

\author{Elizabeth A.\ Den Hartog}
\affiliation{%
Department of Physics, University of Wisconsin-Madison,
Madison, WI 53706, USA}

\author{Vinicius M.\ Placco}
\affiliation{%
NSF's NOIRLab, 
Tucson, AZ 85719, USA}

\author{Rebecca Surman}
\affiliation{%
Department of Physics, University of Notre Dame, 
Notre Dame, IN 46556, USA}
\affiliation{%
Joint Institute for Nuclear Astrophysics -- Center for the
Evolution of the Elements (JINA-CEE), USA}

\author{Timothy C.\ Beers}
\affiliation{%
Department of Physics, University of Notre Dame, 
Notre Dame, IN 46556, USA}
\affiliation{%
Joint Institute for Nuclear Astrophysics -- Center for the
Evolution of the Elements (JINA-CEE), USA}

\author{Rana Ezzeddine}
\affiliation{%
Department of Astronomy, University of Florida, Bryant Space Science Center,  
Gainesville, FL 32611, USA}
\affiliation{%
Joint Institute for Nuclear Astrophysics -- Center for the
Evolution of the Elements (JINA-CEE), USA}

\author{Anna Frebel}
\affiliation{%
Department of Physics and Kavli Institute for Astrophysics and Space Research, 
Massachusetts Institute of Technology, 
Cambridge, MA 02139, USA}
\affiliation{%
Joint Institute for Nuclear Astrophysics -- Center for the
Evolution of the Elements (JINA-CEE), USA}

\author{Terese T.\ Hansen}
\affiliation{%
Department of Astronomy, 
Stockholm University, 
Stockholm, Sweden}

\author{Kohei Hattori}
\affiliation{%
National Astronomical Observatory of Japan,
Mitaka, Tokyo 181-0015, Japan}
\affiliation{%
Institute of Statistical Mathematics, 
Tachikawa, Tokyo 190-0014, Japan}

\author{Erika M.\ Holmbeck}
\affiliation{%
Carnegie Observatories,
Pasadena, CA 91101, USA}
\affiliation{%
Joint Institute for Nuclear Astrophysics -- Center for the
Evolution of the Elements (JINA-CEE), USA}

\author{Charli M.\ Sakari}
\affiliation{%
Department of Physics and Astronomy, San Francisco State University,
San Francisco, CA 94132, USA}

\begin{abstract}

We present a nearly complete 
rapid neutron-capture process (\rpro) chemical inventory
of the metal-poor ([Fe/H] = $-1.46 \pm 0.10$) 
\rpro-enhanced ([Eu/Fe] = $+1.32 \pm 0.08$) 
halo star HD~222925.
This abundance set is
the most complete for any object beyond the solar system,
with a total of 63 metals detected and seven with upper limits.
It comprises 42 elements from $31 \leq Z \leq 90$,
including elements rarely detected in \rpro-enhanced stars, such as
Ga, Ge, As, Se, Cd, In, Sn, Sb, Te, W, Re, Os, Ir, Pt, and Au.
We derive these abundances from
an analysis of 404 absorption lines in 
ultraviolet spectra collected using 
the Space Telescope Imaging Spectrograph 
on the Hubble Space Telescope
and previously analyzed optical spectra.
A series of appendices discusses
the atomic data and quality of fits for these lines.
The \rpro\ elements from Ba to Pb, 
including all elements at the third \rpro\ peak,
exhibit remarkable agreement with the Solar \rpro\ residuals,
with a standard deviation of the differences of only 
0.08~dex (17\%).
In contrast, deviations among the lighter elements from Ga to Te
span nearly 1.4~dex, and they show distinct trends
from Ga to Se,
Nb through Cd, and In through Te.
The \rpro\ contribution to Ga, Ge, and As is small,
and Se is the lightest element whose
production is dominated by the \rpro.
The lanthanide fraction,
$\log X_{\rm La}$ = $-1.39 \pm 0.09$,
is typical for \rpro-enhanced stars
and higher than that of the kilonova from
the GW170817 neutron-star merger event.
We advocate adopting this pattern
as an alternative to the Solar \rpro-element residuals
when confronting
future theoretical models of heavy-element nucleosynthesis
with observations.

\end{abstract}

\keywords{%
Nucleosynthesis (1131);
R-process (1324);
Stellar abundances (1577);
Ultraviolet astronomy (1736);
Spectral line identification (2073)
}

\section{Introduction}
\label{intro}

The rapid neutron-capture process, or \rpro,
is one of the main ways that stars and their remnants produce
the heaviest elements
(atomic number, $Z$, $> 30$).
Recent theoretical and observational advances
generally agree that rare, prolific events
are responsible for much of the \rpro\ material 
found in the Sun and other stars
in and around the Milky Way
(e.g., \citealt{hotokezaka15,ji16nat,abbott17ejecta,siegel19}).
There are many open questions about the \rpro,
including one that has persisted for decades:\
\textit{which elements were produced by the \rpro, and in what amounts?}

The neutron-star merger GW170817 remains the only 
\rpro\ nucleosynthesis event to have been observed in the act, 
but the detailed abundance pattern of that event remains unknown.
The kilonova that followed the merger was linked to 
\rpro\ nucleosynthesis
by comparing the
evolving photometric colors with theoretical predictions
for radiative transfer rates in representative \rpro\ ions
with high opacities
(e.g., \citealt{kasen17}).
Only the \rpro\ element strontium (Sr, $Z = 38$) 
has possibly been identified in the ejecta of that merger event
\citep{watson19}.
The spectral lines of all other \rpro\ elements---including
silver, platinum, and gold---are 
blurred by the high expansion velocity of the ejecta
($\approx$0.1--0.2$c$; e.g., \citealt{chornock17,smartt17})
and thus cannot be unambiguously identified and
translated into abundances.

Instead, the standard approach 
to identifing which elements were produced by the \rpro\ and in what amounts
derives from models calibrated to Solar 
isotopic abundances measured from
Type-I carbonaceous chondrite meteorites.
A model based on stellar evolution and Galactic chemical evolution
(e.g., \citealt{arlandini99}),
or an analytic model (e.g., \citealt{cameron82,kappeler89}), 
is fit to abundances of
isotopes that can only be produced by the
slow neutron-capture process (\spro).
The \spro\ contribution to all other isotopes is inferred,
and all residual abundances are ascribed to the \rpro.
This set of ``\rpro\ residuals'' is 
frequently used as the observational ground truth that
\rpro\ models aim to reproduce
(e.g., \citealt{wanajo01,kratz07,wu16}).
This approach has been justified by the
unexpected observation of a near-perfect match between the 
Solar \rpro\ residuals and the heavy-element abundances
found in a rare few percent
of old, metal-poor stars in the Milky Way
(e.g., \citealt{cowan95,sneden96,hill02,roederer09b,frebel18,cowan21}).
This similarity gave rise to the 
so-called ``universality'' of the \rpro,
at least for the heaviest stable and
observable \rpro\ elements ($56 \leq Z \leq 82$).

This single template for
the \rpro\ pattern has guided
the vast majority of theoretical explorations of 
\rpro\ nucleosynthesis.
It relies on the
critical assumption that
\rpro\ and \spro\ nucleosynthesis are sufficient to explain
the origin of all heavy elements in the solar system.
Multiple events, processes, and patterns
are hidden within the \rpro\ residuals 
(e.g., \citealt{goriely99,travaglio04,bisterzo17,cote18ipro}).
Thus, an independent assessment of a
detailed inventory of the elements produced in a single \rpro\ event
is highly desirable.

Ongoing efforts by the \textit{R}-Process Alliance (RPA)
have identified an ideal metal-poor halo star for this task, \hdtwo.
Our previous analysis \citep{roederer18c}
of the optical spectrum of \hdtwo\
confirmed that it is a moderately metal-poor 
([Fe/H] $= -1.47$)
field red horizontal-branch star
with a high level of \rpro\ enhancement
([Eu/Fe] $= +1.33$).
One event likely dominated the 
production of \rpro\ elements found in \hdtwo.
This assertion,
commonly applied to other highly \rpro-enhanced metal-poor stars,
was recently rooted on firmer observational 
ground by 
a number of such stars, including \hdtwo, 
being placed into groups, based on their orbital kinematics
and small metallicity dispersions \citep{roederer18d,gudin21}.
The stars in each group are presumed to have formed together, within
a low-mass dwarf galaxy or star cluster that was enriched by one
prolific \rpro\ event.
Subsequent tidal interactions with the much more massive Milky Way
disrupted these systems and deposited their stars
into the stellar halo, where they are found today.

We present high-quality ultraviolet (UV) spectroscopy of
\hdtwo\ conducted with the
Space Telescope Imaging Spectrograph (STIS) on board the
Hubble Space Telescope (HST).~
Several \rpro-enhanced stars have been previously observed
by HST (e.g., \citealt{sneden98,cowan05,barbuy11,roederer12d}),
but none of them share the unique combination of characteristics
found in \hdtwo:\
(1) bright enough in the UV 
($V = 9.02$, \citealt{norris85}; 
GALEX $NUV = 13.41$, \citealt{martin05})
for high-resolution spectroscopy
with decent signal-to-noise ratios (S/N) 
at wavelengths as short as 2000~\AA;
(2) sufficiently metal poor that the UV spectrum is not
overwhelmed by strong lines of Fe-group elements;
(3) sufficiently \rpro\ enhanced that lines of
rarely detected \rpro\ elements may be present;
and
(4) dominated by \rpro\ material produced in a single event.
Our derived \rpro\ abundance pattern of \hdtwo\
thus provides a viable alternative to the 
inferred elemental Solar \rpro\ residuals.

Section~\ref{obs} describes the new STIS observations.
Section~\ref{analysis} describes our abundance analysis methods, and
Section~\ref{results} presents our results.
We discuss these results in Section~\ref{discussion}
and summarize our findings in Section~\ref{conclusions}.
Appendix~\ref{appendix} provides a detailed
discussion of the UV absorption lines and
the atomic data used to derive the abundances from them.

\section{Observations}
\label{obs}

\hdtwo\ was observed with STIS \citep{kimble98,woodgate98}
between 2019 October 03 and 2020 March 20.
Several visits early in this series suffered from 
delayed guide-star (re)acquisitions, which was part of a 
higher than usual failure rate 
across many observing programs.
STScI staff recognized and corrected the issue
by adopting a longer and more reliable guide-star acquisition strategy.
Failed observations were repeated,
and our program was successfully executed.

Observations were made using the E230H
echelle grating,
the 0\farcs2~$\times$~0\farcs09 slit, 
and
the near-UV Multianode Microchannel Array detector. 
This setup produces spectra with a resolving power of
$R \equiv \lambda/\Delta\lambda =$~114,000.
Five central wavelength settings were used
(i2063, i2313, i2563, i2812, and c3012),
resulting in complete wavelength coverage from 1936 to 3145~\AA.~
The observations were made during 47 orbits 
spread across 17~visits, not counting observations that were repeated
because of guide-star acquisition failures.
The integration times for the individual setups were
71,250~s (i2063, distributed across 27 individual observations),
23,291~s (i2313, 9 observations),
12,050~s (i2563, 5 observations),
10,106~s (i2812, 4 observations), and
 3,948~s (c3012, 2 observations),
for a total observing time on target of 120,645~s, or 33.5~hr.

The spectra were processed automatically by the 
CALSTIS software package and downloaded from the
Mikulski Archive for Space Telescopes (MAST).~
We first shift all observations to a common rest velocity.
We then coadd and normalize the spectrum 
using tools in the IRAF ``onedspec'' package, 
taking care to match the continuum level 
where the edges of adjacent orders and settings overlap.
The S/N ratios per pixel
in the final, coadded spectrum
are approximately
10/1 at 2000~\AA, 
20/1 at 2100~\AA,
30/1 at 2200~\AA, and
30/1 to 40/1 between 2200 and 2900~\AA,
before decreasing to 
20/1 at 3000~\AA\ and
15/1 at 3100~\AA.~
The region of the spectrum with $\lambda <$~2000~\AA\
has such a low S/N that we do not use it for our analysis.

We also revisit the optical spectrum of \hdtwo\
presented in \citet{roederer18c}.
That spectrum was collected using the 
Magellan Inamori Kyocera Echelle 
spectrograph (MIKE; \citealt{bernstein03})
mounted on the Landon Clay (Magellan~II) Telescope at 
Las Campanas Observatory, Chile.
It has 
$R =$ 68,000 in the blue (3330 $\leq \lambda \leq$~5000~\AA)
and 
$R =$ 61,000 in the red (5000 $\leq \lambda \leq$~9410~\AA)
with S/N ratios of several hundred per pixel.

\section{Analysis Methods}
\label{analysis}

\subsection{Definitions}
\label{definitions}

We adopt the standard nomenclature 
for elemental abundances and ratios.
The abundance of an element X is defined
as the number of X atoms per 10$^{12}$ H atoms,
$\log\varepsilon$(X)~$\equiv \log_{10}(N_{\rm X}/N_{\rm H})+$12.0.
The abundance ratio of the elements X and Y relative to the
Solar ratio is defined as
[X/Y] $\equiv \log_{10} (N_{\rm X}/N_{\rm Y}) - \log_{10} (N_{\rm X}/N_{\rm Y})_{\odot}$.
We adopt the Solar photospheric abundances of \citet{asplund09}.
By convention,
abundances or ratios denoted with the ionization state
are understood to be
the total elemental abundance, as derived from transitions of
that particular ionization state 
after \citet{saha21} ionization corrections have been applied.

\subsection{Stellar Parameters}
\label{params}

We adopt the same stellar parameters and model atmosphere
derived by \citet{roederer18c}:\
effective temperature (\teff) = 5636 $\pm$~103~K,
log of the surface gravity (\logg) = 2.54 $\pm$~0.17 [cgs units],
microturbulent velocity parameter (\vt) = 2.20 $\pm$~0.20~\kmsec,
and
model metallicity ([M/H]) = $-$1.5 $\pm$~0.1.
\teff\ was calculated by averaging the 
\teff\ values predicted by five different optical and near-infrared colors.
The \logg\ was calculated from fundamental relations,
and it included the parallax measurement from the Gaia mission's
second data release (DR2; \citealt{lindegren18}).~
The parallax measurement from Gaia's 
early third data release (EDR3; \citealt{gaia21edr3})
is effectively identical to the DR2 value, and \logg\ would change by 
$< 0.01$~dex using the EDR3 value instead of the DR2 value.
The \vt\ parameter was derived by requiring no dependence between the
line strength and the abundance derived from Fe~\textsc{i} lines.
Finally, the [M/H] value approximately matched the Fe abundance.

\subsection{Spectrum Synthesis}
\label{synthesis}

We derive all abundances by using the MOOG
\citep{sneden73,sobeck11} ``synthesis'' driver
to compare synthetic spectra to the observed spectrum.
MOOG assumes that local thermodynamic equilibrium (LTE) holds
in the line-forming layers of the atmosphere.
Line lists for these syntheses are generated using a version
of the LINEMAKE code \citep{placco21linemake}
that includes updates to the atomic data for UV transitions.
LINEMAKE starts with the \citet{kurucz11} line compendia
and supplements or replaces individual lines 
with atomic data---transition probabilities, hyperfine splitting, etc.---%
recommended by the Wisconsin Atomic Transition Probability group,
the National Institute of Standards and Technology (NIST)
Atomic Spectra Database (ASD),
refinements to Fe~\textsc{i} line lists by 
\citet{peterson15} and \citet{peterson17},
or our own assessments of literature data.
The fitting uncertainties reported in Table~\ref{linetab}
are usually dominated by continuum placement or blending features,
so these are larger than would be expected
based on the well-resolved line profiles.

\startlongtable
\begin{deluxetable*}{ccccccc}
\tablecaption{Line Atomic Data, References, and Derived Abundances
\label{linetab}}
\tabletypesize{\small}
\tablehead{
\colhead{Species} &
\colhead{$\lambda$} &
\colhead{E$_{\rm low}$} &
\colhead{\loggf} &
\colhead{Reference} &
\colhead{$\log\varepsilon$(X)} &
\colhead{Uncertainty} \\
\colhead{} &
\colhead{(\AA)} &
\colhead{(eV)} &
\colhead{} &
\colhead{} &
\colhead{} &
\colhead{(dex)} 
}
\startdata
Be~\textsc{ii} & 3130.422 & 0.00 & $-$0.18 &  1 &$<-$0.90 &\nodata\\ 
Be~\textsc{ii} & 3131.067 & 0.00 & $-$0.48 &  1 &$<-$0.80 &\nodata\\ 
B~\textsc{i}   & 2088.889 & 0.00 & $-$1.02 &  1 & $<$1.10 &\nodata\\ 
B~\textsc{i}   & 2089.570 & 0.00 & $-$0.72 &  1 & $<$0.90 &\nodata\\ 
B~\textsc{i}   & 2496.796 & 0.00 & $-$0.80 &  1 & $<$0.40 &\nodata\\ 
C~\textsc{i}   & 2964.846 & 0.00 & $-$7.20 &  1 &    6.70 & 0.34 \\ 
Al~\textsc{i}  & 2118.332 & 0.00 & $-$1.56 &  1 &    4.32 & 0.19 \\ 
Al~\textsc{i}  & 2129.678 & 0.00 & $-$1.38 &  1 &    4.28 & 0.23 \\ 
Al~\textsc{i}  & 2199.180 & 0.00 & $-$2.60 &  2 &    4.27 & 0.19 \\ 
\enddata
\tablereferences{%
 1 = NIST \citep{kramida20};
 2 = NIST \citep{kramida20} for \loggf\ value and 
       VALD \citep{piskunov95,pakhomov19} for HFS;
 3 = \citet{trabert99} for \loggf\ value and \citet{roederer21} for HFS;
 4 = \citet{biemont93};
 5 = \citet{theodosiou89};
 6 = \citet{lawler13};
 7 = \citet{wood13};
 8 = \citet{wood14v} for \loggf\ value and HFS;
 9 = \citet{sobeck07};
10 = \citet{gurell10};
11 = \citet{lawler17};
12 = \citet{denhartog11} for \loggf\ value and HFS;
13 = \citet{belmonte17};
14 = \citet{denhartog19};
15 = \citet{lawler15} for \loggf\ value and \citet{kurucz11} for HFS;
16 = \citet{lawler18} for \loggf\ value and this study for HFS;
17 = \citet{wood14ni};
18 = \citet{fedchak99};
19 = NIST \citep{kramida20} for \loggf\ value and \citet{kurucz11} for HFS;
20 = \citet{roederer12b} for \loggf\ value;
21 = This study;
22 = \citet{li99};
23 = \citet{holmgren75};
24 = \citet{morton00};
25 = \citet{biemont11};
26 = \citet{ljung06};
27 = \citet{nilsson10};
28 = \citet{nilsson08};
29 = \citet{sikstrom01};
30 = \citet{johansson94};
31 = \citet{xu04};
32 = \citet{curtis00in};
33 = \citet{oliver10};
34 = \citet{hartman10};
35 = \citet{roederer12a};
36 = \citet{denhartog06};
37 = \citet{lawler01tb};
38 = \citet{wickliffe00};
39 = \citet{lawler08};
40 = \citet{wickliffe97tm};
41 = \citet{biemont98} for \loggf\ value and \citet{roederer12b} for HFS/IS;
42 = \citet{roederer10b} for \loggf\ value and \citet{denhartog20} for HFS;
43 = \citet{quinet99} for \loggf\ value and \citet{denhartog20} for HFS;
44 = \citet{lawler09} for \loggf\ value and \citet{denhartog20} for HFS;
45 = \citet{denhartog21hf};
46 = \citet{lawler07};
47 = \citet{quinet09};
48 = \citet{nilsson08w};
49 = \citet{kling00};
50 = Roederer et al., in preparation;
51 = \citet{palmeri05re};
52 = \citet{quinet06};
53 = \citet{kramida20} for \loggf\ value and this study for HFS/IS;
54 = \citet{ivarsson04};
55 = \citet{denhartog05} for \loggf\ value only;
56 = \citet{denhartog05} for \loggf\ value and this study for HFS/IS;
57 = \citet{denhartog05} for \loggf\ value and HFS/IS;
58 = \citet{quinet08};
59 = \citet{zhang18};
60 = \citet{hannaford81} for \loggf\ value and \citet{demidov21} for HFS;
61 = \citet{quinet07} for \loggf\ value and \citet{roederer20} for HFS/IS;
}
\tablecomments{%
The two Al~\textsc{i} transitions at 
2204.619 (E.P.\ = 0.01~eV, \loggf\ = $-2.29$, NIST grade C) and 
2204.660~\AA\ form a single line in our spectrum.
The two Cu~\textsc{i} transitions at 
2024.323 and 2024.337~\AA\ form a single line in our spectrum;
HFS patterns are known for both lines, and the NIST ASD
lists B grades for both \loggf\ values.
A complete machine-readable version of Table~\ref{linetab} is
available online.
A short version is shown here to illustrate its form and content.}
\end{deluxetable*}

There are several strong absorption lines that
severely depress the continuum for a few \AA\ on either side
of each line.
Low-lying levels of 
Fe~\textsc{i} and Fe~\textsc{ii} are mostly responsible for
this effect, but similar behavior is observed among
Mg~\textsc{i} and Mg~\textsc{ii}, 
Si~\textsc{i}, Cr~\textsc{ii}, and Mn~\textsc{ii} lines.
Most of these lines are found between 2320 and 2640~\AA,
although a few lines
are found at shorter wavelengths.
The $^{2}S$--$^{2}P^{\rm o}$
Mg~\textsc{ii} resonance doublet
depresses the continuum by a few percent or more for
about 35~\AA\ on either side of 2800~\AA.~
The \loggf\ values and damping constants are known
for these transitions, as are the
abundances of these elements.
Analysis of other lines in these regions 
proceeds with caution, and only after 
rescaling the observed spectrum locally
to match the predicted wings of the strong lines.

\subsection{Iron (Fe, $Z = 26$) and the UV Continuum}
\label{fe}

We verify that MOOG is modeling the UV continuum reliably
by comparing the abundances derived from
optical and UV lines of Fe.
We synthesize 50~\AA\ regions of the spectrum,
match our syntheses to the observed spectrum,
and search for reasonably unblended Fe lines.
Strong lines whose damping
wings depress the continuum are not considered for this analysis.

We derive abundances from Fe~\textsc{i} lines
when the line is reasonably unblended and has a \loggf\ value
listed in \citet{belmonte17} or the NIST ASD with a 
grade of C or better (uncertainty $<$~25\%, or 0.12~dex).
We derive abundances from Fe~\textsc{ii} lines
when the line is similarly unblended and has a \loggf\ value
listed in \citet{denhartog19} or the NIST ASD with
a grade of C or better.
A total of 70 Fe~\textsc{i} and 42 Fe~\textsc{ii} lines
meet these criteria; they are listed in Table~\ref{linetab}.
The uncertainties listed in Table~\ref{linetab} are statistical and
reflect the goodness of the line fit and the \loggf\ uncertainty.
The weighted mean abundance derived from these UV Fe~\textsc{i} lines
is [Fe/H] $= -$1.61 $\pm$~0.02 ($\sigma$ = 0.18~dex), which 
matches the value derived by \citet{roederer18c} from 124
optical lines,
$-$1.58~$\pm$~0.01 ($\sigma$ = 0.08~dex).
Likewise, the weighted mean abundance derived from these 
UV Fe~\textsc{ii} lines
is [Fe/H] $= -$1.46 $\pm$~0.03 ($\sigma$ = 0.11~dex),
which also matches the value derived by \citeauthor{roederer18c}\
from 10 optical lines,
$-$1.41 $\pm$~0.03 ($\sigma$ = 0.08~dex).\footnote{%
\citet{roederer18c} used the NIST ASD \loggf\ scale for Fe~\textsc{ii}
lines.
Subsequent laboratory analysis by \citet{denhartog19}
showed better agreement with the \citet{melendez09fe} scale.
The value reported in the text here, [Fe/H] $= -$1.41, has been
corrected by $+$0.06~dex to the \loggf\ scale of
\citeauthor{melendez09fe}\ and \citeauthor{denhartog19}, 
which is the scale adopted for the UV Fe~\textsc{ii} lines.}

We draw three important conclusions from this test.
First, the optical and UV abundance scales are in excellent agreement,
which suggests that the continuum is being consistently
modeled within MOOG.~
Secondly, our data reduction and continuum normalization procedures
have yielded a UV spectrum that is approximately ``correct,''
except possibly in regions near strong lines,
as noted in Section~\ref{synthesis}.
Finally, the difference between the Fe abundance derived from 
Fe~\textsc{i} and Fe~\textsc{ii} lines is small but significant,
about $+$0.15 $\pm$~0.04~dex.
A small non-LTE (NLTE) overionization correction
is applied to the abundance derived from Fe~\textsc{i} lines.
\citet{roederer18c} used the INSPECT database 
\citep{bergemann12,lind12} to estimate a correction of $+$0.12~dex,
which agrees well.

\subsection{Other $Z \leq$~30 Elements}
\label{othersynthesis}

We verify that other elements yield consistent abundance results
when derived from optical and UV lines.
We also detect lines of some species that are 
not detected in the optical spectrum of \hdtwo.
Each species is discussed in detail in Appendix~\ref{appendix}.
The abundances derived from each line are presented in
Table~\ref{linetab}, 
along with the wavelength ($\lambda$), 
energy of the lower level of the transition (E$_{\rm low}$),
\loggf\ value, and reference for the
\loggf\ value and any hyperfine splitting (HFS)
or isotope shifts (IS) included in the synthesis.
The weighted mean
abundances derived from UV lines are presented in 
Table~\ref{abundtab}.
The final set of recommended elemental abundances in \hdtwo\ 
is presented in Table~\ref{finalabundtab}.

\startlongtable
\begin{deluxetable}{cccccc}
\tablecaption{Mean Abundances in HD~222925 Derived from the 
STIS/E230H UV Spectrum
\label{abundtab}}
\tabletypesize{\small}
\tablehead{
\colhead{Species} &
\colhead{$\log\varepsilon_{\odot}$(X)\tablenotemark{a}} &
\colhead{$\log\varepsilon$(X)} &
\colhead{[X/Fe]\tablenotemark{b}} &
\colhead{Unc.} &
\colhead{N$_{\rm lines}$} \\
\colhead{} &
\colhead{} &
\colhead{} &
\colhead{} &
\colhead{(dex)} &
\colhead{} 
}
\startdata
Be~\textsc{ii} &    1.38 &$<-$0.90 &$<-$0.82 &\nodata&  2 \\
B~\textsc{i}   &    2.70 & $<$0.40 &$<-$0.84 &\nodata&  3 \\
C~\textsc{i}   &    8.43 &    6.70 & $-$0.27 & 0.30 &  1 \\
Al~\textsc{i}  &    6.45 &    4.23 & $-$0.76 & 0.07 &  8 \\
Al~\textsc{ii} &    6.45 &    4.77 & $-$0.22 & 0.13 &  1 \\
Si~\textsc{i}  &    7.51 &    6.02 & $-$0.03 & 0.14 &  5 \\
Si~\textsc{ii}\tablenotemark{c} 
               &    7.51 &    6.48 & $+$0.43 & 0.18 &  4 \\
P~\textsc{i}   &    5.41 &    4.13 & $+$0.18 & 0.15 &  3 \\
S~\textsc{i}\tablenotemark{c}
               &    7.12 &    5.98 & $+$0.32 & 0.19 &  3 \\
Ca~\textsc{ii} &    6.34 &    5.24 & $+$0.36 & 0.16 &  2 \\
Ti~\textsc{i}  &    4.95 &    3.65 & $+$0.16 & 0.10 &  5 \\
Ti~\textsc{ii} &    4.95 &    3.77 & $+$0.28 & 0.06 & 21 \\
V~\textsc{ii}  &    3.93 &    2.72 & $+$0.25 & 0.14 &  9 \\
Cr~\textsc{i}  &    5.64 &    3.88 & $-$0.30 & 0.07 & 15 \\
Cr~\textsc{ii} &    5.64 &    4.16 & $-$0.02 & 0.06 & 20 \\
Mn~\textsc{ii} &    5.43 &    3.84 & $-$0.13 & 0.12 &  2 \\
Fe~\textsc{i}  &    7.50 &    5.89 & $-$1.61 & 0.18 & 70 \\
Fe~\textsc{ii} &    7.50 &    6.04 & $-$1.46 & 0.11 & 42 \\
Co~\textsc{i}  &    4.99 &    3.39 & $-$0.14 & 0.08 & 11 \\
Co~\textsc{ii} &    4.99 &    3.48 & $-$0.05 & 0.13 &  7 \\
Ni~\textsc{i}  &    6.22 &    4.59 & $-$0.17 & 0.07 & 17 \\
Ni~\textsc{ii} &    6.22 &    4.74 & $-$0.02 & 0.10 &  6 \\
Cu~\textsc{i}  &    4.19 &    1.93 & $-$0.80 & 0.12 &  3 \\
Cu~\textsc{ii} &    4.19 &    2.09 & $-$0.64 & 0.10 &  6 \\
Zn~\textsc{i}  &    4.56 &    3.15 & $+$0.05 & 0.11 &  2 \\
Ga~\textsc{ii} &    3.04 &    1.26 & $-$0.32 & 0.27 &  1 \\
Ge~\textsc{i}  &    3.65 &    1.46 & $-$0.73 & 0.11 &  5 \\
As~\textsc{i}  &    2.30 &    1.01 & $+$0.17 & 0.23 &  1 \\
Se~\textsc{i}  &    3.34 &    2.62 & $+$0.74 & 0.22 &  1 \\
Y~\textsc{ii}  &    2.21 &    1.06 & $+$0.31 & 0.12 &  4 \\
Zr~\textsc{ii} &    2.58 &    1.76 & $+$0.64 & 0.07 & 22 \\
Nb~\textsc{ii} &    1.46 &    0.73 & $+$0.73 & 0.11 &  9 \\
Mo~\textsc{ii} &    1.88 &    1.36 & $+$0.94 & 0.09 & 12 \\
Ru~\textsc{ii} &    1.75 &    1.26 & $+$0.97 & 0.19 &  2 \\
Cd~\textsc{i}  &    1.71 &    0.34 & $+$0.09 & 0.17 &  1 \\
In~\textsc{ii} &    0.80 &    0.51 & $+$1.17 & 0.21 &  1 \\
Sn~\textsc{ii} &    2.04 &    1.39 & $+$0.81 & 0.20 &  1 \\
Sb~\textsc{i}  &    1.01 &    0.37 & $+$0.82 & 0.17 &  1 \\
Te~\textsc{i}  &    2.18 &    1.63 & $+$0.91 & 0.14 &  2 \\
Gd~\textsc{ii} &    1.07 &    0.80 & $+$1.19 & 0.10 &  6 \\
Tb~\textsc{ii} &    0.30 &    0.31 & $+$1.47 & 0.19 &  1 \\
Dy~\textsc{ii} &    1.10 &    1.01 & $+$1.37 & 0.17 &  1 \\
Er~\textsc{ii} &    0.92 &    0.67 & $+$1.21 & 0.12 &  3 \\
Tm~\textsc{ii} &    0.10 &    0.06 & $+$1.42 & 0.13 &  3 \\
Yb~\textsc{ii} &    0.84 &    0.63 & $+$1.25 & 0.19 &  1 \\
Lu~\textsc{ii} &    0.10 & $-$0.10 & $+$1.26 & 0.11 &  7 \\
Hf~\textsc{ii} &    0.85 &    0.31 & $+$0.92 & 0.10 & 15 \\
Ta~\textsc{ii} & $-$0.12 &$<-$0.30 &$<+$1.28 &\nodata& 1 \\
W~\textsc{ii}  &    0.85 &    0.02 & $+$0.63 & 0.11 &  6 \\
Re~\textsc{ii} &    0.26 &    0.16 & $+$1.36 & 0.15 &  2 \\
Os~\textsc{i}  &    1.40 &    1.19 & $+$1.25 & 0.14 &  4 \\
Os~\textsc{ii} &    1.40 &    1.10 & $+$1.16 & 0.12 &  5 \\
Ir~\textsc{i}  &    1.38 &    1.26 & $+$1.34 & 0.12 &  4 \\
Ir~\textsc{ii} &    1.38 &    1.58 & $+$1.66 & 0.23 &  2 \\
Pt~\textsc{i}  &    1.62 &    1.45 & $+$1.29 & 0.10 &  8 \\
Pt~\textsc{ii} &    1.62 &    1.48 & $+$1.32 & 0.26 &  1 \\
Au~\textsc{i}  &    0.92 &    0.53 & $+$1.07 & 0.22 &  1 \\
Pb~\textsc{ii} &    2.04 &    1.14 & $+$0.56 & 0.14 &  1 \\
Bi~\textsc{i}  &    0.65 & $<$0.80 &$<+$1.61 &\nodata& 1 \\
\enddata
\tablenotetext{a}{%
\citet{asplund09}
}
\tablenotetext{b}{%
[Fe/H] is given for Fe~\textsc{i} and Fe~\textsc{ii}.
For all other species, the [X/Fe] ratios are referenced to the
Fe abundance derived from Fe~\textsc{ii} lines; i.e., [Fe/H] $= -$1.46.
}
\tablenotetext{c}{%
Newly derived from the MIKE optical spectrum.}
\end{deluxetable}

\startlongtable
\begin{deluxetable}{ccccccc}
\tablecaption{Recommended Metal Abundances in HD~222925
\label{finalabundtab}}
\tabletypesize{\small}
\tablehead{
\colhead{$Z$} &
\colhead{El.} &
\colhead{$\log\varepsilon$(X)} &
\colhead{[X/H]} &
\colhead{Unc.\tablenotemark{a}} &
\colhead{[X/Fe]} &
\colhead{Unc.\tablenotemark{b}} \\
\colhead{} &
\colhead{} &
\colhead{} &
\colhead{} &
\colhead{(dex)} &
\colhead{} &
\colhead{(dex)} 
}
\startdata
 3 & Li &$<$0.80 & \nodata&\nodata&\nodata  &\nodata\\
 4 & Be &$<-$0.90& $-$2.28 &\nodata&$<-$0.82&\nodata\\
 5 & B  &$<$0.40 & $-$2.30 &\nodata&$<-$0.84&\nodata\\
 6 & C  &   7.11 & $-$1.32 & 0.15 & $+$0.14 & 0.17 \\
 7 & N  &   6.45 & $-$1.38 & 0.20 & $+$0.08 & 0.21 \\
 8 & O  &   7.65 & $-$1.04 & 0.13 & $+$0.42 & 0.07 \\ 
11 & Na &   4.49 & $-$1.75 & 0.07 & $-$0.29 & 0.07 \\
12 & Mg &   6.43 & $-$1.17 & 0.08 & $+$0.29 & 0.05 \\
13 & Al &   4.78 & $-$1.67 & 0.13 & $-$0.21 & 0.13 \\
14 & Si &   6.47 & $-$1.04 & 0.11 & $+$0.42 & 0.11 \\
15 & P  &   4.13 & $-$1.28 & 0.18 & $+$0.18 & 0.15 \\
16 & S  &   5.98 & $-$1.14 & 0.19 & $+$0.32 & 0.19 \\
19 & K  &   3.68 & $-$1.35 & 0.10 & $+$0.11 & 0.04 \\
20 & Ca &   5.14 & $-$1.20 & 0.08 & $+$0.26 & 0.05 \\
21 & Sc &   1.82 & $-$1.33 & 0.13 & $+$0.13 & 0.08 \\
22 & Ti &   3.82 & $-$1.13 & 0.10 & $+$0.33 & 0.05 \\
23 & V  &   2.67 & $-$1.26 & 0.15 & $+$0.20 & 0.08 \\
24 & Cr &   4.16 & $-$1.48 & 0.11 & $-$0.02 & 0.05 \\
25 & Mn &   3.80 & $-$1.63 & 0.19 & $-$0.17 & 0.16 \\
26 & Fe &   6.04 & $-$1.46 & 0.10 & $+$0.00 & 0.10 \\
27 & Co &   3.48 & $-$1.51 & 0.20 & $-$0.05 & 0.11 \\
28 & Ni &   4.74 & $-$1.48 & 0.12 & $-$0.02 & 0.09 \\
29 & Cu &   2.09 & $-$2.10 & 0.10 & $-$0.64 & 0.10 \\
30 & Zn &   3.15 & $-$1.41 & 0.11 & $+$0.05 & 0.05 \\
31 & Ga &   1.26 & $-$1.78 & 0.27 & $-$0.32 & 0.27 \\
32 & Ge &   1.46 & $-$2.19 & 0.13 & $-$0.73 & 0.11 \\
33 & As &   1.01 & $-$1.29 & 0.23 & $+$0.17 & 0.23 \\
34 & Se &   2.62 & $-$0.72 & 0.22 & $+$0.74 & 0.22 \\
37 & Rb &$<$2.10 &$<-$0.42 &\nodata&$<+$1.04&\nodata\\
38 & Sr &   1.98 & $-$0.89 & 0.13 & $+$0.57 & 0.13 \\
39 & Y  &   1.04 & $-$1.17 & 0.10 & $+$0.29 & 0.07 \\
40 & Zr &   1.74 & $-$0.84 & 0.11 & $+$0.62 & 0.08 \\
41 & Nb &   0.71 & $-$0.75 & 0.14 & $+$0.71 & 0.11 \\
42 & Mo &   1.36 & $-$0.52 & 0.10 & $+$0.94 & 0.07 \\
44 & Ru &   1.32 & $-$0.43 & 0.11 & $+$1.03 & 0.11 \\
45 & Rh &   0.64 & $-$0.27 & 0.16 & $+$1.19 & 0.12 \\
46 & Pd &   1.05 & $-$0.52 & 0.15 & $+$0.94 & 0.08 \\
47 & Ag &   0.44 & $-$0.50 & 0.18 & $+$0.96 & 0.13 \\
48 & Cd &   0.34 & $-$1.37 & 0.25 & $+$0.09 & 0.17 \\
49 & In &   0.51 & $-$0.29 & 0.21 & $+$1.17 & 0.21 \\
50 & Sn &   1.39 & $-$0.65 & 0.20 & $+$0.81 & 0.20 \\
51 & Sb &   0.37 & $-$0.64 & 0.17 & $+$0.82 & 0.17 \\
52 & Te &   1.63 & $-$0.55 & 0.17 & $+$0.91 & 0.14 \\
56 & Ba &   1.26 & $-$0.92 & 0.09 & $+$0.54 & 0.06 \\
57 & La &   0.51 & $-$0.59 & 0.09 & $+$0.87 & 0.07 \\
58 & Ce &   0.85 & $-$0.73 & 0.08 & $+$0.73 & 0.07 \\
59 & Pr &   0.22 & $-$0.50 & 0.10 & $+$0.96 & 0.08 \\
60 & Nd &   0.88 & $-$0.54 & 0.09 & $+$0.92 & 0.08 \\
62 & Sm &   0.62 & $-$0.34 & 0.09 & $+$1.12 & 0.08 \\
63 & Eu &   0.38 & $-$0.14 & 0.09 & $+$1.32 & 0.08 \\
64 & Gd &   0.82 & $-$0.25 & 0.09 & $+$1.21 & 0.08 \\
65 & Tb &   0.18 & $-$0.12 & 0.11 & $+$1.34 & 0.09 \\
66 & Dy &   1.01 & $-$0.09 & 0.11 & $+$1.37 & 0.08 \\
67 & Ho &   0.12 & $-$0.36 & 0.15 & $+$1.10 & 0.12 \\
68 & Er &   0.73 & $-$0.19 & 0.10 & $+$1.27 & 0.08 \\
69 & Tm &$-$0.09 & $-$0.19 & 0.10 & $+$1.27 & 0.08 \\
70 & Yb &   0.55 & $-$0.29 & 0.19 & $+$1.17 & 0.19 \\
71 & Lu &$-$0.04 & $-$0.14 & 0.10 & $+$1.32 & 0.09 \\
72 & Hf &   0.32 & $-$0.53 & 0.12 & $+$0.93 & 0.10 \\
73 & Ta &$<-$0.30&$<-$0.18 &\nodata&$<+$1.28&\nodata\\
74 & W  &   0.02 & $-$0.83 & 0.11 & $+$0.63 & 0.11 \\
75 & Re &   0.16 & $-$0.10 & 0.15 & $+$1.36 & 0.15 \\
76 & Os &   1.17 & $-$0.23 & 0.14 & $+$1.23 & 0.09 \\
77 & Ir &   1.28 & $-$0.10 & 0.17 & $+$1.36 & 0.10 \\
78 & Pt &   1.45 & $-$0.17 & 0.14 & $+$1.29 & 0.10 \\
79 & Au &   0.53 & $-$0.39 & 0.23 & $+$1.07 & 0.22 \\
82 & Pb &   1.14 & $-$0.90 & 0.16 & $+$0.56 & 0.14 \\
83 & Bi &$<$0.80 &$<+$0.15 &\nodata&$<+$1.61&\nodata\\
90 & Th &$-$0.06 & $-$0.08 & 0.12 & $+$1.38 & 0.11 \\
92 & U  &$<-$0.50&$<+$0.04 &\nodata&$<+$1.50&\nodata\\
\enddata
\tablecomments{%
Readers interested in the 
details of why these recommendations are made
are encouraged to consult the relevant sections in Appendix~\ref{appendix}.
The C abundance is corrected for stellar evolution effects,
as described in \citet{placco14c}.
The O abundance is adopted from \citet{navarrete15}.
A complete machine-readable version of Table~\ref{finalabundtab} is
available online.
}
\tablenotetext{a}{%
Uncertainty on $\log\varepsilon$(X) and [X/H] abundances.
}
\tablenotetext{b}{%
Uncertainty on [X/Fe] abundance ratios.
This value also approximates the uncertainty
in the abundance ratios of other elements relative to each other;
e.g., [Ba/Eu].
}
\end{deluxetable}

\begin{figure}
\begin{center}
\includegraphics[angle=0,width=3.35in]{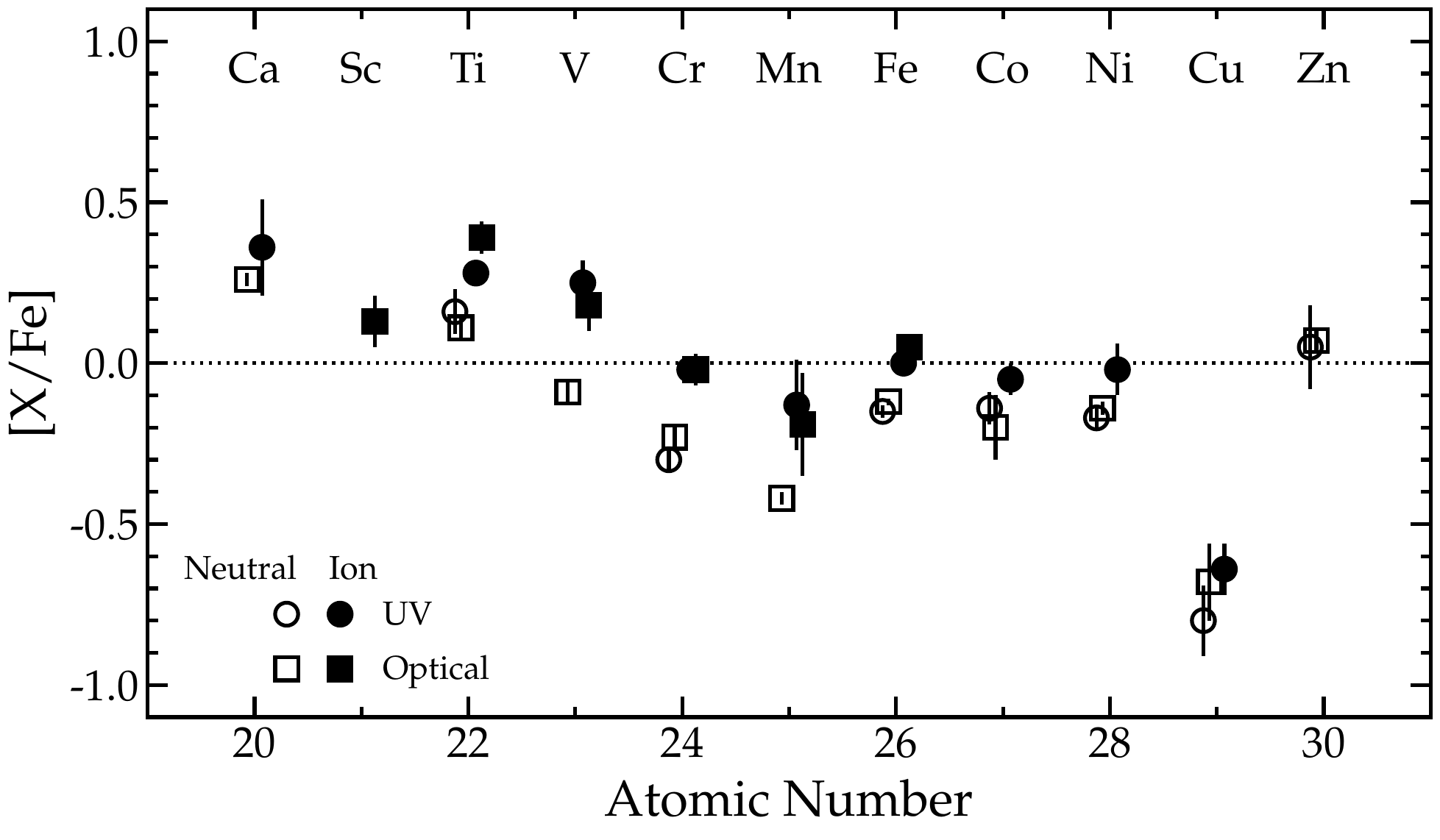}
\end{center}
\caption{
\label{fegroupplot}
Derived abundances for the elements with 20~$\leq Z \leq$~30
in \hdtwo.
The legend identifies the 
meanings of the symbols, and the 
dotted line represents the Solar ratio.
The vertical error bars mark 1$\sigma$ statistical uncertainties.
The points are slightly offset along the horizontal axis
to improve clarity.
}
\end{figure}

Figure~\ref{fegroupplot} illustrates the abundance ratios for
Ca and elements in the Fe group, and it
shows that the abundances derived from the UV spectrum
of \hdtwo\ are fully compatible with those derived from the optical spectrum.
In most cases, the abundances derived from lines of the neutral species
are lower by $\approx$~0.1--0.3~dex than the
abundances derived from lines of the corresponding ions.
NLTE abundance calculations for metal-poor stars
suggest this difference is primarily due to overionization
(e.g., 
Ca:\ \citealt{mashonkina17ca};
Ti:\ \citealt{sitnova20};
Cr:\ \citealt{bergemann10cr};
Mn:\ \citealt{bergemann08};
Fe:\ \citealt{bergemann12};
Co:\ \citealt{bergemann10};
Cu:\ \citealt{korotin18}).
Few NLTE predictions have been made for
the UV transitions examined here
or the Fe-group elements in metal-poor, red horizontal-branch stars.
The observed abundance behaviors
are consistent with the magnitude and direction of the
trends presented in the literature.
The ionized fraction always dominates 
the neutral fraction for atoms of these elements
in the atmosphere of \hdtwo,
so we regard the ions
as more reliable abundance indicators.

Figure~\ref{fegroupplot} also illustrates general
properties among the Fe-group element abundances in \hdtwo.
Sc, Ti, and V
are collectively enhanced relative to Fe.
The correlated abundances among these three elements
continue a trend among metal-poor stars identified by
\citet{sneden16}, \citet{cowan20}, and \citet{ou20}.
Cr, Mn, Co, Ni, and Zn
are found in approximately Solar ratios relative to Fe.
Cu is subsolar relative to Fe, by $\approx$0.6~dex.
All of these ratios are broadly consistent with 
general chemical-evolution patterns found among Milky Way stars
at this metallicity, as summarized by \citet{roederer18c}.
Our results indicate that the elements with $Z \leq$~30 in 
\hdtwo\ are typical for a metal-poor halo star.

\subsection{Elements with $Z \geq 31$}

The UV spectrum of \hdtwo\ is rich in heavy-element absorption lines.
We identify several rarely detected \rpro\ elements,
including Ga, Ge, As, Se, Cd, In, Sn, Sb, Te, W, Re, Os, Ir, Pt, and Au.
We present evidence of their detection in Appendix~\ref{appendix}.
Figures~\ref{specplot1}--\ref{specplot3} illustrate the synthesis
of representative lines of these and other species.

\begin{figure*}
\begin{center}
\includegraphics[angle=0,width=3.2in]{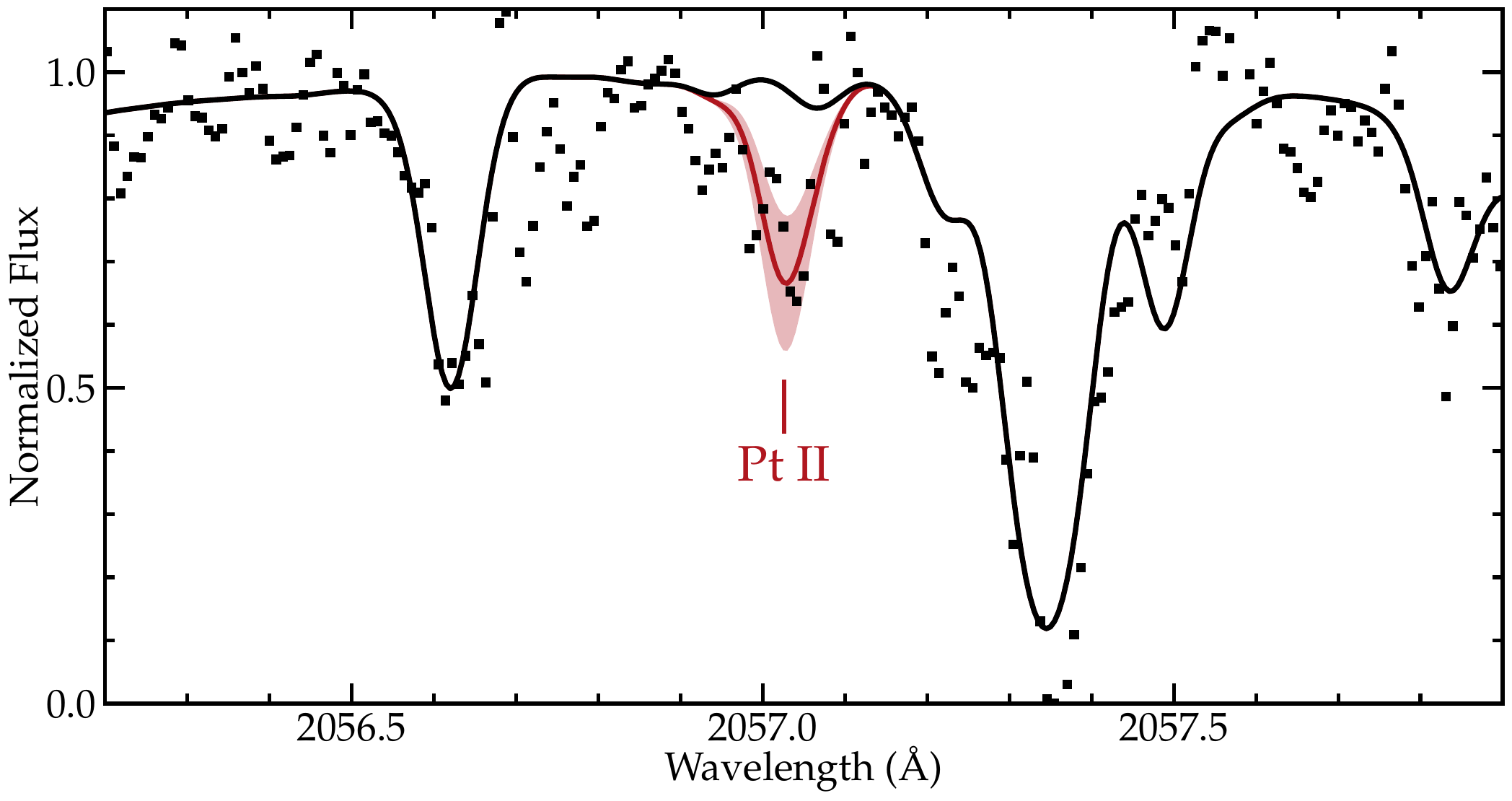} 
\hspace*{0.05in}
\includegraphics[angle=0,width=3.2in]{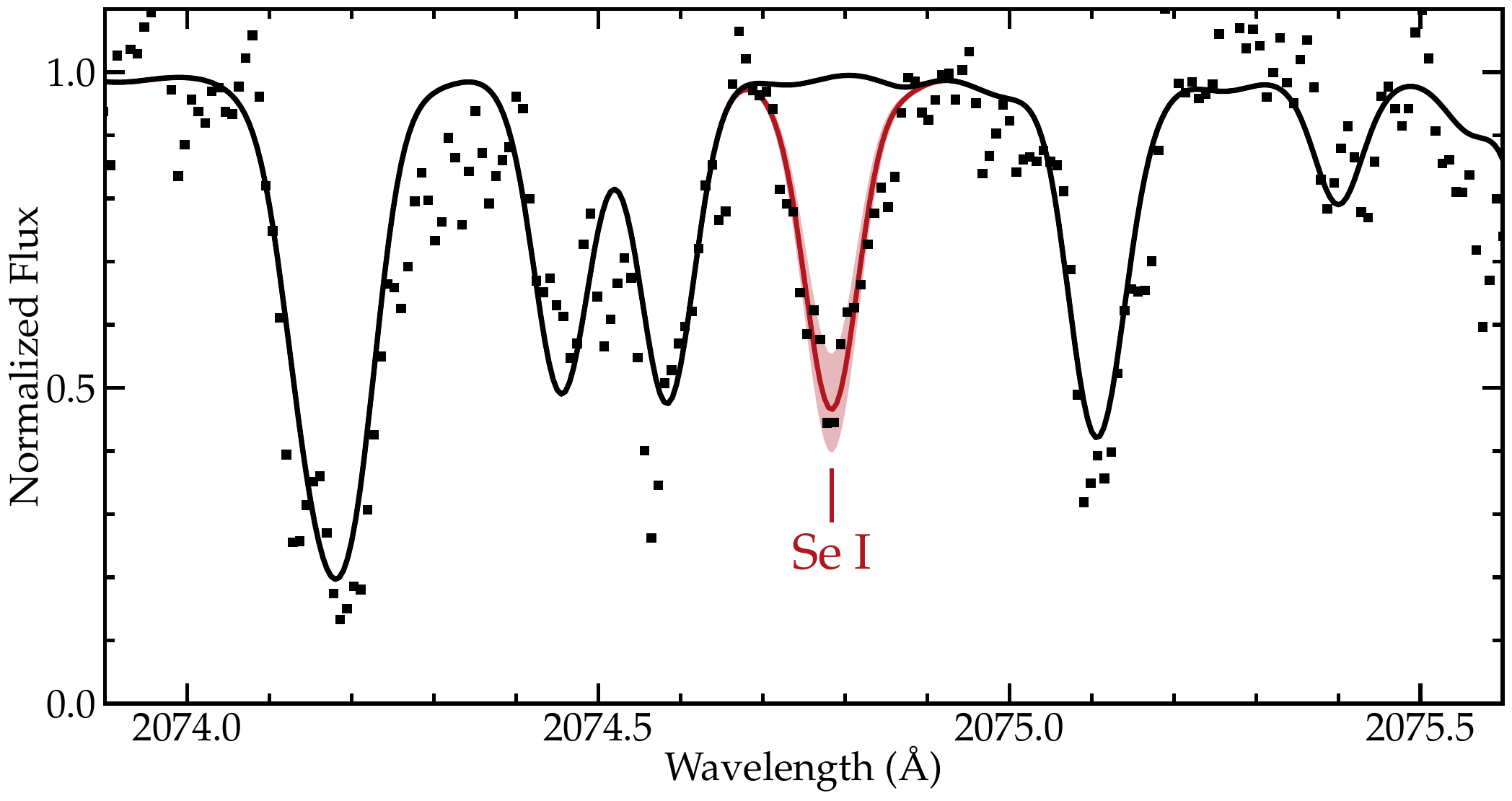} \\
\vspace*{0.05in}
\includegraphics[angle=0,width=3.2in]{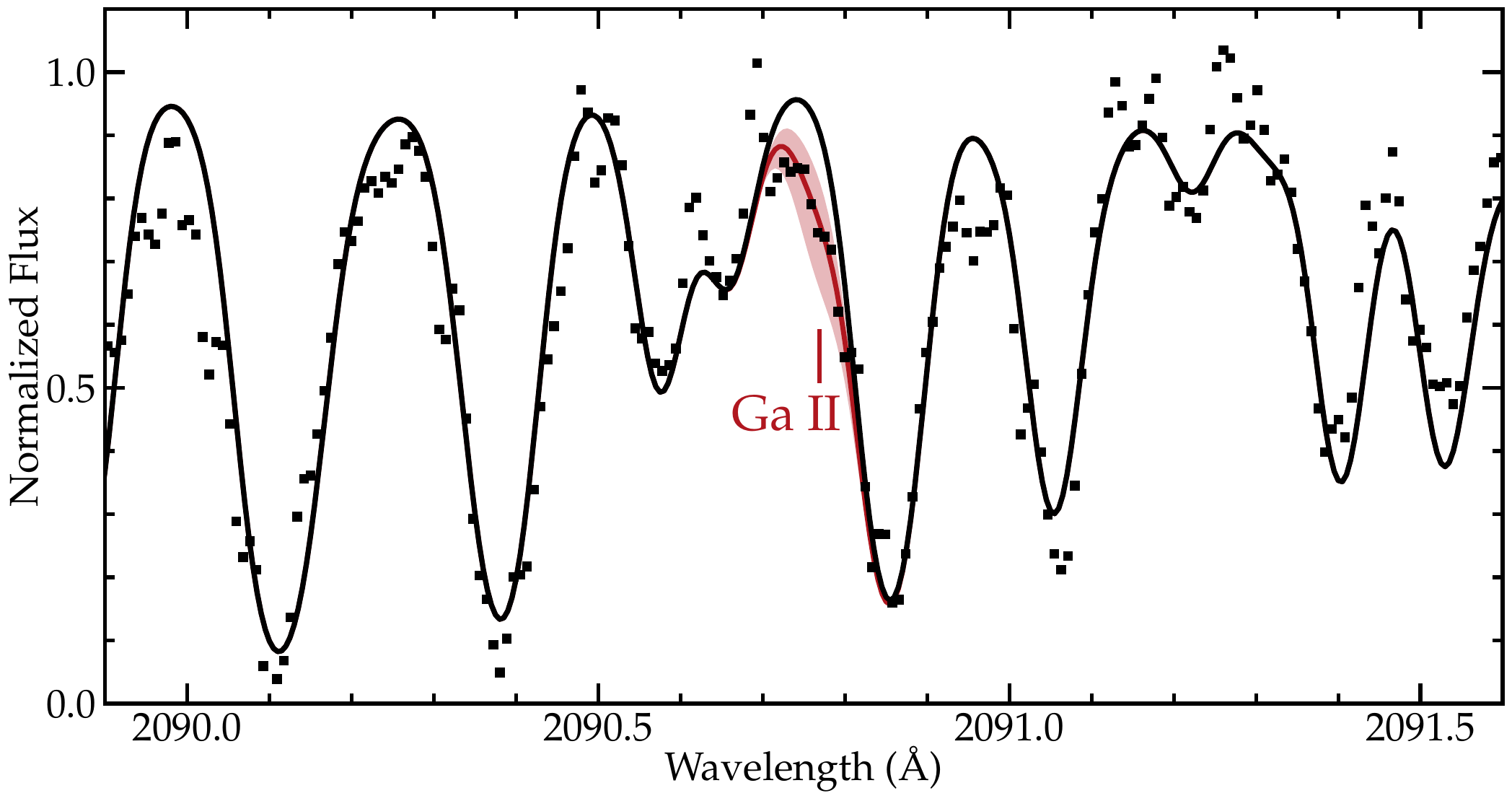} 
\hspace*{0.05in}
\includegraphics[angle=0,width=3.2in]{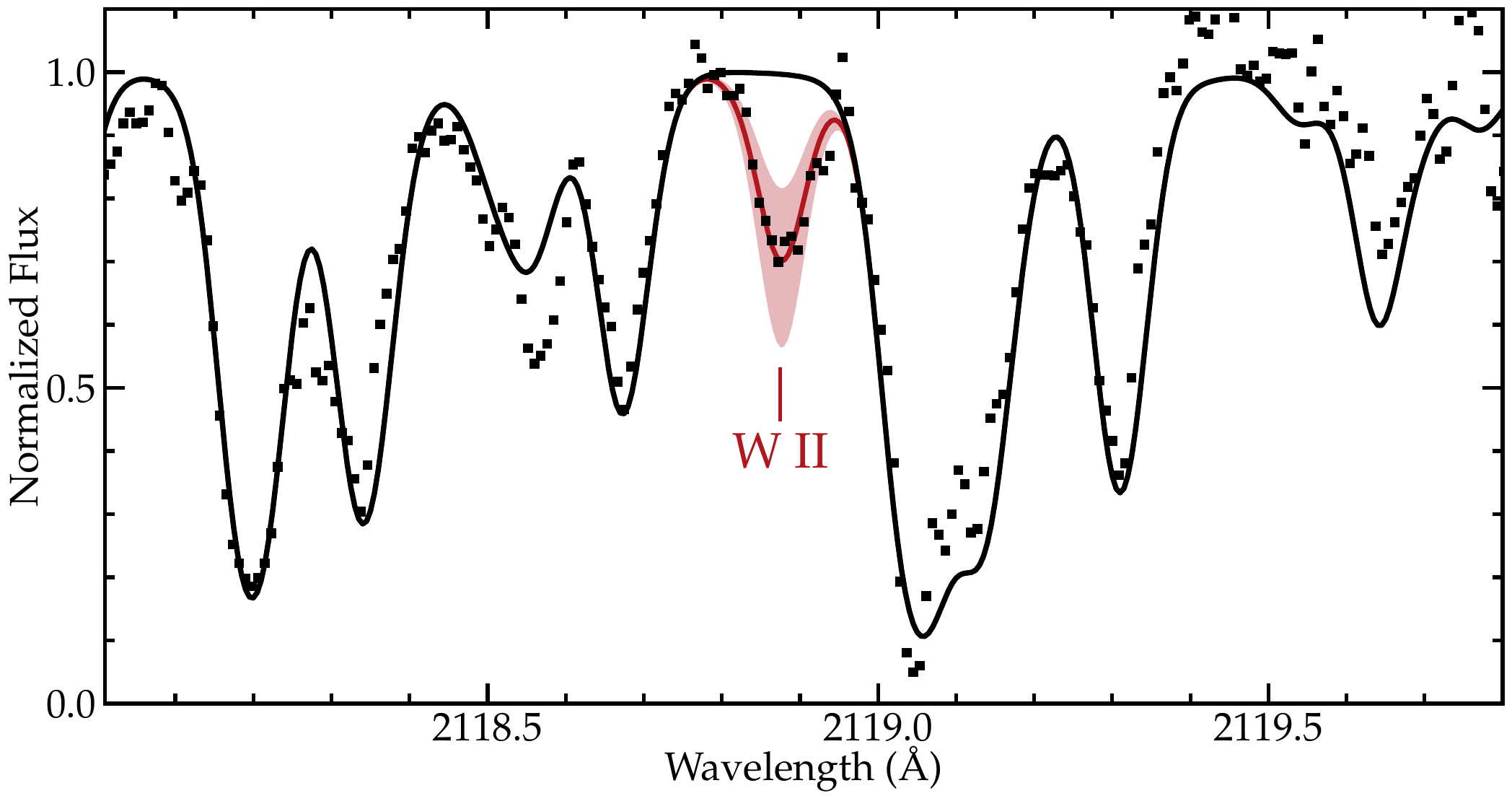} \\
\vspace*{0.05in}
\includegraphics[angle=0,width=3.2in]{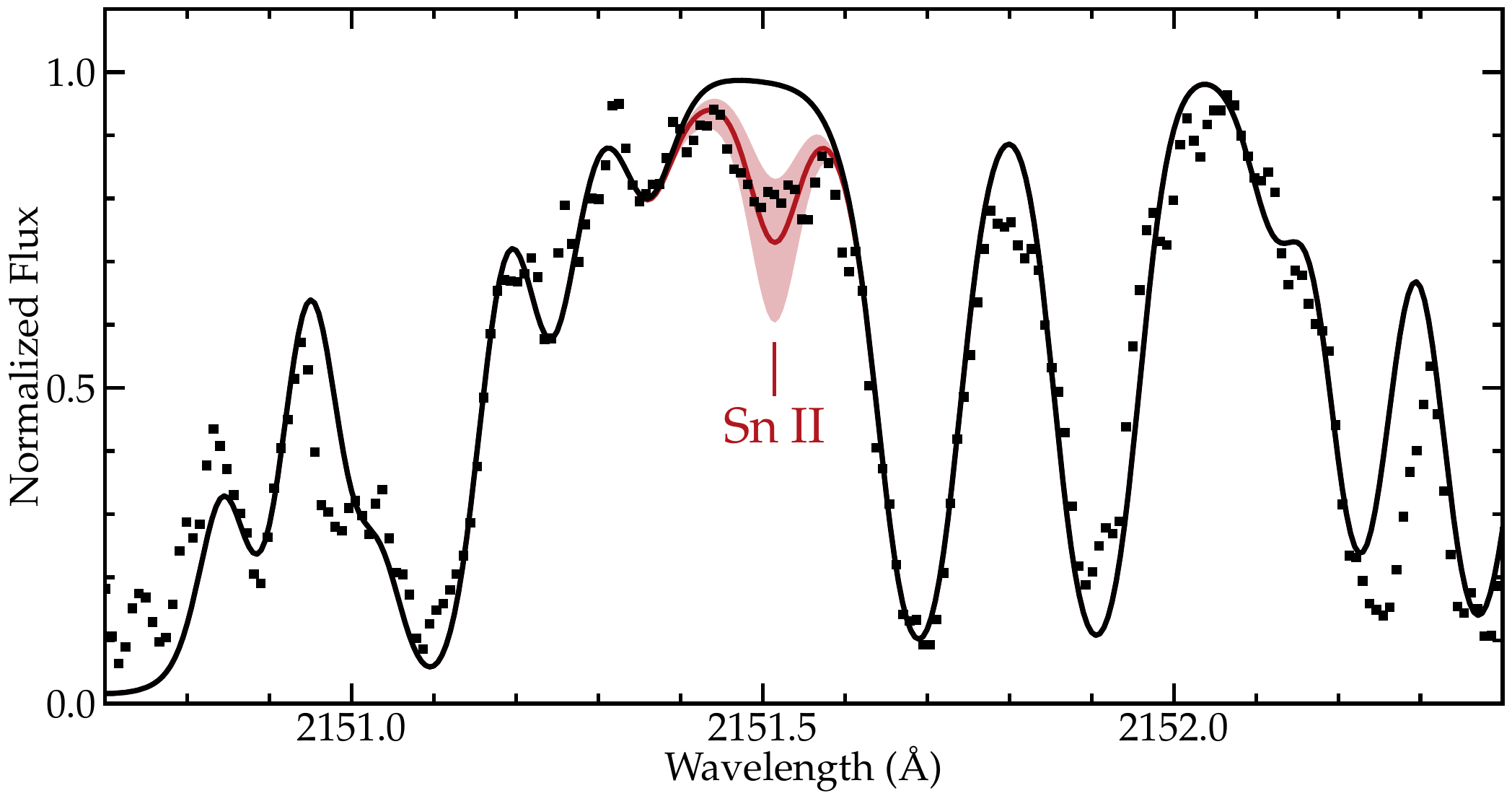} 
\hspace*{0.05in}
\includegraphics[angle=0,width=3.2in]{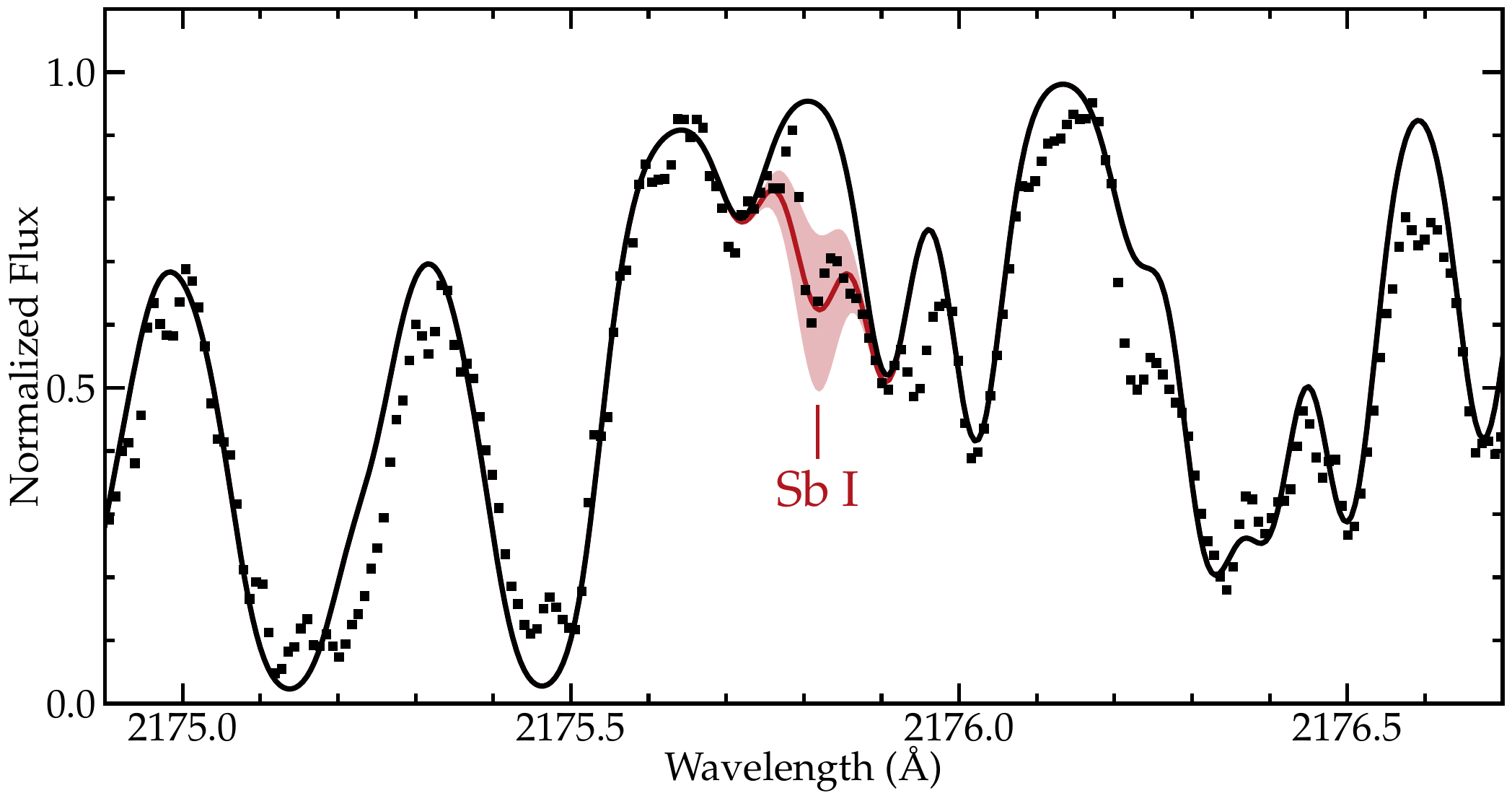} \\
\vspace*{0.05in}
\includegraphics[angle=0,width=3.2in]{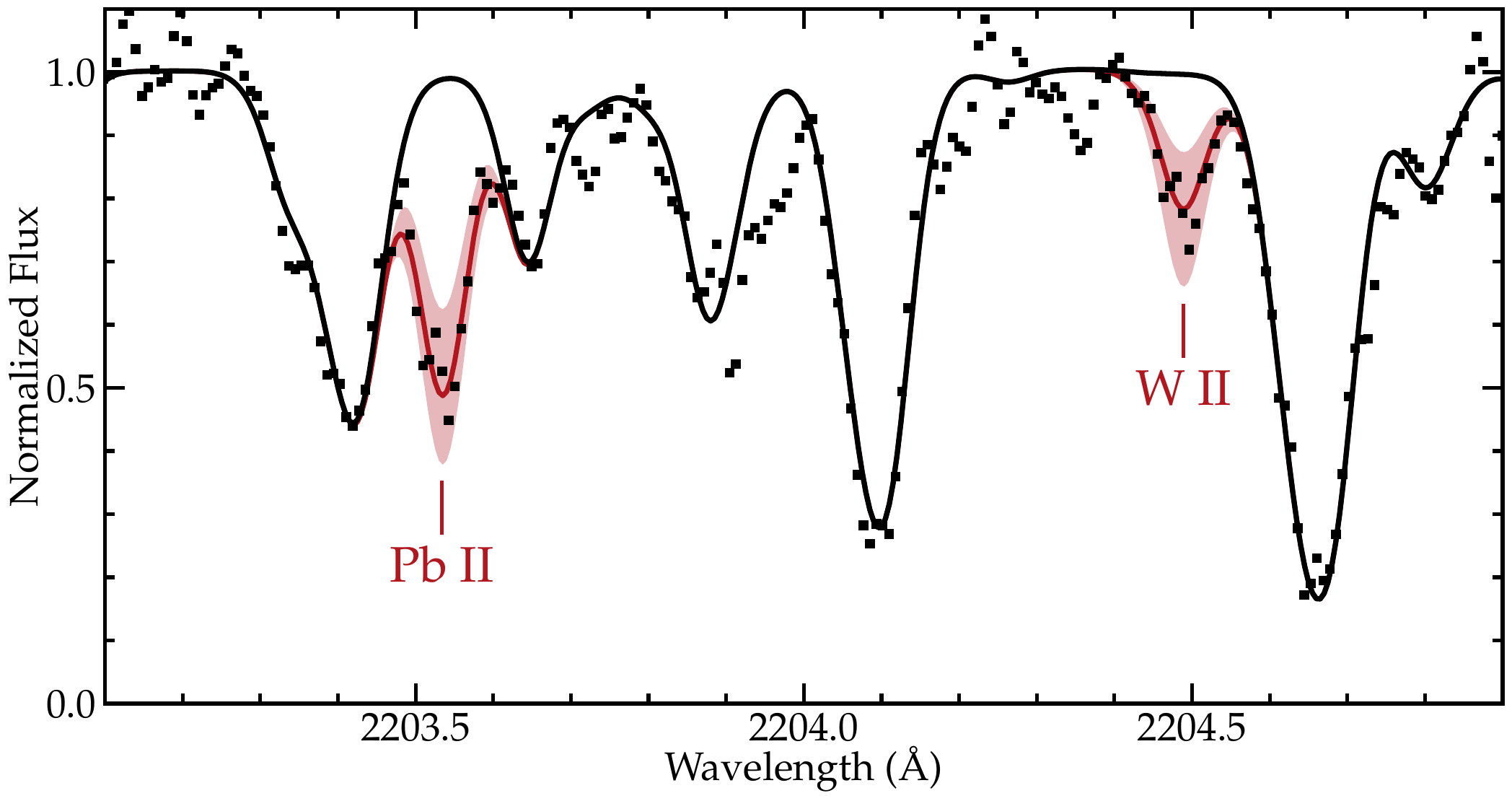} 
\hspace*{0.05in}
\includegraphics[angle=0,width=3.2in]{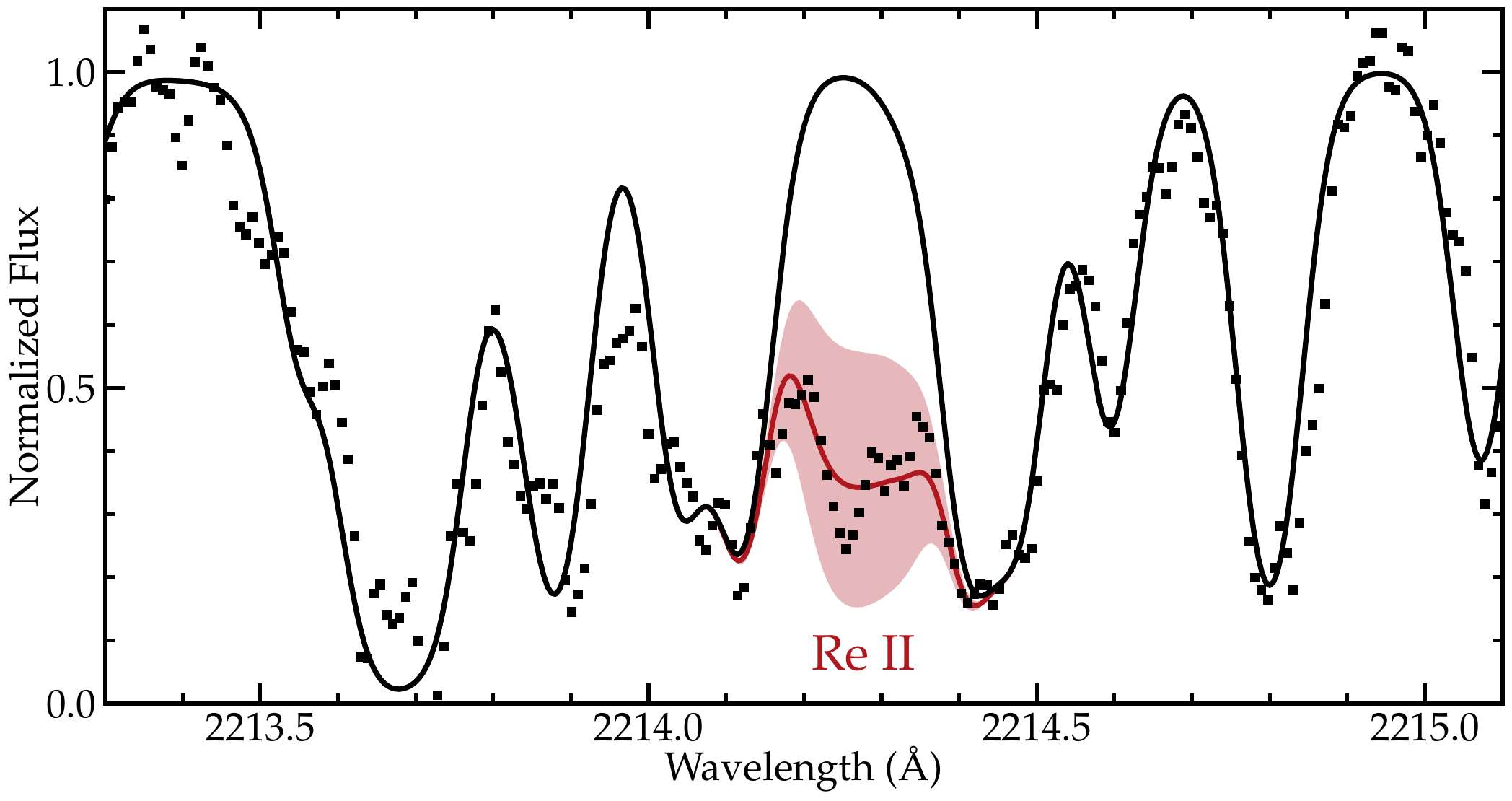} 
\end{center}
\caption{
\label{specplot1}
STIS E230H spectra of \hdtwo\ compared with model spectra
around lines of interest, which are marked and labeled.
The thick red line indicates the best-fit abundance for the line(s)
of interest.
The light red bands indicate variations in this abundance by
$\pm$~0.3~dex ($\approx$1.5--2.0 times the typical uncertainties),
to facilitate visibility.
The thick black line indicates a model that excludes the element of interest.
}
\end{figure*}

\begin{figure*}
\begin{center}
\includegraphics[angle=0,width=3.2in]{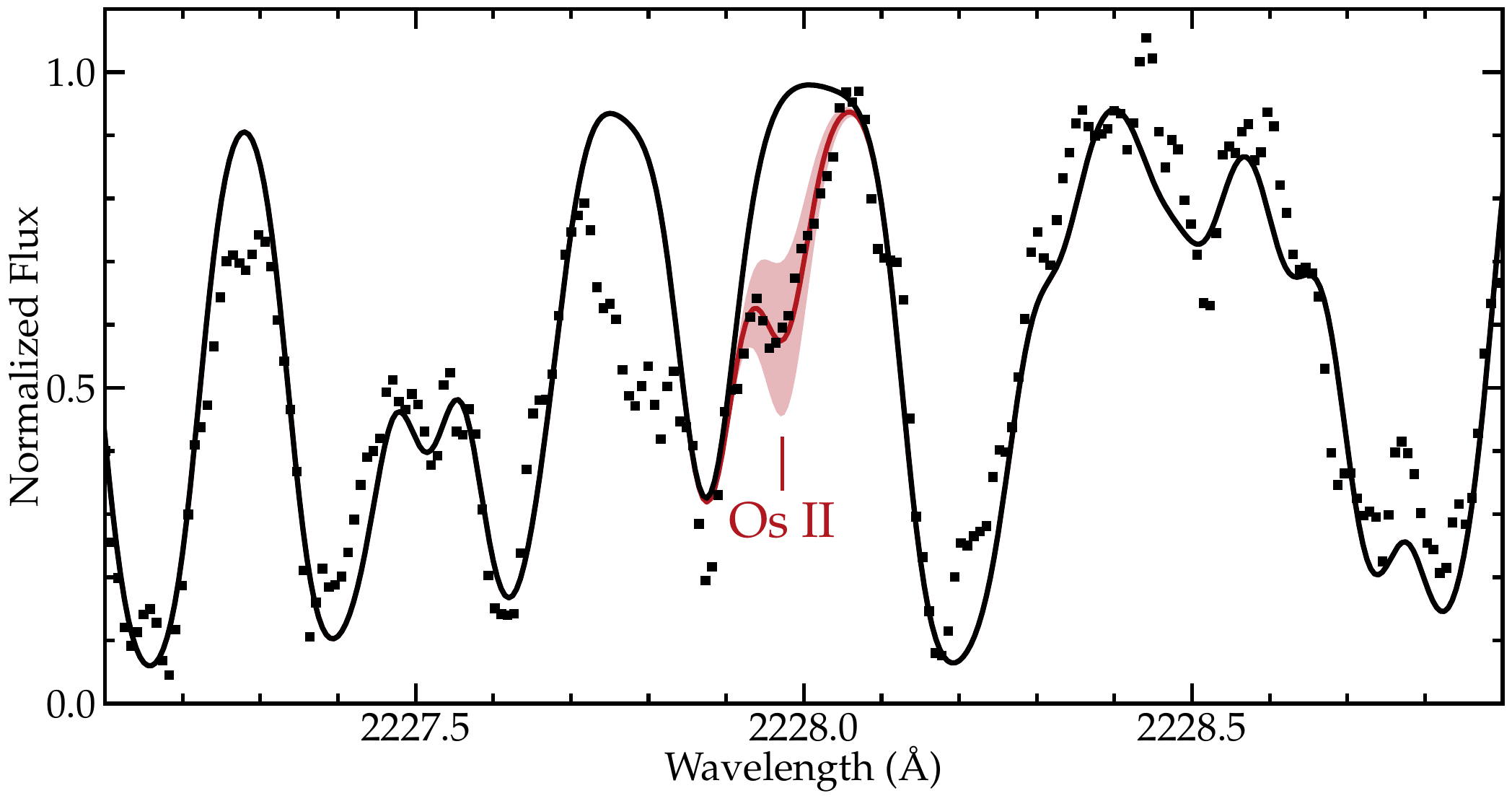} 
\hspace*{0.05in}
\includegraphics[angle=0,width=3.2in]{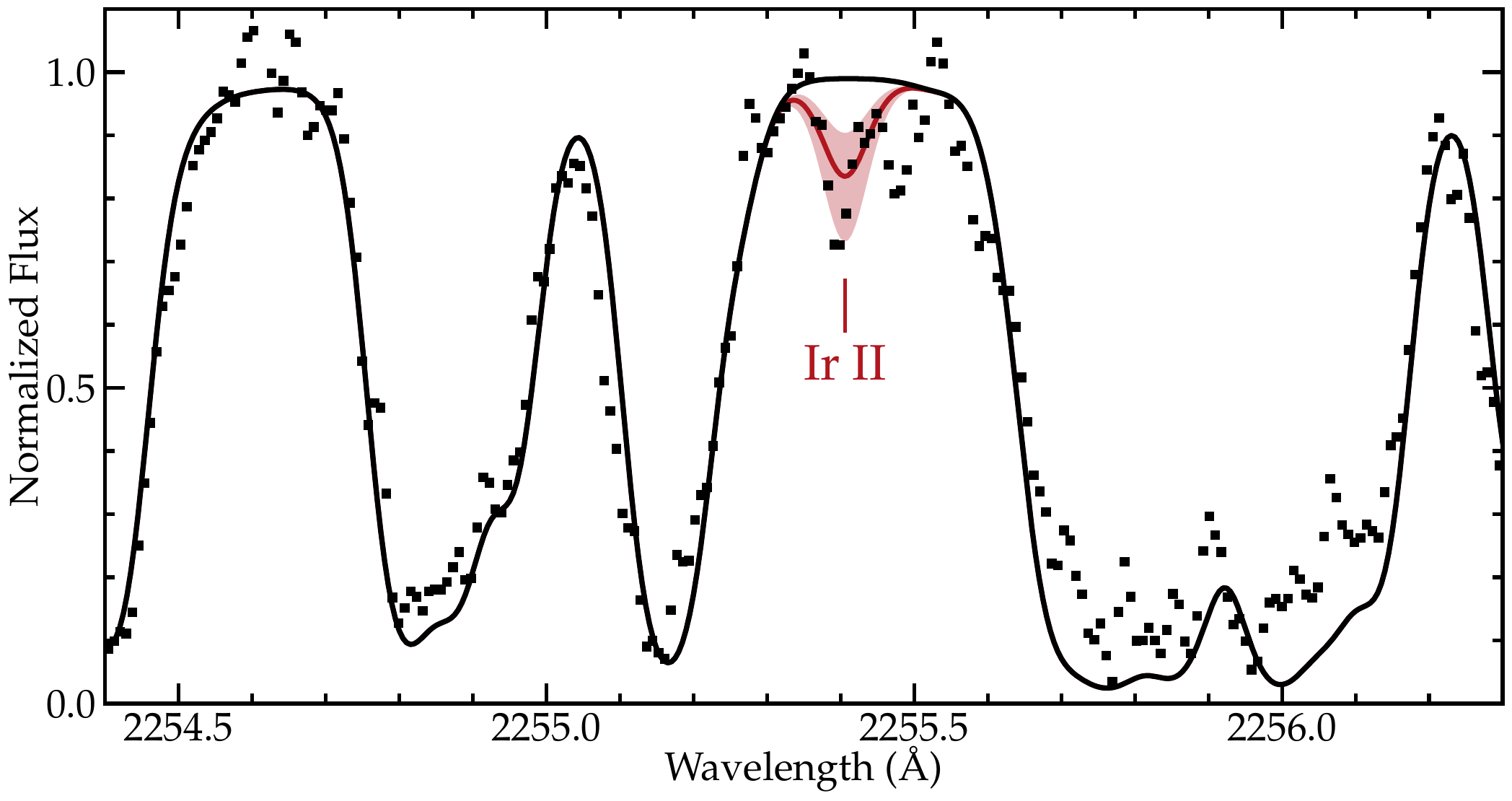} \\
\vspace*{0.05in}
\includegraphics[angle=0,width=3.2in]{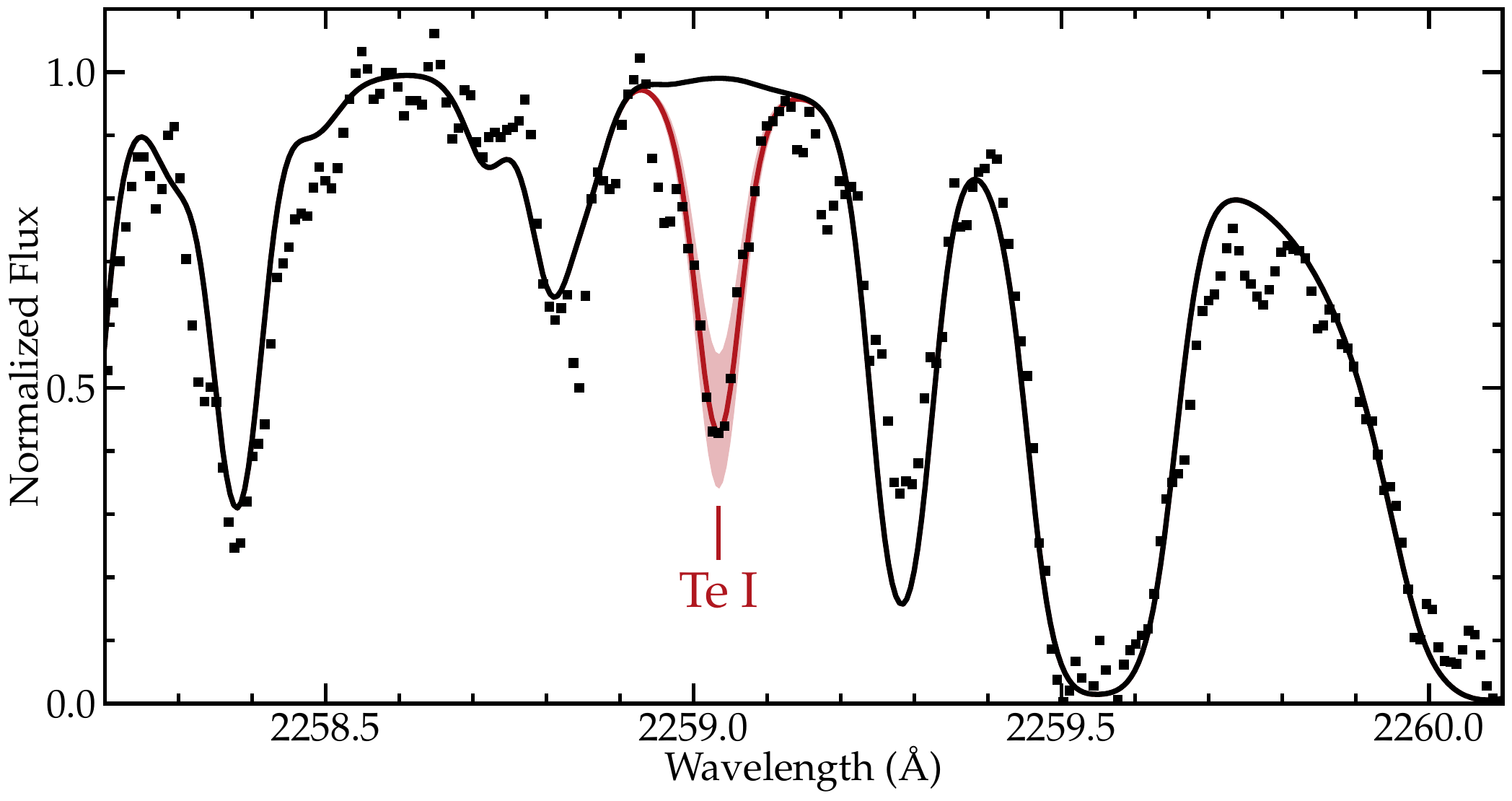} 
\hspace*{0.05in}
\includegraphics[angle=0,width=3.2in]{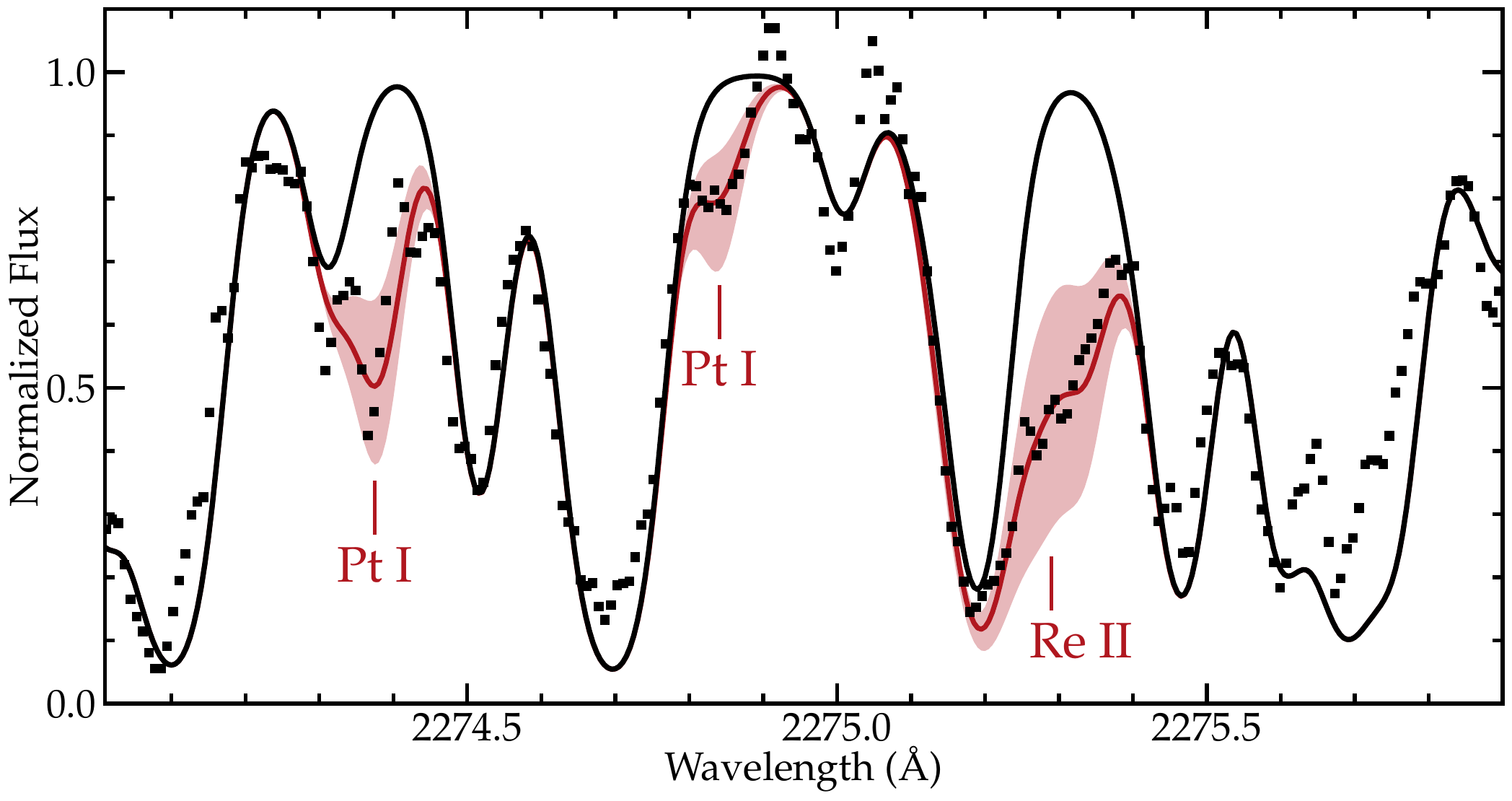} \\
\vspace*{0.05in}
\includegraphics[angle=0,width=3.2in]{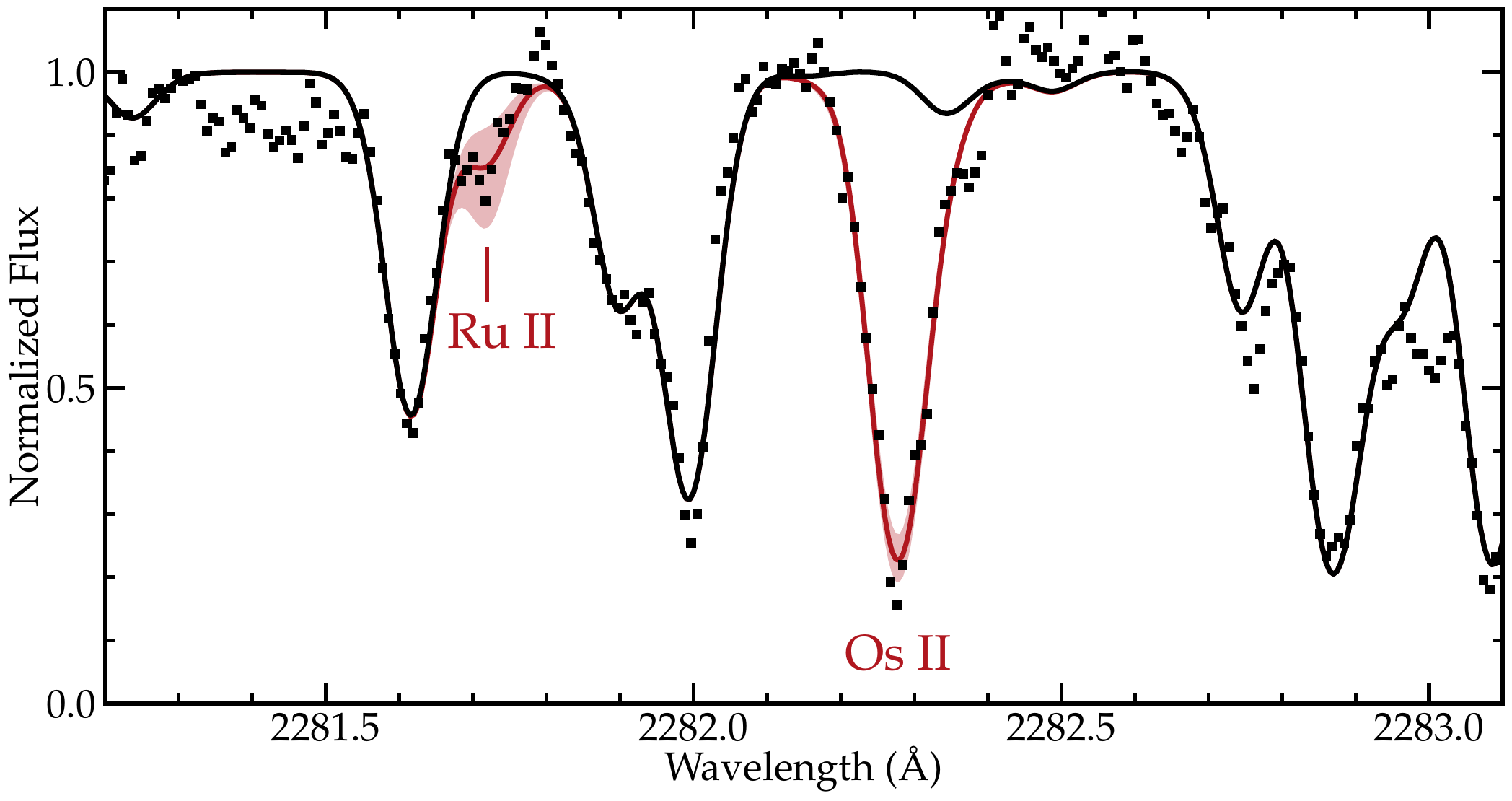} 
\hspace*{0.05in}
\includegraphics[angle=0,width=3.2in]{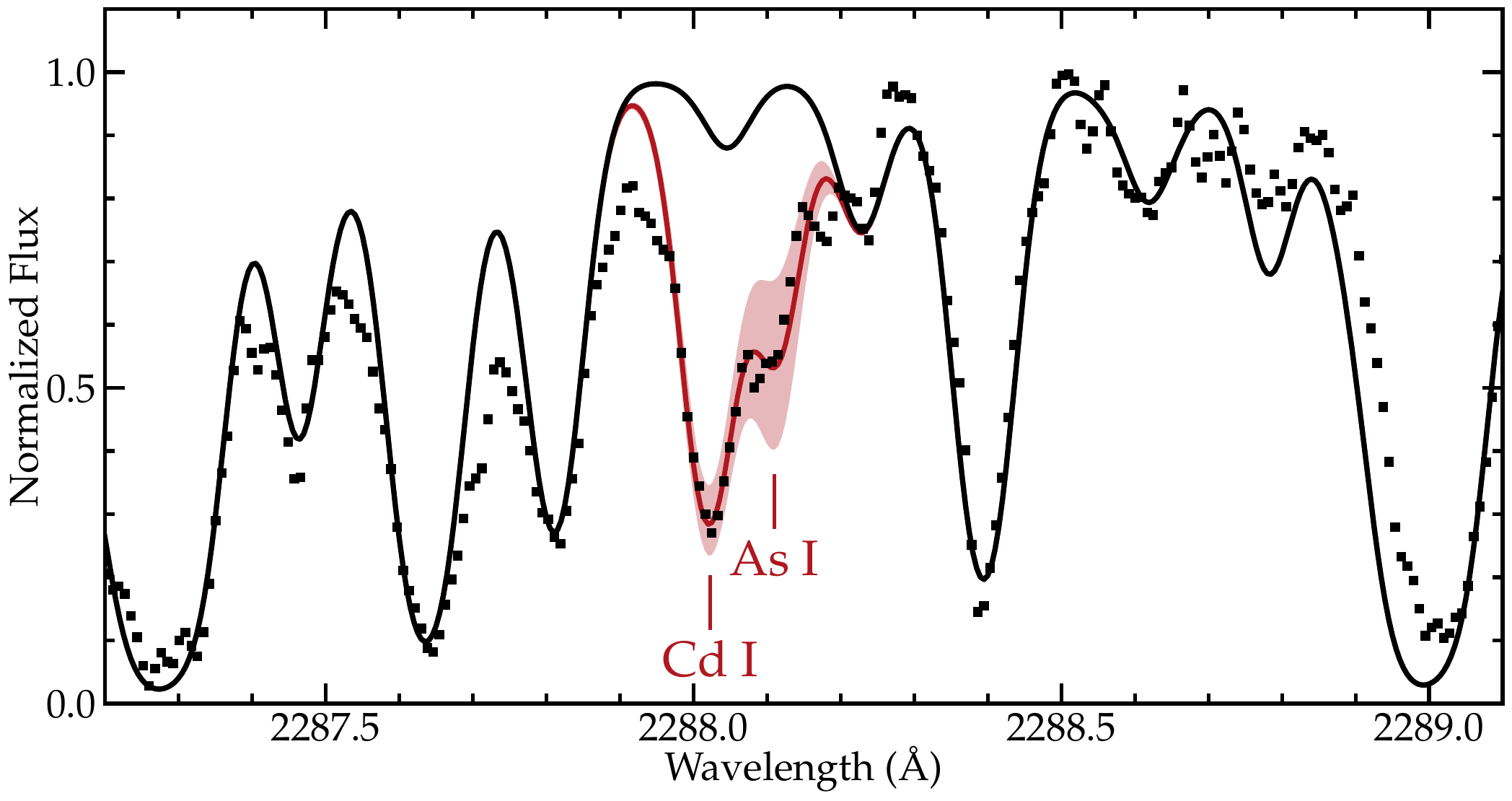} \\
\vspace*{0.05in}
\includegraphics[angle=0,width=3.2in]{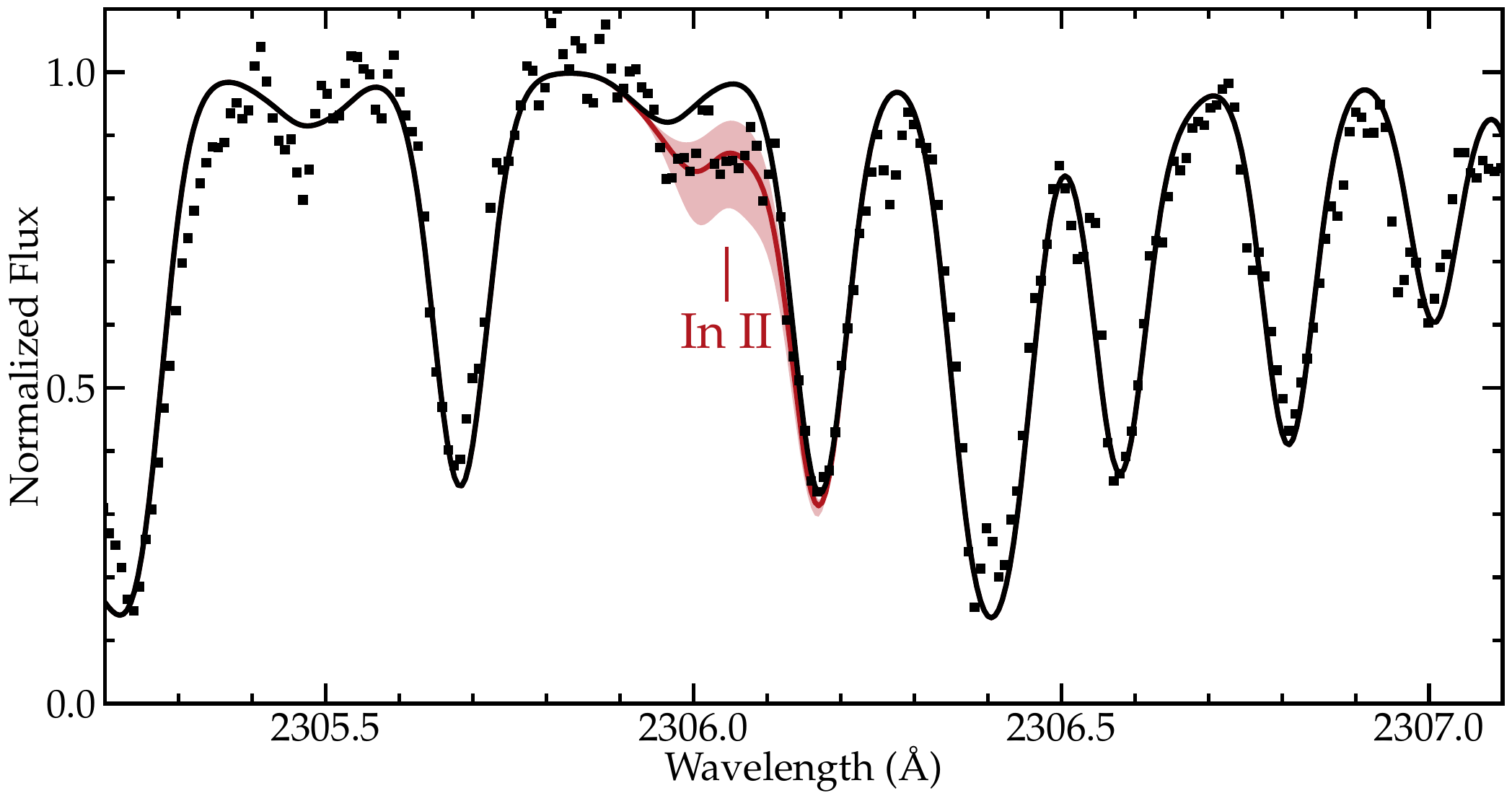} 
\hspace*{0.05in}
\includegraphics[angle=0,width=3.2in]{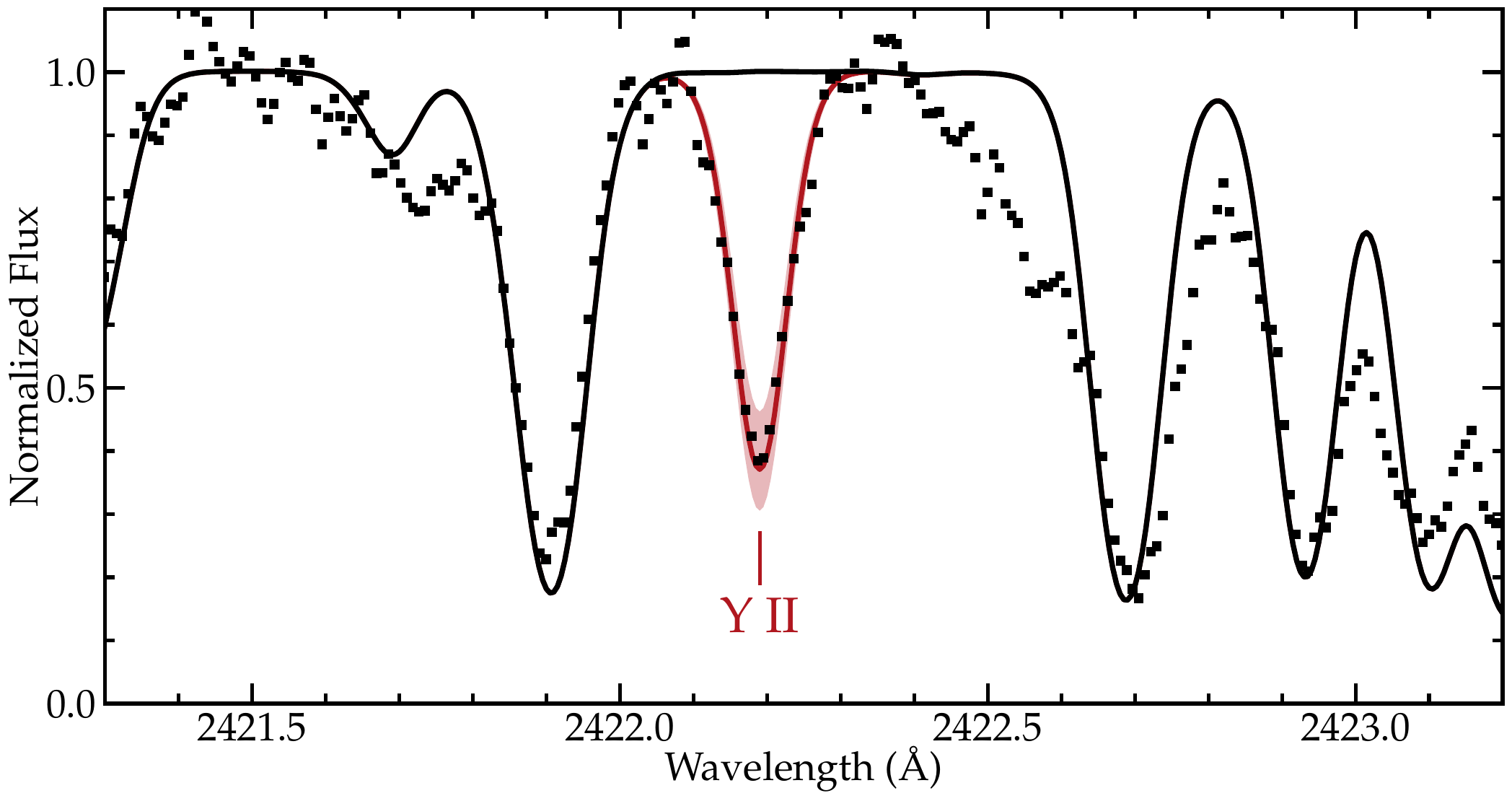} 
\end{center}
\caption{
\label{specplot2}
STIS E230H spectra of \hdtwo\ compared with model spectra
around lines of interest.
The symbols are the same as in Figure~\ref{specplot1}.
}
\end{figure*}

\begin{figure*}
\begin{center}
\includegraphics[angle=0,width=3.2in]{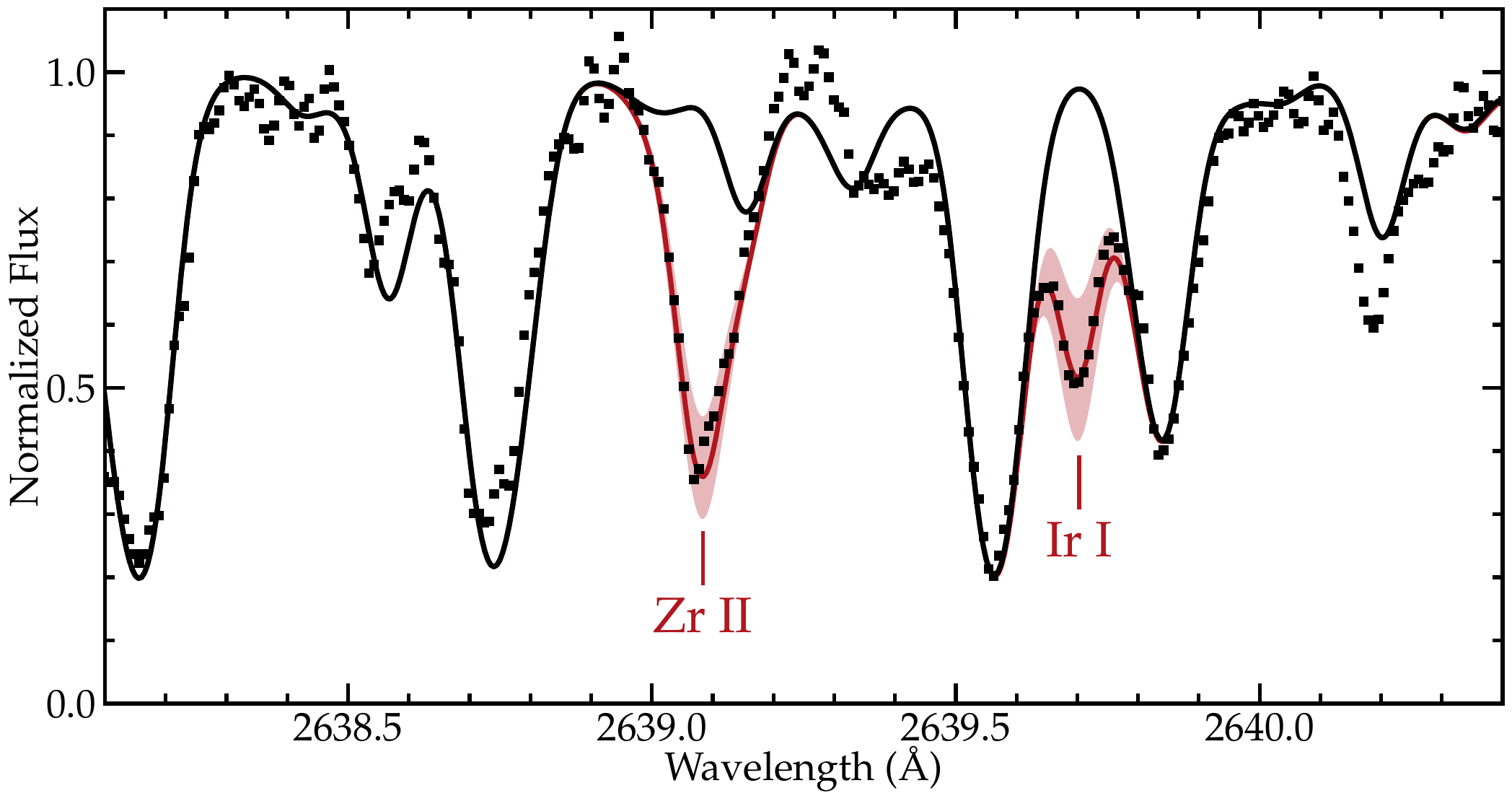} 
\hspace*{0.05in}
\includegraphics[angle=0,width=3.2in]{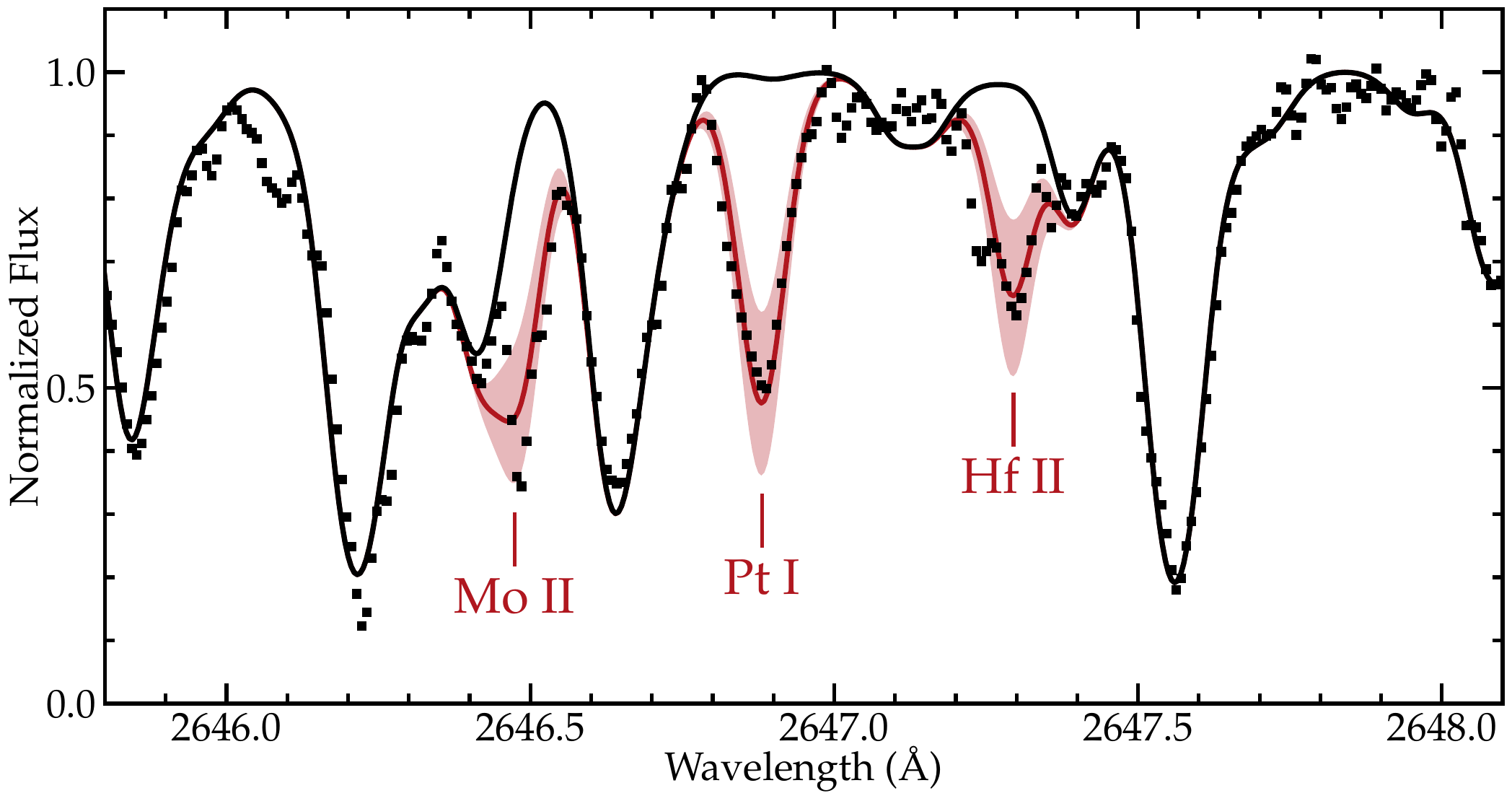} \\
\vspace*{0.05in}
\includegraphics[angle=0,width=3.2in]{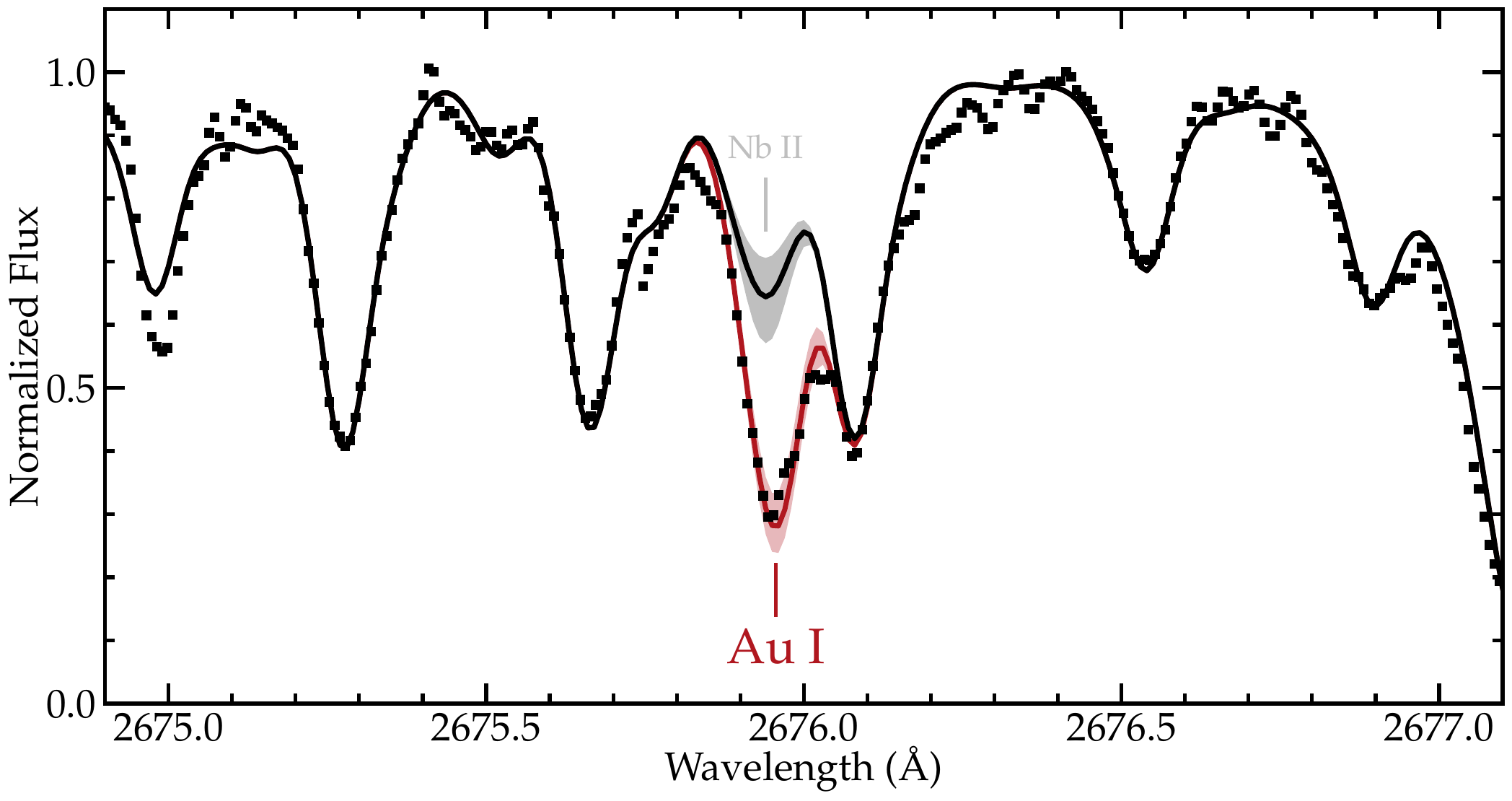} 
\hspace*{0.05in}
\includegraphics[angle=0,width=3.2in]{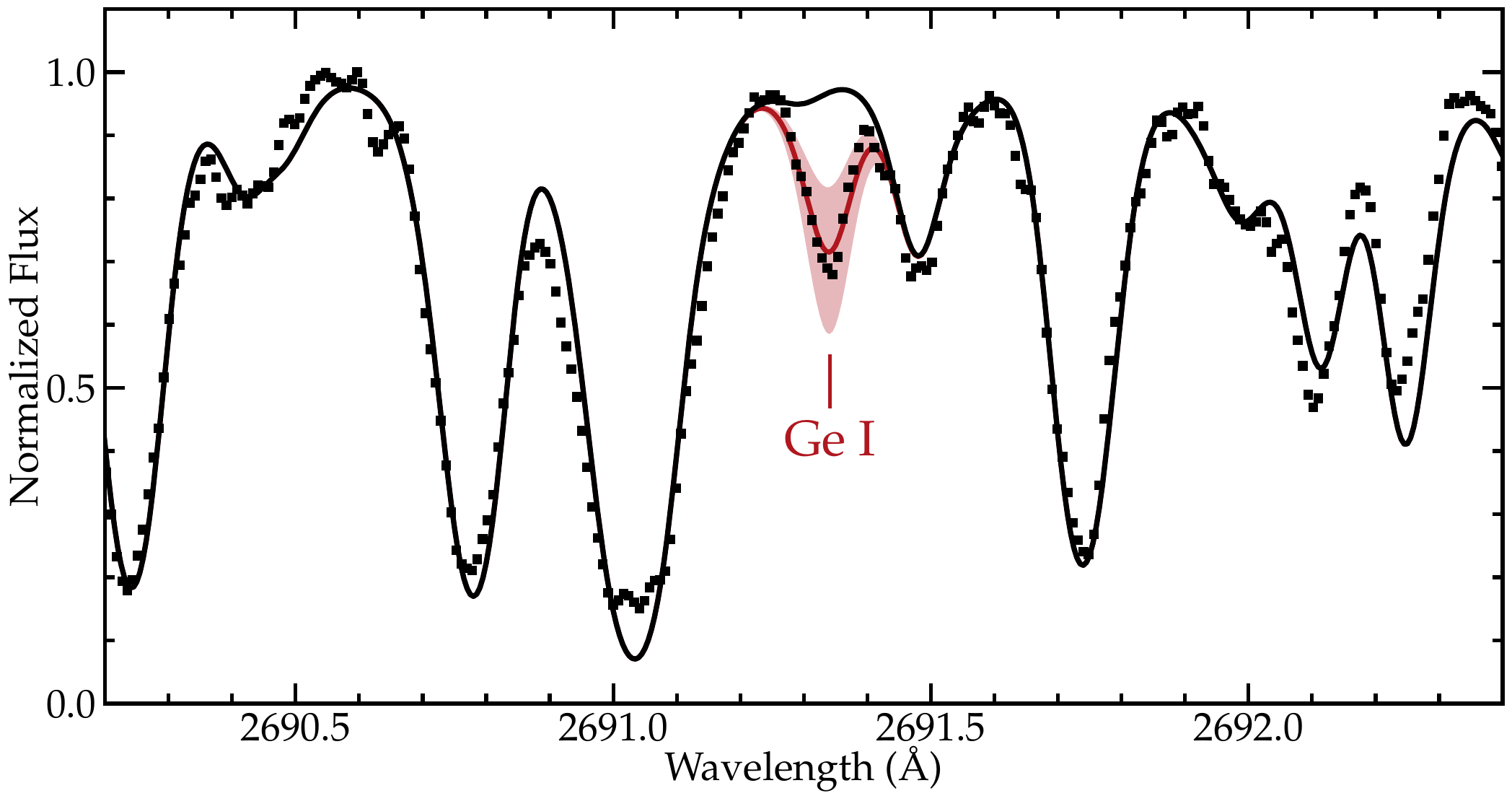} \\
\vspace*{0.05in}
\includegraphics[angle=0,width=3.2in]{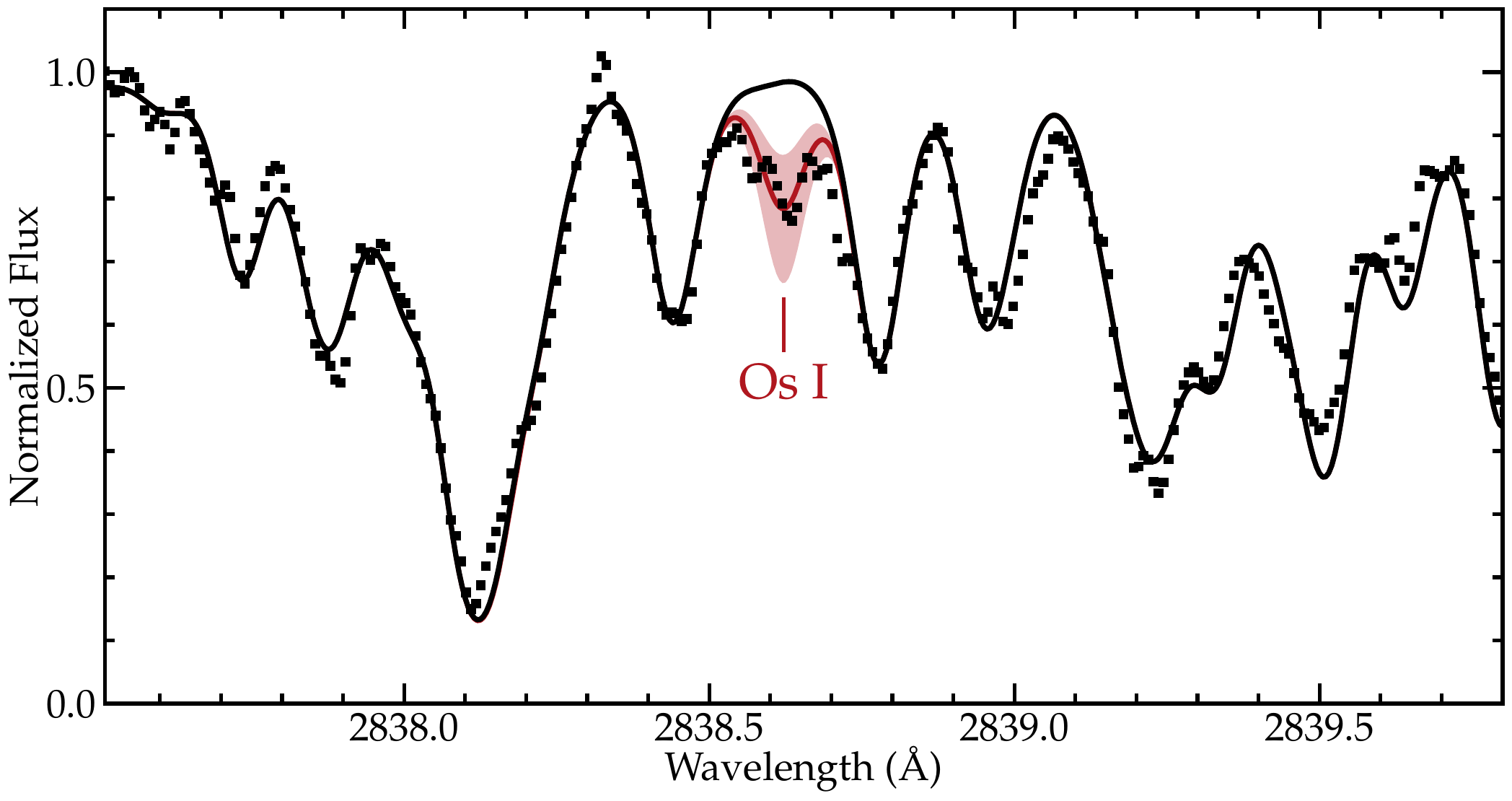} 
\hspace*{0.05in}
\includegraphics[angle=0,width=3.2in]{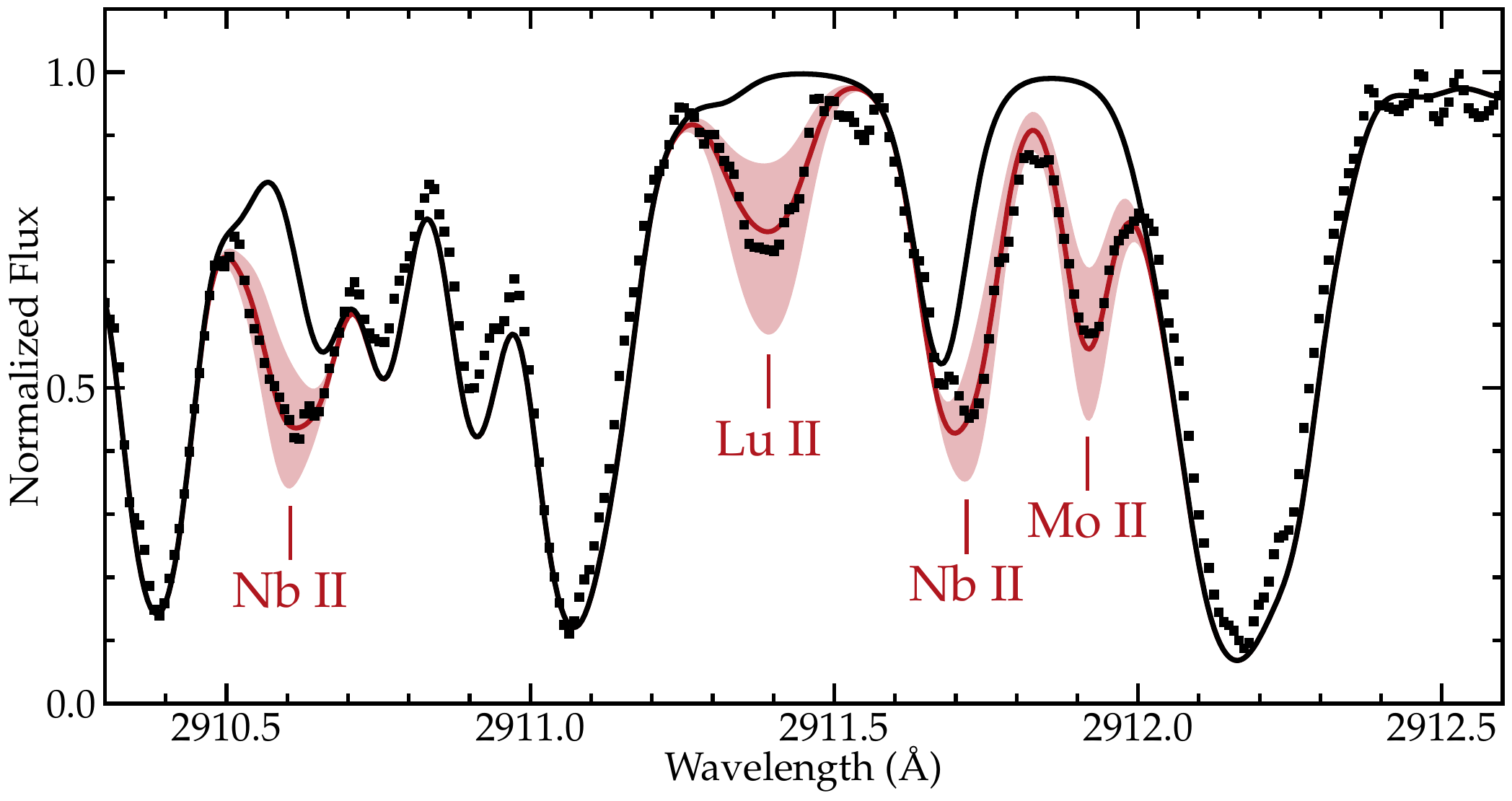} \\
\vspace*{0.05in}
\includegraphics[angle=0,width=3.2in]{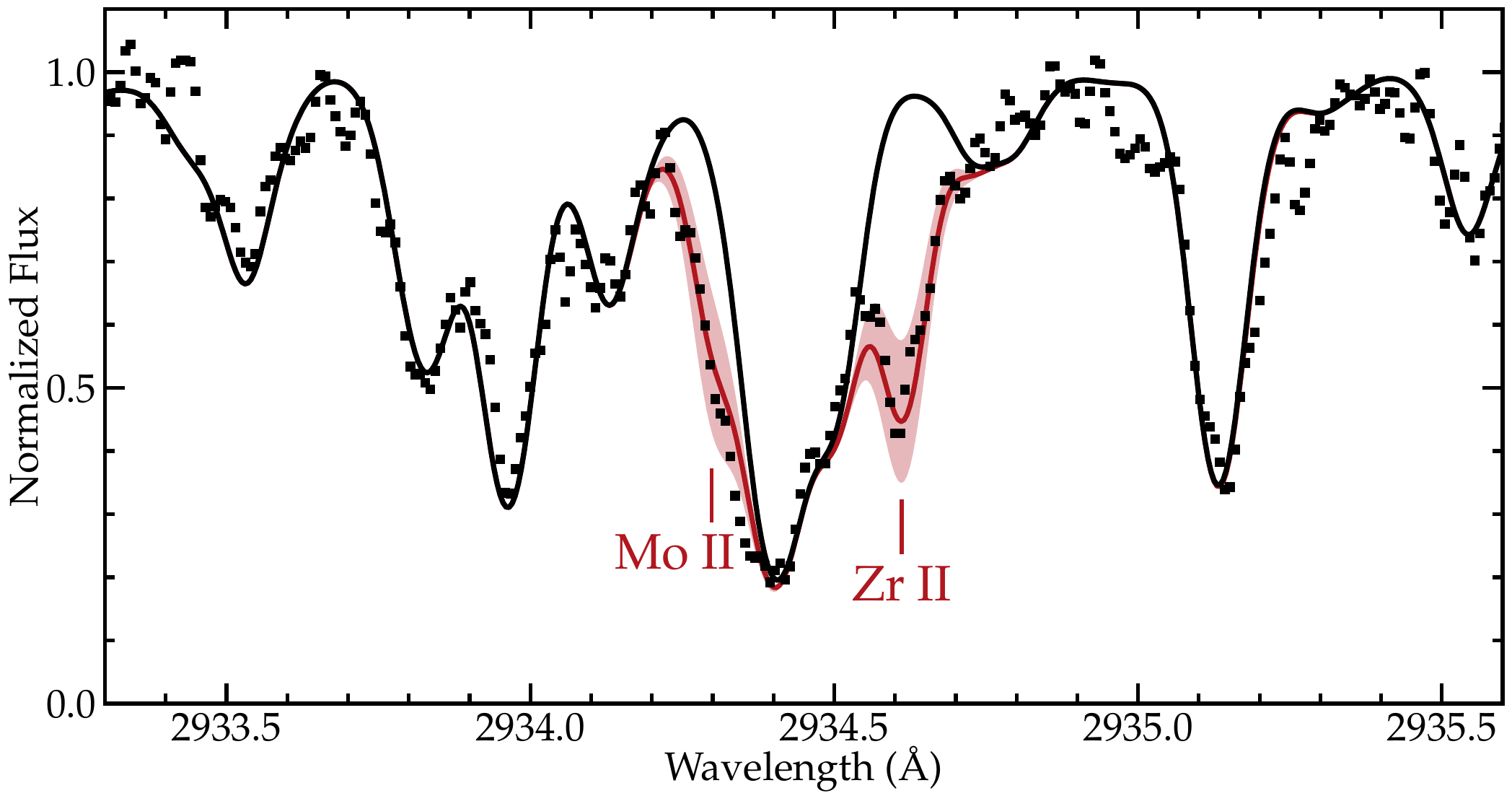} 
\hspace*{0.05in}
\includegraphics[angle=0,width=3.2in]{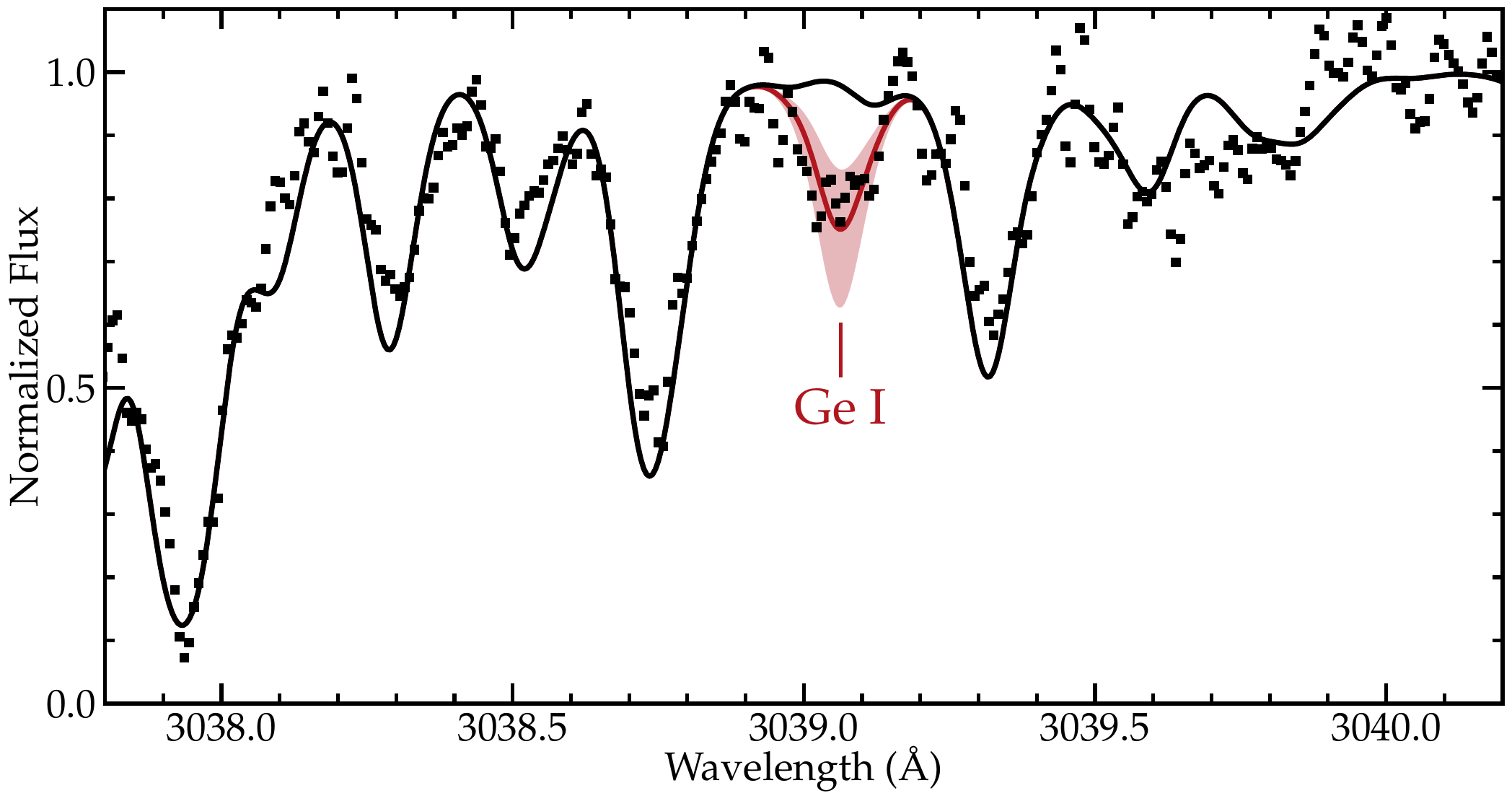} 
\end{center}
\caption{
\label{specplot3}
STIS E230H spectra of \hdtwo\ compared with model spectra
around lines of interest.
The symbols are the same as in Figure~\ref{specplot1}.
The gray band in the second panel on the left  
marks a change in the blending Nb~\textsc{ii} line by 
its uncertainty, $\pm$~0.20~dex.
}
\end{figure*}

Table~\ref{finalabundtab} presents our recommended abundances for
each element.
These values are considered on a case-by-case basis,
including our best attempts to account for
elements detected in multiple ionization states, 
poorly studied NLTE effects, 
missing or poor atomic data,
problematic lines, small numbers of lines detected, 
and other issues.
We encourage readers who are interested in the 
details of how we arrive at these recommended abundances
to consult the relevant sections in Appendix~\ref{appendix}.

\subsection{Uncertainties}

All abundance uncertainties are computed following the method
presented in \citet{roederer18c}.
We draw $10^{3}$ samples from a normal distribution centered
on the adopted value of each model atmosphere parameter,
and interpolate a new model atmosphere for each of those draws.
We approximate the equivalent width of 
each line through a reverse curve-of-growth analysis,
based on the abundance derived from the spectrum synthesis.
We draw $10^{3}$ new equivalent widths for each line,
assuming a normal distribution of uncertainties 
related to the S/N and the goodness of the synthetic spectrum fit.
The \loggf\ of each line is also resampled
$10^{3}$ times from a normal distribution of uncertainties,
based on NIST grades or uncertainties quoted in the original literature.
New abundances are computed for each resample.
The 16th and 84th percentiles of the resulting distributions
are roughly symmetric, and the uncertainties
quoted in Tables~\ref{abundtab} and \ref{finalabundtab}
represent 1$\sigma$ uncertainties in the abundance ratios.

\section{Results}
\label{results}

\subsection{Elements Detected in HD~222925}

A total of 63~elements are detected in \hdtwo, plus H,
which is detectable through the Balmer series lines, and He,
which was detected previously by \citet{navarrete15}.
This tally includes
42~elements with $31 \leq Z \leq 92$ produced by the \rpro.
We also report upper limits on the abundances
of 7 elements, including 4 produced by the \rpro.
The rich UV and optical spectra of \hdtwo\ enable a nearly complete
characterization of the \rpro\ abundance pattern
in a star whose atmosphere retains the heavy-element abundance pattern
of its natal cloud.

These numbers represent a substantial improvement upon previous efforts.
Among \rpro-enhanced stars, for example,
\citet{hill02}, \citet{plez04}, 
and \citet{siqueiramello13} derived abundances
for 37 \rpro\ elements in
\object[BPS CS 31082-001]{CS~31082--001} (54~elements in total, plus H),
\citet{sneden03a} derived abundances
for 31 \rpro\ elements in 
\object[BPS CS 22892-052]{CS~22892--052} (52~elements in total, plus H), and 
\citet{roederer12d,roederer14c,roederer14d} derived abundances
for 35 \rpro\ elements in 
\object[HD 108317]{HD~108317} (52~elements in total, plus H).
Among chemically peculiar stars with strong metal-line spectra,
abundances have been derived 
for 51~elements (plus H) in
the rapidly oscillating peculiar A (roAp) star
\object[HD 101065]{HD~101065}, also known as Przybylski's star
\citep{cowley00},
and 
abundances have been derived for
54~elements (plus H) in the bright metallic-line (Am) star 
\object[Sirius]{Sirius} \citep{landstreet11,cowley16}. 
To the best of our knowledge, \hdtwo\ presents 
both the most complete set of \rpro\ elements
and the most complete set of abundances overall
for any object beyond the solar system.

\begin{figure*}
\begin{center}
\includegraphics[angle=0,width=5.6in]{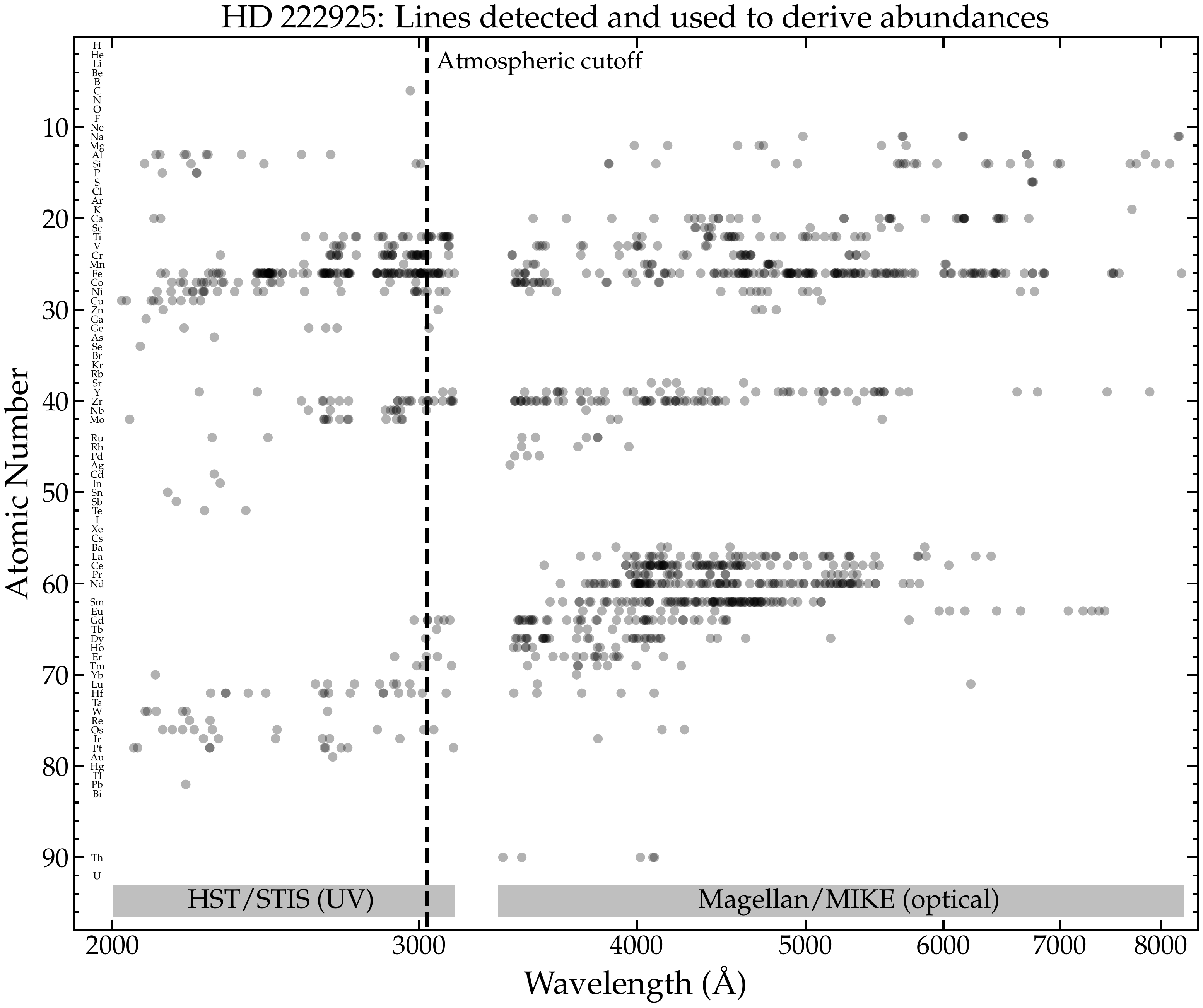} 
\end{center}
\caption{
\label{linesplot}
Illustration of the wavelengths of
the lines that are detected and used to derive the abundances in \hdtwo.
Each dot represents a line.
The atmospheric cutoff is marked and labeled.
The gap between $\approx$3145~\AA\ and $\approx$3330~\AA\ 
reflects the gap between our STIS and MIKE spectra of \hdtwo.
}
\end{figure*}

\begin{figure*}
\begin{center}
\includegraphics[angle=0,width=5.6in]{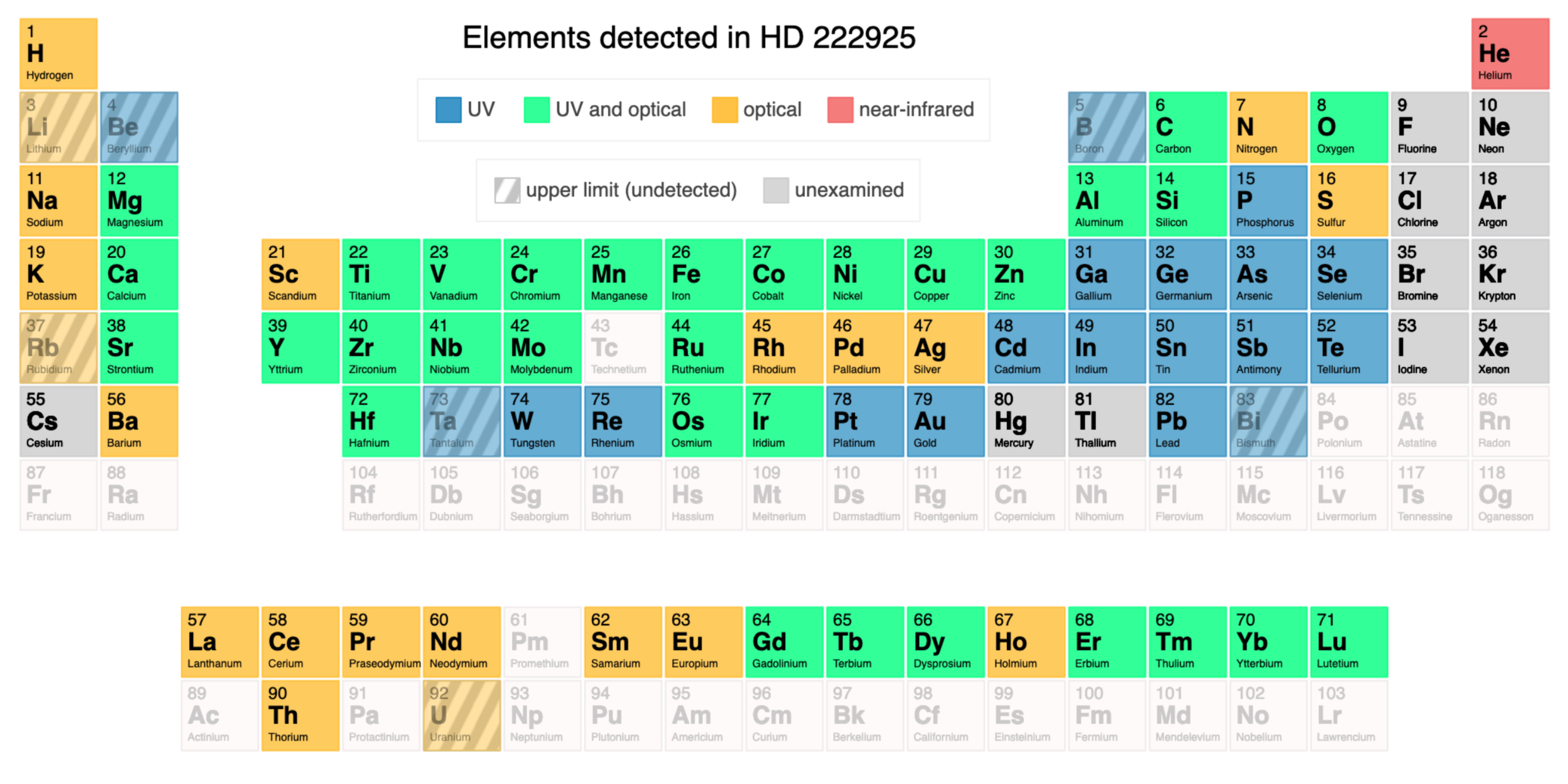}
\end{center}
\caption{
\label{periodictableplot}
Periodic table showing the elements examined in \hdtwo.
Elements with no long-lived isotopes are 
indicated using light gray font.
He was previously detected in near-infrared spectra by \citet{navarrete15}.
}
\end{figure*}

High-resolution UV spectroscopy is essential to this advance.
Figures~\ref{linesplot} and \ref{periodictableplot} 
illustrate this role in two complementary ways.
Figure~\ref{linesplot} illustrates
the wavelengths of the lines that are detected and used 
to derive the abundances in the UV and optical spectra of \hdtwo.
It emphasizes the \textit{number of lines} available in the 
UV spectrum.
Figure~\ref{periodictableplot} shows
a periodic table that emphasizes the \textit{number of elements}
detectable in the UV region of the spectrum.
Many of these elements are detected for the first time
in a highly \rpro-enhanced star.
Abundances of some elements
(e.g., Nb, Lu, Hf, Os, and Ir)
are determined more reliably,
because the UV spectral range
presents a substantial increase in the number of lines available.
Other elements 
(e.g., Al, Co, Ni, Cu, Mo, Ru, and Pb)
benefit because the UV spectral range
permits the detection of the dominant ionization state.
The optical and near-infrared spectral ranges
cannot replicate 
the rich diversity of lines found in the UV spectra of 
stars such as \hdtwo.

\subsection{Elements Unexamined in HD~222925}
\label{unexamined}

Among all stable elements between H and U, 
only 11 remain unexamined in \hdtwo:\
fluorine (F, $Z = 9$),
neon (Ne, $Z = 10$), 
chlorine (Cl, $Z = 17$), 
argon (Ar, $Z = 18$),
bromine (Br, $Z = 35$), 
krypton (Kr, $Z = 36$),
iodine (I, $Z = 53$),
xenon (Xe, $Z = 54$),
cesium (Cs, $Z = 55$), 
mercury (Hg, $Z = 80$), and
thallium (Tl, $Z = 81$).
These elements are labeled as ``unexamined''
in Figure~\ref{periodictableplot}.
Their neutral and ionized species
present no transitions in optical or UV
spectra with a realistic chance of detection.
Elements with no long-lived isotopes also
remain unexamined in \hdtwo:\
technetium (Tc, $Z = 43$),
promethium (Pm, $Z = 61$), 
polonium through actinium (Po--Ac, $Z = 84$--89),
protactinium (Pa, $Z = 91$), and
all transuranic elements ($Z \geq 93$).

We predict that two more elements, I and Hg, may be detectable
in \hdtwo.
The I~\textsc{i} line at 2061.633~\AA\ is blended with a strong
Cr~\textsc{ii} line, but I~\textsc{i} lines at 1830.380 and 1844.453~\AA\
may be strong enough to permit detections.
The Hg~\textsc{ii} resonance line at 1942.273~\AA\ is in a region
of our UV spectrum where the S/N is too low to permit a 
reliable detection,
and the Hg~\textsc{i} resonance line at 2536.521~\AA\ is too blended
to detect or derive an upper limit.
Future UV observations with a more sensitive telescope
should enable detection of the 
I~\textsc{i} $\lambda\lambda$1830 and 1844 and
the Hg~\textsc{ii} $\lambda$1942 lines in \hdtwo.

\subsection{The \textit{R}-process Abundance Pattern}
\label{rpropattern}

\begin{figure*}
\begin{center}
\includegraphics[angle=0,width=5.0in]{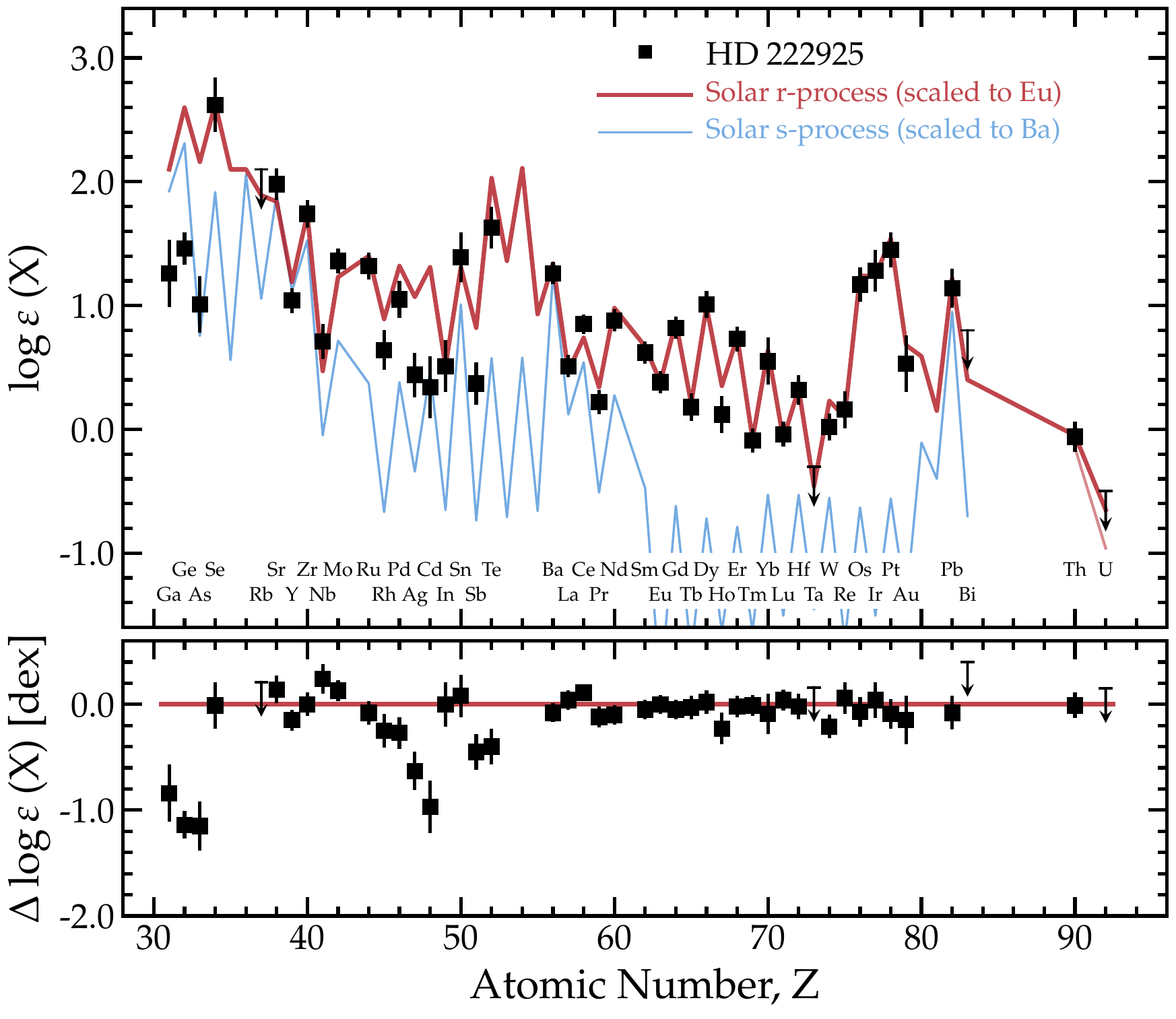}
\end{center}
\caption{
\label{patternplot}
Top:\ the \rpro\ abundance pattern in \hdtwo.
The Solar \spro\ pattern 
(the thin blue line, scaled to match the Ba abundance) and
the \rpro\ residuals 
(the thick red line, scaled to match the Eu abundance)
are shown for comparison 
(\citealt{sneden08}, except Y, which is adopted from \citealt{bisterzo14}).
The detected elements are marked by the filled black squares,
and the upper limits derived from non-detections are marked 
by the arrows.
The light red line at Th and U accounts for 13~Gyr of 
radioactive decay; i.e., 8.5~Gyr of additional decay relative 
to the Solar abundances.
Bottom:\ the difference between the \hdtwo\ abundances
and the Solar \rpro\ residuals when scaled to Eu.
}
\end{figure*}

Figure~\ref{patternplot} illustrates the
heavy-element abundance pattern of \hdtwo.
The red line marks the Solar \rpro\ residuals,\footnote{The 
Solar abundances listed in Table~1 of \citet{sneden08}
should include 
N[s] = 0.055 and N[r] = 0.172 for $^{99}$Ru
and
N[r] = 0.373 for $^{136}$Xe
(J.\ Cowan 2022, private communication).}
scaled downward by 
0.11~dex to match the 
Eu abundance in \hdtwo.
The Solar \rpro\ residuals and the observed pattern
are in near-perfect agreement for the elements with 
$Z \geq 56$.
The mean difference is 
$-0.05 \pm 0.02$~dex ($\sigma = 0.08$~dex, or 17\%).
Ho is the most discrepant, 
0.23 $\pm$~0.12~dex below the scaled Solar \rpro\ residual pattern.
This difference is $< 2\sigma$ significant.
As discussed in Appendix~\ref{ree},
we find no reason to discount our
Ho abundance derivation.

The elements with $Z \leq 52$ behave differently.
The deviations from the \rpro\ residuals for the
lighter elements
span nearly 1.4~dex from As
($-1.15$~dex) to Nb ($+0.24$~dex),
regardless of the overall normalization.
The deviations do not appear to be random, and
three general trends emerge.

First, the lightest elements in this region, Ga, Ge, and As
($31 \leq Z \leq 33$),
are deficient by more than 0.8~dex relative to the 
scaled Solar \rpro\ pattern.
The abundance of Se ($Z = 34$), an element at the first \rpro\ peak,
agrees with the scaled Solar residual pattern.

Secondly, there is an 
overall decrease in the \hdtwo\ pattern relative to the 
\rpro\ residuals for Nb through Cd ($41 \leq Z \leq 48$).
The even-$Z$ element
Cd is also unusual in that its abundance is
comparable to or less than its odd-$Z$ neighbors,
Ag and In ($Z$ = 47 and 49).
As discussed in Appendix~\ref{cadmium}, 
we find no reason to discount our Cd abundance derivation.

Finally,
In, Sn, and Sb ($49 \leq Z \leq 51$), 
detected here for the first time in 
an \rpro-enhanced star, 
lie either on the pattern (In, Sn) or slightly below it (Sb)
when normalizing to Eu.
Te ($Z = 52$), an element at the second \rpro\ peak,
is deficient by 0.40~dex.

\section{Discussion}
\label{discussion}

We begin
by introducing several general environmental and chemical properties
of \hdtwo\ that inform our discussion of its 
heavy-element abundance pattern.
\begin{enumerate}[nolistsep]
\item
\hdtwo\ is on an eccentric orbit with
pericenter near 1.0 $\pm$~0.1~kpc,
apocenter near 16.6 $\pm$~0.6~kpc, and
maximum distance above the Galactic plane of 5.3 $\pm$~0.1~kpc
\citep{roederer18d}.
Its azimuthal angular momentum
is small and retrograde,
$J_{\phi} = -L_{z} = -380 \pm 40$~kpc~\kmsec,
suggesting that \hdtwo\ was accreted by the Milky Way,
perhaps from a satellite or star cluster
associated with the Gaia Sausage/Enceladus galaxy
\citep{belokurov18sausage,helmi18enceladus}
that was accreted at least $\approx$10~Gyr ago 
\citep{feuillet21,montalban21}.
\item
\hdtwo\ does not show evidence of radial velocity variations
over more than 7~yr of observations
(Appendix~\ref{rvappendix}),
indicating it is not likely to be
in a binary or multiple-star system.
\item 
All light ($Z \leq 30$) elements in \hdtwo\
exhibit abundance ratios typical for halo stars
with [Fe/H]~$\approx -1.5$,
indicating that the production 
of these elements was dominated by 
normal Type~II supernovae.
Neither C nor N are enhanced in \hdtwo\
([C/Fe] = $+0.14 \pm 0.17$, 
 [N/Fe] = $+0.08 \pm 0.21$).
The $\alpha$ elements O, Mg, Si, S, Ca, and Ti
are enhanced relative to Fe, 
[$\alpha$/Fe] = $+0.32 \pm 0.03$.
As discussed in Appendix~\ref{hydrogenhelium},
only the He abundance
is potentially anomalous in \hdtwo.
\item
The metallicity of \hdtwo, [Fe/H] = $-1.46 \pm 0.10$, 
is higher than most known \rpro-enhanced stars,
which suggests that multiple supernovae have contributed
to the lighter elements in \hdtwo.
We follow \citet{roederer18c} in 
assuming that a single \rpro\ event dominated
the heavy-element enrichment of the gas from which \hdtwo\ formed.
As discussed there,
any \spro\ contributions from
asymptotic giant branch (AGB) 
stars are minimal compared to the \rpro\ material.
\end{enumerate}

\subsection{The First \textit{R}-process Peak}
\label{start}

The first \rpro\ peak is expected to occur around mass number
$A \sim 80$, Se to Kr, comprising
the stable $\beta$-decay products
of radioactive nuclei at the $N$ = 50 closed neutron shell
along the \rpro\ path.
The elements just below the first \rpro\ peak, Ga, Ge, and As, are 
deficient by more than 1~dex relative to Se:\
\logeps{Se/Ga} = $+$1.36 $\pm$~0.35,
\logeps{Se/Ge} = $+$1.16 $\pm$~0.25, and 
\logeps{Se/As} = $+$1.61 $\pm$~0.32.
We investigate in this section
the \rpro\ contributions to each of these elements.

\begin{figure}
\begin{center}
\includegraphics[angle=0,width=3.35in]{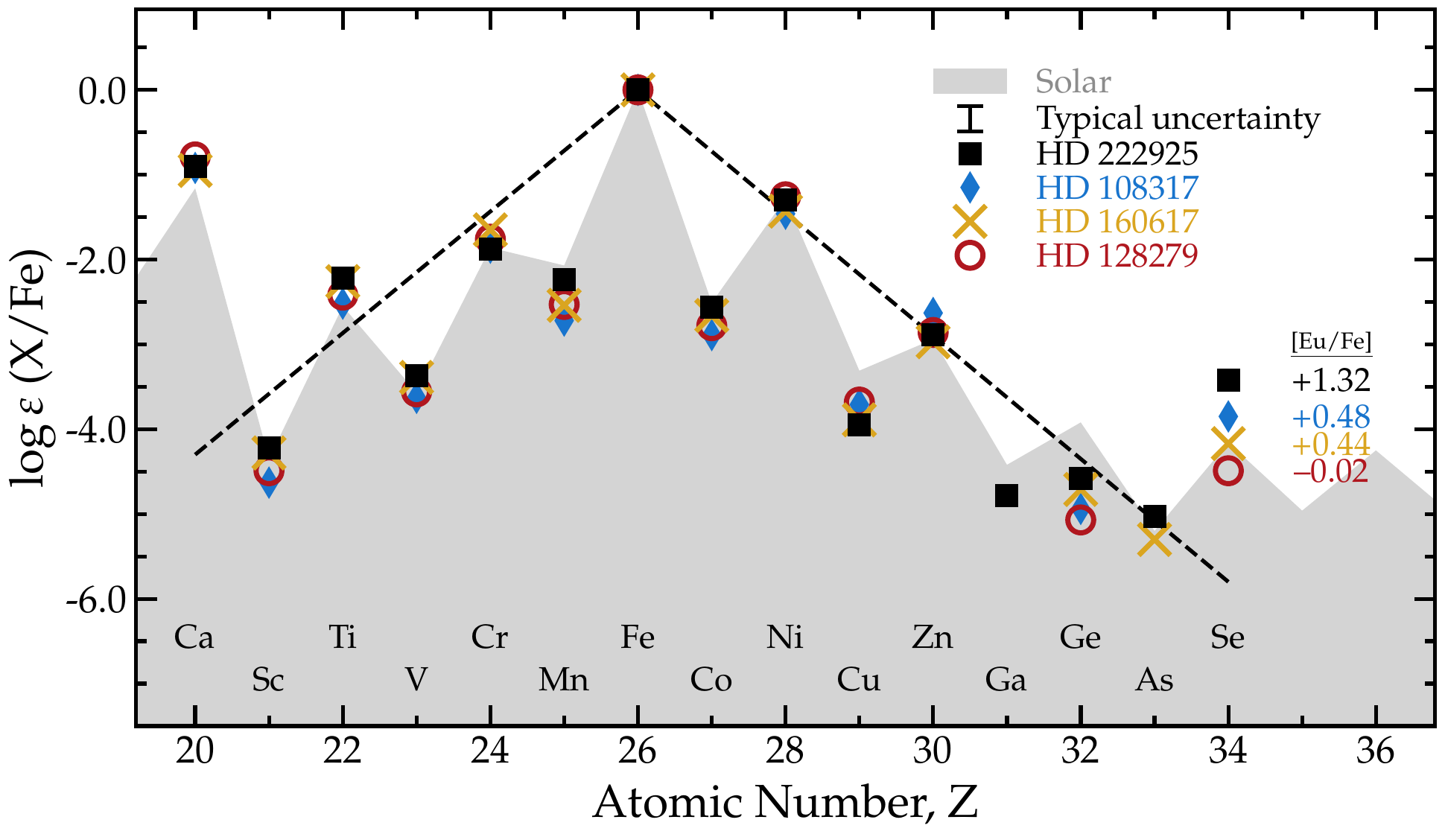}
\end{center}
\caption{
\label{transironplot}
Abundances near the Fe peak in the solar system and four metal-poor stars.
The abundance patterns are normalized to 
\logeps{Fe} = 0.0.
[Eu/Fe] ratios are indicated for the four metal-poor stars
next to their Se abundances.
The data are taken from
\citet{roederer12d,roederer14d} for 
\mbox{HD~108317} and \mbox{HD~128279} and
\citet{roederer12b} for
\mbox{HD~160617}.
The dashed lines approximately follow the decline in the
abundances of the even-$Z$ elements
on either side of the Fe peak.
The Se abundances in the metal-poor stars, especially \hdtwo,
are far in excess of this extrapolation.
}
\end{figure}

Figure~\ref{transironplot} illustrates
the Fe group and first few trans-Fe-group elements
from four \rpro-enhanced stars, along with
the Solar abundances.  
These are the only four \rpro-enhanced stars in which Se has been detected.
In massive stars, Fe-group elements are produced during 
explosive $^{28}$Si burning, which produces the 
familiar shape of the Fe peak.
The even-$Z$ elements fall off in abundance 
away from Fe, as indicated by the dashed lines.
At the light end, Ca sits above the line,
indicating that another process,
$\alpha$-capture, dominates its production.
Se exhibits a similar upturn relative to the
downward trend of even-$Z$ abundances heavier than Fe.
We follow \citet{roederer14d} in proposing 
that this excess of Se is due to \rpro\ nucleosynthesis.
The four stars shown in Figure~\ref{transironplot} 
have a range of [Eu/Fe] ratios, and thus \rpro\ enhancement levels,
ranging from $-$0.02 to $+$1.32~dex.
There is a clear sequence in the Se abundances in that the
highest [Se/Fe] ratio is found in \hdtwo, the star
with the highest [Eu/Fe] ratio, and the
lowest [Se/Fe] ratio is found in 
\object[HD 128279]{HD~128279}, the star with the lowest
[Eu/Fe] ratio.

We conclude 
that the \rpro\ dominates production
in the mass range that includes Se, $76 \leq A \leq 80$.
Some of the Ga, Ge, and As in \hdtwo\
could also have originated in the \rpro, and
some may have been produced by
(neutron-rich) $\alpha$-rich freezeout from
nuclear statistical equilibrium
in the supernovae that produced the Fe-group elements
\citep{woosley92}.
This result is compatible with the finding by \citet{cowan05}
that Ge abundances in metal-poor stars
correlate more closely with Fe than Eu.

\begin{figure}
\begin{center}
\includegraphics[angle=0,width=3.35in]{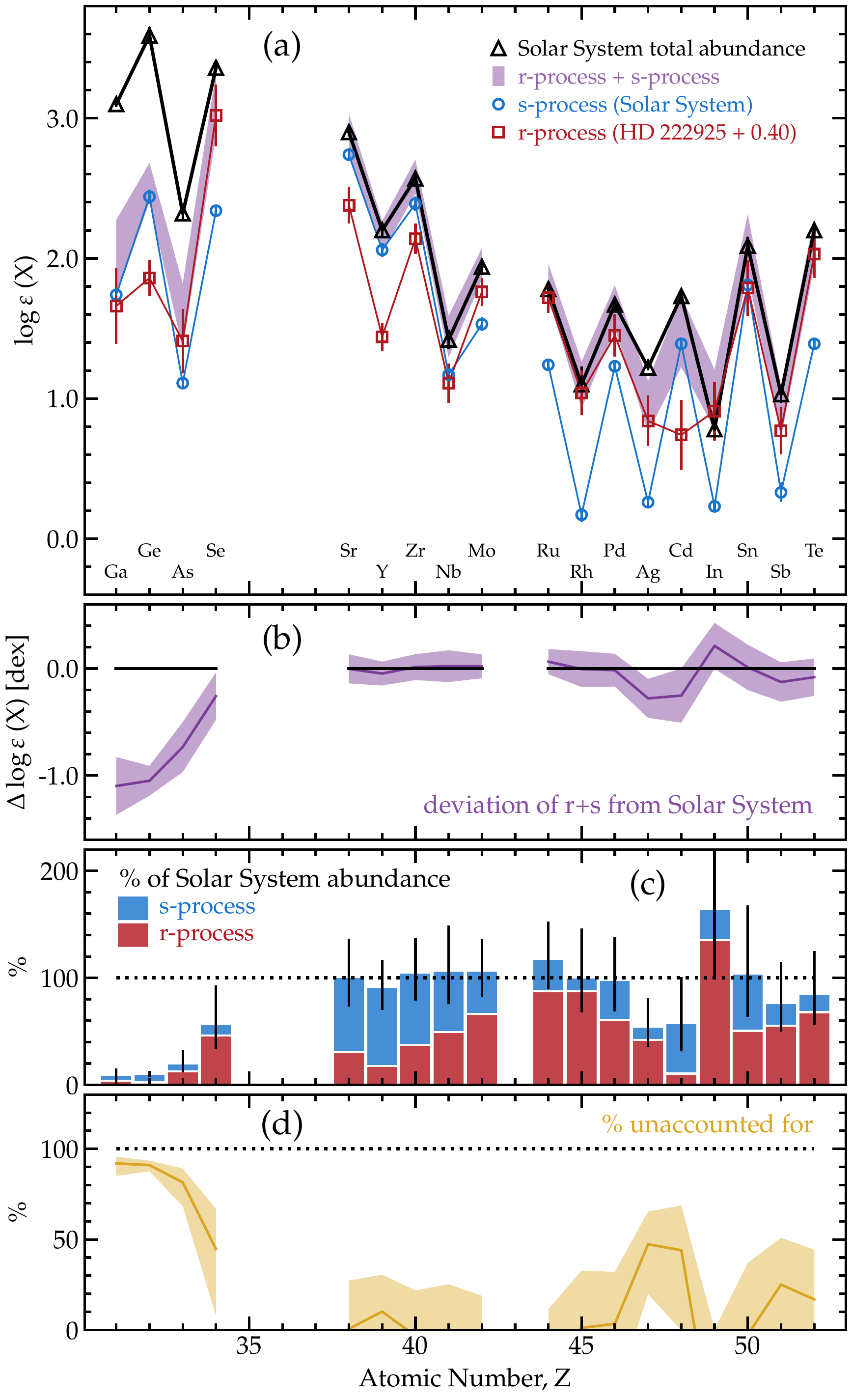}
\end{center}
\caption{
\label{solarfracplot}
Panel (a):\
abundances of elements from Ga to Te in
the solar system (\citealt{lodders09}; black triangles), 
excluding elements that are not detected in \hdtwo.
The AGB \spro\ contribution to the Solar abundances
(\citealt{bisterzo11} for Ga through Se,
\citealt{bisterzo14} for all others)
is shown with blue circles;
the \rpro\ abundances derived from \hdtwo,
scaled by $+0.40$~dex, are shown as red squares;
and
the sum of the \rpro\ and \spro\ abundances,
with $\pm 1\sigma$ uncertainties,
are shown as the purple band.
Panel (b):\
the difference between the summed \rpro\ and \spro\ abundances
(the purple line and band)
and the total Solar abundances (the black line).
Panel (c):\
percentage contributions by the \rpro\ (red) and \spro\ (blue)
to the total Solar abundances.
The dotted line marks 100\%.
Panel (d):\
percentage contributions (yellow band) that are 
unaccounted for by either the \rpro\ or \spro.
The dotted line marks 100\%.
}
\end{figure}

This result has implications for the
origin of Ga, Ge, As, and Se in the solar system.
We present a toy model 
that assumes that all \rpro\ material in the solar system
originated in events identical to the one that 
enriched \hdtwo.
Figure~\ref{solarfracplot}
illustrates the total Solar elemental abundance distribution
\citep{lodders09} from Ga to Te.
We adopt the \spro\ abundances from 
\citet{bisterzo11,bisterzo14},
who calculated the \spro\ fraction of each
element from AGB stellar models
and Galactic chemical-evolution models.
Our model substitutes the \hdtwo\ 
\rpro\ pattern
in place of the traditional \rpro\ residuals.
\hdtwo\ is not physically related to the solar system,
so the relative scaling of the \spro\ and \rpro\
patterns is not known a priori.
We scale the \rpro\ pattern upward to the maximum 
extent permitted,
without overpredicting any Solar abundance
beyond its uncertainty:\
for a shift of $+0.40$~dex, 
In is overproduced by 0.21 $\pm$~0.21~dex.
Note that this shift is unrelated to the
$-$0.11~dex shift applied
in Section~\ref{rpropattern}
to match the 
Solar \rpro\ abundance of Eu to the
\hdtwo\ Eu abundance.
Our toy model
sets an \textit{upper limit} on the allowable \rpro\ contribution
to the Solar abundances 
of lighter \rpro\ elements from Ga to Te.

The purple band in panel (a) of
Figure~\ref{solarfracplot} represents the 
sum of the 
scaled \rpro\ abundance pattern from \hdtwo\ and the
\spro\ abundance pattern
from the \citet{bisterzo14} Galactic chemical evolution model,
hereafter the ``\rs'' pattern.
It includes uncertainties from the \spro\ model,
observational uncertainties in the \rpro\ abundance pattern,
and uncertainties in the Solar abundances.
Panel (b) of Figure~\ref{solarfracplot} illustrates that this
\rs\ pattern provides an acceptable fit to most
of the Solar abundances.
Panel (c) of Figure~\ref{solarfracplot}
shows the percentages of the Solar abundances
contributed by each of the \rpro\ and \spro.
We emphasize that the \rpro\ percentages
in panel (c)
represent upper limits on the
contribution to the solar system 
from  \rpro\ events,
such as the one that enriched the gas from which \hdtwo\ formed.

Panel (d)
of Figure~\ref{solarfracplot} illustrates the
elements for which
the \rpro\ and AGB \spro\ contributions are 
insufficient to account for the Solar abundances.
The maximum \rpro\ contributions to
Ga, Ge, and As are 
$4^{+4}_{-2}$\%, 
$1.6^{+0.6}_{-0.4}$\%, and 
$13^{+9}_{-5}$\%, respectively.
This finding is consistent with previous work 
(e.g., \citealt{frohlich06apj,pignatari10,wanajo11,
roederer12c,niu14,kobayashi20}),
which showed that the origins of the Solar Ga, Ge, and As
are dominated by other nucleosynthesis processes.
The \rpro\ accounts for roughly half of the Se abundance,
$48^{+31}_{-20}$\%,
indicating that substantial \rpro\ contributions start at Se.

\subsection{The Second \textit{R}-process Peak}

The second \rpro\ peak is expected to occur around mass number
$A \sim 130$, Te to Xe, comprising 
the stable $\beta$-decay products
of radioactive nuclei at the $N$ = 82 closed neutron shell
along the \rpro\ path.
The Te abundance lies 0.40~dex lower than the 
Solar \rpro\ residuals when the Solar pattern
is scaled to match Eu.
The Sn abundance in \hdtwo\ is nearly as high as the
Te abundance, \logeps{Te/Sn} = $+0.24 \pm 0.24$, whereas the
Solar \rpro\ residuals anticipate a value of
$+$0.36 or $+$0.72
(\citealt{bisterzo14} and \citealt{sneden08}, respectively).
Conversely, the \logeps{Sn/Cd} ratio in \hdtwo\
is high, $+1.05 \pm 0.26$, whereas the 
Solar \rpro\ residuals anticipate a value of
$+$0.31 or $+$0.00.
We caution that the Sn abundance is derived from a single line,
and our identification of this line is
less secure than others (Section~\ref{tin}).

These discrepancies
could signal the need for further improvements
in nuclear structure models, and
nuclear masses in particular, for
nuclei just below the $A \sim 130$ peak
(see \citealt{kratz14}).
For example, the Sn and Sb behavior is consistent with 
a recent measurement by \citet{li21} that
improved the uncertainty on the mass of 
the neutron-rich isotope $^{123}$Pd.
That measurement implies a slight decrease in the
\rpro\ abundance of the 
$A = 123$ isobar, 
which is linked to one of only two stable Sb isotopes.
It also corresponds to a slight enhancement in the
$A = 122$ isobar,
and we propose that the impact of this enhancement 
on the Sn abundance may be minimal,
because $^{122}$Sn comprises only one of six stable
Sn isotopes accessible to the \rpro.

Another possible interpretation of these discrepancies 
could be that the second \rpro\ peak is shifted 
to lower mass numbers by $\approx$2--4 mass units 
compared to the Solar pattern.
Simulations of \rpro\ nucleosynthesis 
suggest that the shape and placement of the second peak 
depends sensitively on the astrophysical conditions.
The electron fraction and entropy of the outflow 
determine the position of the \rpro\ path on the nuclear chart,
and the initial placement of the peak 
is set by the neutron-richness of the $N = 82$ closed shell nuclei 
populated along the path \citep{meyer97}.
How quickly the temperature and density drop as a function of time 
determines how rapidly the \rpro\ freezes out of the
$(n,\gamma)$-$(\gamma,n)$ equilibrium, 
with the final placement and width of the peak 
set by late-time neutron capture 
\citep{surman01,surman09,arcones11}.
The neutron-richness of the conditions 
also determines whether fissioning nuclei are reached; 
if so, the second peak is shaped in part 
by the deposition of fission products 
\citep{eichler15,vassh19,lemaitre21,sprouse21}. 
Therefore, the discrepancy noted here is intriguing,
and calls for new comparisons between models and observations
for elements around the second \rpro\ peak.

\subsection{The Lanthanides}
\label{lanthanides}

The lanthanide elements span $57 \leq Z \leq 71$, 
La to Lu, which are also known as rare-earth elements.
The lanthanide fraction,
$X_{\rm La}$, 
is the mass ratio between the lanthanides 
and all \rpro\ elements.
We calculate $X_{\rm La}$ = $0.041 \pm 0.008$,
or $\log X_{\rm La}$ = $-1.39 \pm 0.09$,
for \hdtwo.
This value matches the means of the distributions
calculated by \citet{ji19} for
other highly \rpro-enhanced stars,
$\log X_{\rm La}$ = $-1.55 \pm 0.3$, using
the same Solar \rpro\ distribution from 
\citet{sneden08}, or
$\log X_{\rm La}$ = $-1.44 \pm 0.3$, using
one from \citet{arnould07} that
yields slightly reduced contributions to Ga and Ge.

This calculation requires the abundances of
elements that are undetected or unexamined in \hdtwo.
We estimate their abundances by
extrapolating from neighboring elements
using the Solar \rpro\ residuals.
We rely less on these extrapolations
than previous work did.
Our estimate of the lanthanide fraction 
in a metal-poor star
is the first wherein a majority of the 
mass of \rpro\ elements have been detected directly.
The elements detected in \hdtwo\ comprise $\approx$70\%
of the mass of the \rpro\ elements.
Without UV spectra we would have only been able to 
detect elements that comprise $\approx$25\%
of the mass of \rpro\ elements.
The availability of high-quality UV spectra
thus greatly reduces the systematic uncertainties
in the calculation of $X_{\rm La}$.

The lanthanide elements dominate the opacity
in \rpro-rich kilonovae emerging from merging pairs of 
neutron stars (e.g., \citealt{kasen13}),
so kilonova light curves are sensitive to the
lanthanide fraction of material ejected from the mergers.
The lanthanide fraction in \hdtwo\ 
is higher, by a factor of $\sim$~6,
than that of the kilonova associated with GW170817,
$\log X_{\rm La} \approx -2.2 \pm 0.5$ \citep{ji19}.
This difference maintains the tension 
identified by \citeauthor{ji19} between
the lanthanide fraction in highly \rpro-enhanced stars
and this particular kilonova.
This tension could signal the
operation of another dominant source of
\rpro\ elements in the early universe,
such as magnetorotational hypernovae (e.g., \citealt{yong21nature}),
if observations of future kilonovae 
fail to detect events
with lanthanide fractions higher than that found
in the GW170817 event.

\subsection{The Third \textit{R}-process Peak, Lead, 
and Actinides}
\label{thirdpeak}

The third \rpro\ peak is expected to occur around mass number
$A \sim 195$, Os to Pt, comprising 
the stable $\beta$-decay products
of radioactive nuclei at the $N$ = 126 closed neutron shell
along the \rpro\ path.
The third-peak element abundances in \hdtwo\
are in superb agreement with the 
Solar \rpro\ residuals.
Our detections of W, Re, and Au confirm that the
Solar \rpro\ residuals accurately reflect the 
rise and fall of the third-peak abundances,
as well as the placement of the third peak near $A \sim 195$.

Pb is the heaviest stable element detected in \hdtwo.
Pb is unique among \rpro\ elements
because it mostly ($>$~85\%; \citealt{cowan99})
formed through the decay of 
radioactive nuclei with $A > 209$,
including the long-lived isotopes of Th and U.~
\citet{roederer20} used the \mbox{Th/Pb}
chronometer to calculate an age of
8.2 $\pm$~5.8~Gyr for the \rpro\ material in \hdtwo.
This calculation is relatively insensitive to the
details of the \rpro\ model, because of 
the close link between Pb and Th production.
See \citeauthor{roederer20}\ for further discussion of the
prospects for using the \mbox{Th/Pb} ratio
as an age indicator, provided
that the Pb~\textsc{ii} line can be detected in
UV spectra of more \rpro-enhanced stars.

\subsection{Future Prospects}

\begin{figure}
\begin{center}
\includegraphics[angle=0,width=3.35in]{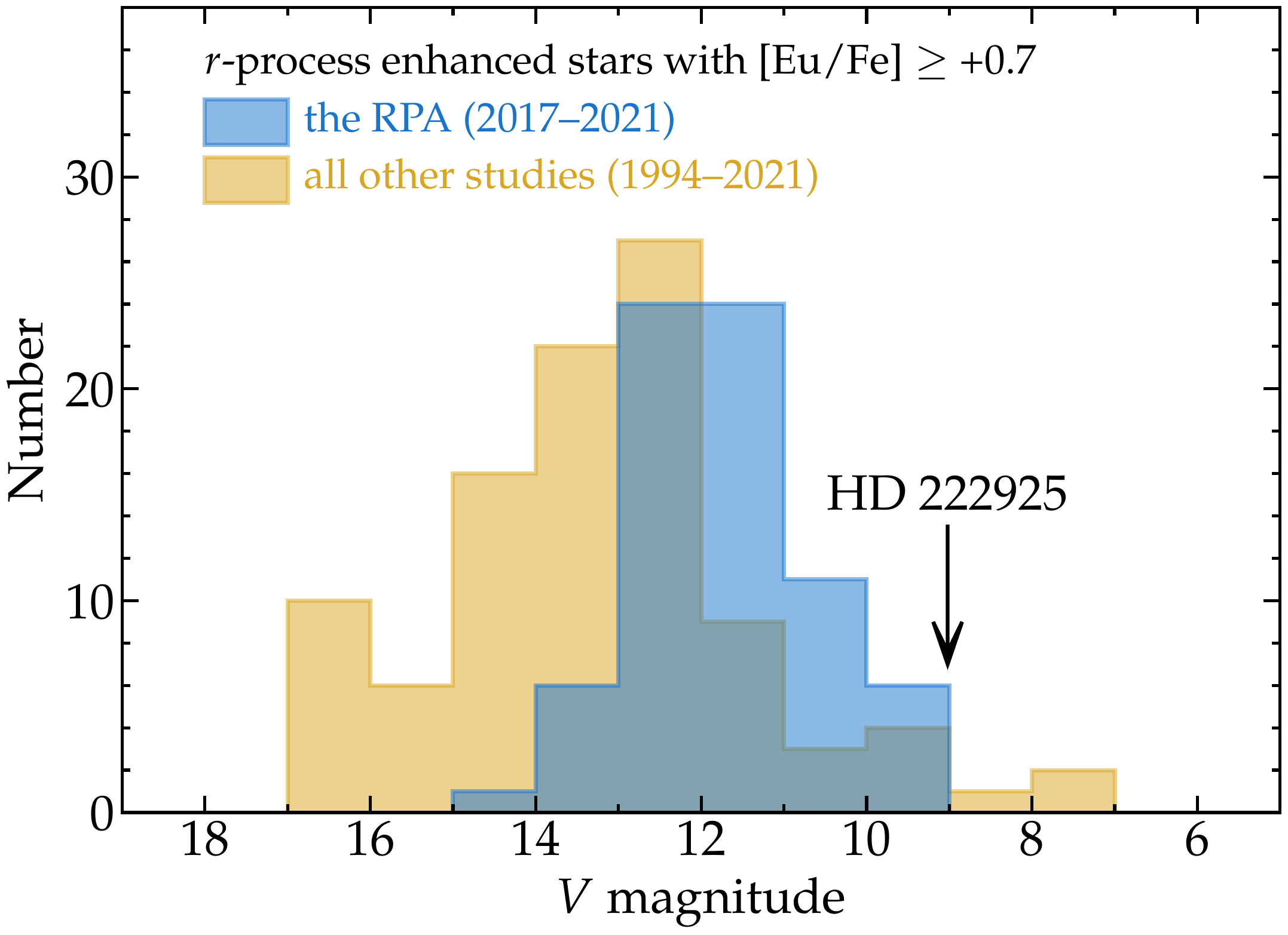} 
\end{center}
\caption{
\label{maghistplot}
Histogram showing the magnitude distribution of
highly \rpro-enhanced stars,
here defined as those with [Eu/Fe] $\geq +0.7$ \citep{holmbeck20}.
The three stars brighter than \hdtwo\ in the $V$ band,
\object[115444]{HD~115444} \citep{westin00},
\object[120559]{HD~120559} \citep{hansen12}, and
\object[221170]{HD~221170} \citep{ivans06},
are all fainter than \hdtwo\ in the GALEX $NUV$ band.
Data for these 181~stars have been compiled from more than
40~literature sources, including 
\citet{barklem05heres}, \citet{ezzeddine20}, \citet{hansen18},
\citet{holmbeck20}, \citet{howes16}, \citet{ishigaki13}, 
\citet{jacobson15smss}, \citet{rasmussen20}, 
\citet{roederer14e}, \citet{sakari18north}, and \citet{yong21}.
}
\end{figure}

More than a quarter century
has passed since the discovery of the 
first highly \rpro-enhanced star,
\object[BPS CS 22892-052]{CS~22892--052}, by \citet{sneden94}.
\hdtwo\ remains the only star in this class that is
bright enough in the UV
for high-quality spectroscopy at wavelengths as short as
2000~\AA\ to be practicable.
Identifying additional bright \rpro-enhanced stars
is one of the major goals of the RPA.~
More than 60 highly \rpro-enhanced stars are already known
with $V > 12$, as shown in Figure~\ref{maghistplot}.
A majority of these stars have been identified by the RPA
in only the last 5~yr.
All of these stars would potentially be observable by the
large UV-optical-infrared flagship mission recently recommended 
by the Astro2020 Decadal Survey
\citep{astro2020}
for launch in the 2040s.
While awaiting that transformative development,
we encourage continued investment to
maximize the operational lifetimes of STIS and HST.~

\section{Conclusions}
\label{conclusions}

We have collected 
new high-resolution UV spectra of the
bright, \rpro-enhanced, metal-poor star \hdtwo.
We use a standard LTE abundance analysis to derive 
abundances from 404~lines in these spectra,
and we merge our results with ones
derived from 908~lines in an optical spectrum.
We summarize our main results and findings as follows.
\begin{enumerate}[nolistsep]
\item
\hdtwo\ presents the most complete chemical inventory
of any object beyond the solar system.
We detect 63~metals in \hdtwo,
including 42~elements produced by the \rpro.
These detections include a number of \rpro\ elements 
that are rarely detected in stars,
including Ga, Ge, As, Se, Cd, In, Sn, Sb, Te, W, Re, Os, Ir, Pt, and Au.
We report upper limits on the abundances of seven other elements
not detected in our spectra.
We critically evaluate these abundances 
and produce a set of recommended abundances,
which is presented in Table~\ref{finalabundtab}.
\item
The \rpro\ elements with $31 \leq Z \leq 52$ 
do not generally match the Solar \rpro\ residuals,
regardless of how they are scaled
(Figure~\ref{patternplot}).
\item
The \rpro\ contributes a small amount to Ga, Ge, and As
($31 \leq Z \leq 33$).
In the solar system,
less than $\approx$4\%, $\approx$2\%, and $\approx$13\%, respectively,
of these elements originated in 
\rpro\ nucleosynthesis events like the one that enriched 
the gas from which \hdtwo\ formed
(Figure~\ref{solarfracplot}).
\item
Se ($Z = 34$) is the lightest element with a substantial
contribution from the \rpro\ in both \hdtwo\ and the solar system
(Figures~\ref{transironplot} and \ref{solarfracplot}).
\item
There is a gradual downward trend relative to the
Solar \rpro\ residuals from Nb through Cd ($41 \leq Z \leq 48$;
Figure~\ref{patternplot}).
\item
The elements Sb and Te ($Z = 51$ and 52, respectively)
at the second \rpro\ peak
are deficient by $\approx$0.4~dex
relative to the Solar \rpro\ residuals when scaled to Eu.
The elements just below the second \rpro\ peak,
In and Sn ($Z = 49$ and 50, respectively),
match the Solar \rpro\ residuals
(Figure~\ref{patternplot}).
This behavior could indicate that
the second \rpro\ peak is shifted to lower mass numbers 
by $\approx$2--4 mass units.
Improvements in nuclear structure models and new experiments
will play an important role in interpreting this behavior.
\item
The \rpro\ elements with $Z \geq 56$ (Ba and heavier)
present a near-perfect abundance match to the
Solar \rpro\ residuals when scaled to match the Eu abundance
(Figure~\ref{patternplot}).
This agreement includes all elements surrounding the
third \rpro\ peak from Re to Au ($75 \leq Z \leq 79$)
and Pb ($Z = 82$), which is formed mainly
through radioactive decay of heavier isotopes.
\item
The \hdtwo\ lanthanide fraction,
$\log X_{\rm La}$ = $-1.39 \pm 0.09$, is normal for 
highly \rpro-enhanced stars \citep{ji19},
and it relies far less than previous estimates
on extrapolations using the Solar \rpro\ residuals.
This lanthanide fraction is higher, by a factor of
$\sim$6, than that inferred from the kilonova
observed after the merger of two neutron stars 
in the GW170817 gravitational-wave event
(Section~\ref{lanthanides}).
\end{enumerate}
\hdtwo\ exhibits no remarkable characteristics 
in its chemical abundance pattern, 
other than the overall enhancement of \rpro\ elements.  
Thus, it may be considered as reflecting the yields 
of the dominant \rpro\ source(s) in the early universe.
This nearly complete \rpro\ abundance template 
provides an alternative to the elemental Solar \rpro\ residuals 
for confronting
future theoretical models of heavy-element nucleosynthesis
with observations; 
we eagerly anticipate the results of those new comparisons.

\acknowledgments

We appreciate the work of P.\ Royle, S.\ Deustua, and the
staff at the STScI for their attention to the observing challenges
encountered in October, 2019, and for their 
sustained dedication to improving the quality of STIS observations.
This project would not have been possible without their persistence.
I.U.R.\ thanks C.\ Cowley for a discussion of chemically peculiar stars,
J.\ Cowan and M.\ Pignatari for discussions of the Solar abundance pattern,
and
R.\ Peterson for sharing results in advance of publication.
We thank the referee for helpful comments on our work.
We 
acknowledge generous support 
provided by NASA through grants GO-15657 and GO-15951
from the Space Telescope Science Institute, 
which is operated by the Association of Universities 
for Research in Astronomy, Incorporated, 
under NASA contract NAS5-26555.
We acknowledge support 
awarded by the U.S.\ National Science Foundation (NSF):\
grants 
PHY~14-30152 (Physics Frontier Center/JINA-CEE),
OISE~1927130 
(International Research Network for Nuclear Astrophysics/IReNA),
AST~1716251 (A.F.), and 
AST~1815403 (I.U.R.).
I.U.R.\ acknowledges support from the NASA
Astrophysics Data Analysis Program, grant 80NSSC21K0627.
R.S.\ acknowledges support from grant
DE-FG02-95-ER40934
awarded by the U.S.\ Department of Energy.
The work of V.M.P.\ is supported by NOIRLab, 
which is managed by AURA under a cooperative agreement with 
the NSF.~
This research has made use of NASA's
Astrophysics Data System Bibliographic Services;
the arXiv preprint server operated by Cornell University;
the SIMBAD and VizieR
databases hosted by the
Strasbourg Astronomical Data Center;
the ASD hosted by NIST;
the MAST at STScI; 
and
Image Reduction and Analysis Facility (IRAF) software packages
distributed by the National Optical Astronomy Observatories,
which are operated by AURA,
under cooperative agreement with the NSF.~

\facility{HST (STIS), Magellan (MagE, MIKE)}

\software{IRAF \citep{tody93},
LINEMAKE \citep{placco21linemake},
matplotlib \citep{hunter07},
MOOG \citep{sneden73,sobeck11},
numpy \citep{vanderwalt11},
scipy \citep{jones01}}

\appendix
\restartappendixnumbering

\section{Discussion of Individual Lines and Species}
\label{appendix}

In this Appendix, we discuss individual lines of interest,
with an emphasis on
the availability of atomic data
relevant to the derivation of stellar abundances.
We also discuss the agreement among abundance indicators
and upper limits derived from nondetections.

\subsection{Hydrogen (H, $Z = 1$) and Helium (He, $Z = 2$)}
\label{hydrogenhelium}

We detect H directly through the Balmer series transitions
in the optical spectrum of \hdtwo.
The H$^{-}$ ion also forms the dominant source of continuous opacity
in the optical and UV.~
\citet{navarrete15} detected the near-infrared He~\textsc{i} line at 
10830~\AA\ in \hdtwo.
Those authors did not derive the He abundance, but they
noted that the equivalent width of this line,
$\approx$230~m\AA,
is about 5--6 times larger 
than that of other stars in their field-star sample.
It is unclear why the He~\textsc{i} line is so strong.
\citeauthor{navarrete15} postulated that it could be
related to chromospheric activity.
\hdtwo\ exhibits other signs of chromospheric activity,
including emission in the cores of the 
Mg~\textsc{ii} resonance doublet at 2800~\AA,
similar to other old, metal-poor, red giants \citep{dupree07}.
Further study of this He~\textsc{i} line would be worthwhile.

\subsection{Lithium (Li, $Z = 3$), Beryllium (Be, $Z = 4$),
 and Boron (B, $Z = 5$)}

We detect no lines of the light element Li.
We adopt an upper limit on its abundance from
\citet{roederer18c},
\logeps{Li} $< 0.80$,
derived from nondetection of 
the Li~\textsc{i} line at 6707.80~\AA.~
This fragile element is diluted and then destroyed 
when a star, such as \hdtwo, evolves into a red giant and 
horizontal-branch star (e.g., \citealt{charbonnel95}).
Our upper limit is compatible with observations of
other evolved, metal-poor stars, which show
\logeps{Li} $\leq 1$ along the upper red giant branch
(e.g., \citealt{gratton00,lind09n6397,kirby16li}).

We detect no lines of the light elements Be or B, either.
Upper limits from several lines are reported in Table~\ref{linetab}.
Absorption is detected at 2089.56~\AA, near the wavelength of a
B~\textsc{i} line at 2089.570~\AA.~
No absorption that can plausibly be attributed to B~\textsc{i} 
is detected at the wavelengths of other B~\textsc{i} lines
at 2088.889, 2496.769, or 2497.722~\AA, however,
so we conclude that B~\textsc{i} is undetected in our spectrum of \hdtwo.
We report upper limits in Table~\ref{linetab}.

\subsection{Carbon (C, $Z = 6$)}
\label{carbon}

We detect a reasonably unblended line of C~\textsc{i} at 2964.846~\AA.~
The NIST ASD lists a \loggf\ for this line with a D grade. 
The abundance derived from this line, \logeps{C} = 6.70 $\pm$~0.34,
is in good agreement with the C abundance derived from 
the CH ``G'' band, \logeps{C} = 6.65 $\pm$~0.17.
Given the large uncertainty 
in the \loggf\ value of the C~\textsc{i} UV line,
our recommended C abundance relies on the 
abundance derived from the CH ``G'' band.
We correct this value by $+$0.46~dex for
stellar evolution effects, as described in \citet{placco14c}.
The corrected value is listed in Table~\ref{finalabundtab}.

\subsection{Nitrogen (N, $Z = 7$)}

No atomic or molecular N features are detected in our UV spectrum
of \hdtwo.
Our recommended N abundance is based on that derived from
NH molecular bands detected by \citet{roederer18c}.

\subsection{Oxygen (O, $Z = 8$)}

A number of OH features are detected redward
of $\approx$2810~\AA\ in our UV spectrum of \hdtwo.
Our syntheses yield \logeps{O} $\approx +7.9$ or so
from these features when using the \citet{kurucz11} OH line list.
The [O/Fe] ratio, $+$0.67, is 0.25~dex higher than 
that found by \citet{navarrete15},
[O/Fe] = $+$0.42 $\pm$~0.07, based on a careful
NLTE differential analysis (relative to the Sun)
of the O~\textsc{i} triplet near 7770~\AA.~
The UV OH transitions are known to overestimate the
O abundance when derived using 1D LTE model atmospheres
(e.g., \citealt{bessell15}).
In the absence of a 3D hydrodynamical model atmosphere for \hdtwo,
we recommend the [O/Fe] ratio derived by \citeauthor{navarrete15} 

\subsection{Sodium (Na, $Z = 11$) and Magnesium (Mg, $Z = 12$)}

We identify no Na~\textsc{i}, Mg~\textsc{i}, or Mg~\textsc{ii} lines
with reliable \loggf\ values that are unblended and
sufficiently---but not too---strong in our spectrum of \hdtwo.
Our recommended abundances for Na and Mg
are based on the optical lines studied by \citet{roederer18c},
including NLTE corrections to the Na abundance
based on \citet{lind11}.

\subsection{Aluminum (Al, $Z = 13$)}

We detect 8 UV Al~\textsc{i} lines that are reasonably unblended
and have reliable \loggf\ values.
These lines, which arise from the ground level,
yield a consistent abundance,
\logeps{Al} = 4.23 $\pm$~0.07.
This abundance, however, is about 0.6~dex lower than the abundance derived 
from 3 high-excitation ($>$ 3.1~eV) Al~\textsc{i} lines
in the optical spectrum,
\logeps{Al} = 4.79 $\pm$~0.22.
We also detect an unblended Al~\textsc{ii} line at 2669.155~\AA,
as discussed in detail in \citet{roederer21}.
This line has a reliable \loggf\ value (NIST grade A+).~
The abundance derived from this line,
\logeps{Al} = 4.77 $\pm$~0.13, agrees with that from the
high-excitation Al~\textsc{i} lines.
Our result affirms the calculations of \citet{mashonkina16al},
who found that inelastic collisions with H effectively 
coupled the highly excited levels of neutral Al with
the ground state of the ion, which dominates the ionization balance.
\citet{nordlander17al} found a similar result
and confirmed the abundance behavior for several UV Al~\textsc{i} lines
in two metal-poor stars.
Our recommended Al abundance in \hdtwo\ averages
the abundances from the
three high-excitation Al~\textsc{i} lines and the one
Al~\textsc{ii} line.

\subsection{Silicon (Si, $Z = 14$)}

The abundances derived from Si~\textsc{i} lines in our study and
\citet{roederer18c} exhibit similar, though not identical, behavior to Al.
Si has a much higher first ionization potential (FIP; 8.15~eV)
than Al (5.99~eV),
so a larger fraction of Si atoms are neutral in the
atmosphere of \hdtwo.
Seventeen of the 18 Si~\textsc{i} lines examined by \citeauthor{roederer18c}\
originate from highly excited levels ($>$~4.9~eV).
The exception, the Si~\textsc{i} line at 4102~\AA, 
originates from a moderately excited level at 1.91~eV.~
These lines yield \logeps{Si} = 6.38 $\pm$~0.07.
In contrast, the 5 UV lines of Si~\textsc{i} originate from 
levels with E$_{\rm low}$ ranging from 0.01 to 0.78~eV.
These lines yield \logeps{Si} = 6.02 $\pm$~0.14,
which is notably lower.
A similar discrepancy was noticed
in the metal-poor, red giant
\object[BD +44 493]{BD~$+$44$^{\circ}$493}
\citep{roederer16d}.

We present 4 new detections of Si~\textsc{ii} lines in 
the MIKE optical spectrum of \hdtwo.
We also search for Si~\textsc{ii} lines in our STIS UV spectrum,
but these lines are too blended or too weak 
to yield reliable abundances.
The 4 optical Si~\textsc{ii} lines yield
\logeps{Si} = 6.48 $\pm$~0.05,
which is
in agreement with the abundance derived from 
the high-excitation Si~\textsc{i} lines.

\citet{mashonkina16al} found that departures from LTE
were minimal in the line-forming layers of a moderately metal-poor dwarf,
although that study only considered the 
moderately or highly excited levels of neutral Si.
We suspect that the low-lying levels of neutral Si
may be overpopulated in our LTE calculations,
so the UV Si~\textsc{i} lines may yield low abundances,
though not to the same degree as found in Al.
New NLTE calculations for these levels would be welcome.
Our recommended Si abundance in \hdtwo\ 
is based on only the optical Si~\textsc{ii} and 
high-excitation Si~\textsc{i} lines.

\subsection{Phosphorus (P, $Z = 15$)}

We detect several P~\textsc{i} lines in \hdtwo.
Some of them are saturated, blended, 
and not useful as abundance indicators,
but three P~\textsc{i} lines yield reliable abundances.
These lines arise from moderately excited levels ($\approx$1.41~eV).
The derived [P/Fe] ratio ($+$0.18 $\pm$~0.15) is in good 
agreement with other stars of similar metallicity \citep{roederer14f}.
Our recommended P abundance is based on these UV lines.

\subsection{Sulphur (S, $Z = 16$)}

We detect three lines of the S~\textsc{i} multiplet~8
in our optical MIKE spectrum.
Each of these three lines is comprised of three
fine-structure components.
Theoretical \loggf\ values are reasonably consistent
for these fine-structure components:\
the standard deviations of the total \loggf\ value
of each line from different sources are $\approx$0.03~dex
(\citealt{wiese69,biemont93}; 
see \citealt{caffau05} and \citealt{scott15}).
We conservatively adopt twice this value as the \loggf\ uncertainty.
Table~\ref{linetab} lists the center-of-gravity wavelengths
based on the \citet{kurucz11} fine-structure components
and the combined \loggf\ values from \citeauthor{biemont93} 
We also apply a small $-0.05$~dex NLTE abundance correction,
following the calculations presented by \citet{korotin17}.
Our recommended S abundance is based on
these three S~\textsc{i} lines.

\subsection{Potassium (K, $Z = 19$)}

No K~\textsc{i} or K~\textsc{ii} lines are strong enough to detect
in our UV spectrum of \hdtwo.
Our recommended K abundance is based on one optical
K~\textsc{i} line, including NLTE corrections
from \citet{takeda02}.

\subsection{Calcium (Ca, $Z = 20$)}

We detect 2 lines of Ca~\textsc{ii} in the UV spectrum of \hdtwo.
We adopt the \loggf\ values for these lines 
from \citet{theodosiou89} as recommended by
\citet{denhartog21ca}.
That study also reassessed the \loggf\ values for optical
Ca~\textsc{i} lines, and we recalculate the Ca abundance derived
from the 34~Ca~\textsc{i} lines analyzed by \citet{roederer18c}.
The revised \loggf\ scale only changes the mean \logeps{Ca} abundance
by $+$0.02~dex, but the standard deviation decreases substantially, from
0.15 to 0.09~dex.
The abundances derived from the Ca~\textsc{i} and Ca~\textsc{ii} lines
are in good agreement, as shown in Figure~\ref{fegroupplot}.
Our recommended Ca abundance
is based on the weighted average of all 36 lines.

\subsection{Other Fe-group Elements}

Lines of neutral and ionized atoms are detected in our UV 
spectrum of \hdtwo\ for most of the Fe-group elements.
These elements have relatively low FIPs,
ranging from 6.56~eV (scandium, Sc, $Z = 21$) 
to 7.90~eV (Fe; \citealt{morton03}).
These atoms are primarily found in their ionized states
in the atmosphere of \hdtwo.
Zinc (Zn, $Z = 30$) is an exception;
its FIP is much higher (9.39~eV),
and substantial fractions of both neutral and ionized Zn
are present.

We detect no UV lines of Sc~\textsc{i} or Sc~\textsc{ii}
with reliable \loggf\ values and HFS patterns, 
and our recommended Sc abundance is
based on that derived from Sc~\textsc{ii} lines in the optical spectrum.

We detect and analyze several UV Ti~\textsc{i} and Ti~\textsc{ii} lines.
Our recommended titanium (Ti, $Z = 22$) abundance is based on
the abundances derived from the optical and UV Ti~\textsc{ii} lines.
The abundances derived from these two ionization stages differ by
0.28 $\pm$~0.06~dex (optical lines) or 
0.12 $\pm$~0.08~dex (UV lines),
after Saha corrections are applied.
These small differences are compatible with previous studies of
Ti in other stars,
although it is unclear why the difference derived from optical lines
is larger than the difference derived from UV lines.
NLTE calculations, like those presented in \citet{sitnova20},
could help to reconcile the Saha discrepancy.

There are no strong and unblended V~\textsc{i} lines 
with reliable \loggf\ values and HFS patterns in our spectrum,
but we detect and analyze several V~\textsc{ii} lines.
Our recommended vanadium (V, $Z = 23$) abundance is based 
on the abundances derived from the optical and UV V~\textsc{ii} lines.

We detect and analyze 15 lines of Cr~\textsc{i} and 
20 lines of Cr~\textsc{ii} in the UV spectrum.
Our recommended chromium (Cr, $Z = 24$) abundance is based 
on the abundances derived from the optical and UV Cr~\textsc{ii} lines.

There are no Mn~\textsc{i} lines in the UV spectrum that were
covered by our preferred laboratory source, \citet{denhartog11}.
We analyze two UV Mn~\textsc{ii} lines from \citeauthor{denhartog11},
in addition to the three optical Mn~\textsc{ii} lines analyzed by
\citet{roederer18c}.
Our recommended manganese (Mn, $Z = 25$) abundance is based on
the abundances derived from these five Mn~\textsc{ii} lines.

The cases of Fe (Section~\ref{fe})
and cobalt (Co, $Z = 27$; Appendix~\ref{cobalttext})
are sufficiently complex to warrant separate discussions.

We detect and analyze multiple lines of Ni~\textsc{i} and Ni~\textsc{ii}
in our UV spectrum of \hdtwo.
Our recommended nickel (Ni, $Z = 28$) abundance is derived from
the UV Ni~\textsc{ii} lines.

Cu~\textsc{i} and Cu~\textsc{ii} lines are detected in the UV spectrum.
No HFS $A$ constants are available for the Cu~\textsc{ii} lines,
but these lines are reasonably weak and so the abundances derived from them
are minimally affected by this shortcoming
(see Section~3.1 of \citealt{roederer12b}).
Following \citet{korotin18} and \citet{roederer18a},
the recommended copper (Cu, $Z = 29$) abundance is based on
the abundances derived from the UV Cu~\textsc{ii} lines.

The Zn~\textsc{ii} resonance doublet at 2025 and 2062~\AA\
is detected in our UV spectrum of \hdtwo, but
these lines are too blended to yield a reliable zinc (Zn, $Z = 30$) 
abundance.
Two Zn~\textsc{i} lines are detected in the UV spectrum and
useful for analysis.
Our recommended Zn abundance is based on
the abundances derived from the optical and UV Zn~\textsc{i} lines,
which have been found to yield consistent results with
the Zn~\textsc{ii} lines when both species are considered
\citep{roederer18a}.

\subsection{Cobalt (Co, $Z = 27$)}
\label{cobalttext}

A total of 11 Co~\textsc{i} lines with
\loggf\ values and HFS constants reported by \citet{lawler15}
are detected and reasonably unblended.
The mean abundance derived from these lines,
\logeps{Co} = 3.39 $\pm$~0.05, is in good agreement with 
the Co abundance derived from 21 optical Co~\textsc{i} lines,
\logeps{Co} = 3.33 $\pm$~0.10.
Seven Co~\textsc{ii} lines 
with \loggf\ values reported by \citet{lawler18}
are detected and reasonably unblended.
Table~\ref{cohfstab}
presents the complete line component patterns for these lines,
computed using the HFS $A$ constants from 
\citet{ding20}, or from
\citet{lawler18} and \citeauthor{ding20} 
averaged together, for levels in common.
HFS constants are also presented in 
the Vienna Atomic Line Database 
(VALD; \citealt{pakhomov19}), and \citet{fu21} present 
new measurements that confirm the odd-parity HFS $A$ constants we use.
Table~\ref{cohfstab} includes two other Co~\textsc{ii} lines,
at 2214.793 and 2314.975~\AA,
that are detected but
too saturated to be useful as abundance indicators.
The abundance derived from the seven UV Co~\textsc{ii} lines,
\logeps{Co} = 3.48 $\pm$~0.05,
is in fair agreement with the abundance
derived from Co~\textsc{i} lines.

\begin{deluxetable}{ccccccc}
\tablecaption{Hyperfine Structure
Line Component Patterns for Co~\textsc{ii} Lines
\label{cohfstab}}
\tabletypesize{\small}
\tablehead{
\colhead{Wavenumber} &
\colhead{$\lambda_{\rm air}$} &
\colhead{$F_{\rm upper}$} &
\colhead{$F_{\rm lower}$} &
\colhead{Component Position} &
\colhead{Component Position} &
\colhead{Strength} \\
\colhead{(cm$^{-1}$)} &
\colhead{(\AA)} &
\colhead{} &
\colhead{} &
\colhead{(cm$^{-1}$)} &
\colhead{(\AA)} &
\colhead{}
}
\startdata
45709.627 & 2187.0388 & 5.5 & 4.5 &    0.259000 & $-$0.012393 & 0.300000 \\
45709.627 & 2187.0388 & 4.5 & 4.5 &    0.079700 & $-$0.003814 & 0.097222 \\
45709.627 & 2187.0388 & 4.5 & 3.5 &    0.040100 & $-$0.001919 & 0.152778 \\
\enddata
\tablecomments{%
Energy levels from the NIST ASD and the index of air \citep{peck72}
are used to compute the
center-of-gravity wavenumbers and air wavelengths, $\lambda_{\rm air}$,
and component positions are given relative to those values.
Strengths are normalized to sum to 1 for each line.
A complete machine-readable version of Table~\ref{cohfstab} is
available online. 
A short version is shown here to illustrate its form and content.
}
\end{deluxetable}

Previous work \citep{cowan20} has noted the difficulty of 
interpreting abundances derived from Co~\textsc{i} and \textsc{ii} lines.
In that study of three metal-poor ([Fe/H] $\approx -$3) 
dwarf stars, the
Co~\textsc{i} lines yielded a higher abundance ($\approx +$0.4~dex)
than the Co~\textsc{ii} lines.
That behavior is opposite what
would be expected if NLTE overionization is responsible
for the offset.
In contrast, in \hdtwo,
the seven Co~\textsc{ii} lines 
exhibit the expected behavior,
yielding a Co abundance higher by $+$0.11~dex
than the Co~\textsc{i} lines.
NLTE calculations (see \citealt{bergemann10}) have not been
performed for the UV lines examined here, and 
we encourage new calculations.

We suggest that the seven Co~\textsc{ii} lines with HFS
are the best Co abundance indicators in \hdtwo.

\subsection{Gallium (Ga, $Z = 31$)}
\label{gallium}

Our spectrum covers the Ga~\textsc{ii} line at 2090.768~\AA,
which is illustrated in Figure~\ref{specplot1}.
This line has not previously been examined in cool stars.
Excess absorption with $\approx$20\% continuum depth
is detected at the correct wavelength in our spectrum of \hdtwo.

This line connects the
$3d^{10}4s^{2}$ $^{1}$S$_{0}$ ground state to the
$4s4p$ $^{3}$P$^{\rm o}_{1}$ excited state.
There are three modern calculations of the upper-level radiative lifetime,
which can be related to the transition probability through the
branching fraction (BF) and level degeneracy.
These results,
from \citet{mcelroy05}, \citet{liu06}, and \citet{chen10},
agree to within $\approx$11\%, and
we average them together to 
yield \loggf\ = $-3.25 \pm 0.05$ for this line.
We directly adopt the HFS pattern and IS of the two stable Ga isotopes,
$^{69}$Ga and $^{71}$Ga, from \citet{karlsson00}.
Ga has a low FIP, 6.00~eV,
and Ga$^{+}$ is the dominant ionization state in the
atmospheres of cool stars.
We adopt an \rpro\ isotope mix from \citet{sneden08};
this line yields the same abundance
if we instead adopt a Solar isotope mix.

How likely is this identification to be correct?
There are no other obvious species that might
absorb at this wavelength
in the \citet{kurucz11} lists or NIST ASD.~
There are also no newly identified Fe~\textsc{i} lines
at this wavelength 
in the \citet{peterson15} or
\citet{peterson17} catalogs.
Ideally, we would verify the identity of this line
using other Ga~\textsc{ii} lines in the spectrum.
Unfortunately, there are no other Ga~\textsc{ii} lines
of comparable or greater strength that
would be expected in our spectrum.
This spin-forbidden line is the only Ga~\textsc{ii} line in the NIST ASD
with $\lambda >$~1500~\AA\ 
that is connected to the ground level of Ga$^{+}$.

As emphasized by \citet{peterson21}
in a similar context,
there are dozens of predicted and potentially detectable
unidentified Fe~\textsc{i} (and Fe~\textsc{ii}) 
lines within several \AA,
whose energy levels, and thus wavelengths, are poorly known.
We test this scenario by checking this region of the spectrum in
several other archival STIS spectra.
These spectra include the more metal-rich dwarf 
\object[HD76932]{HD~76932} 
(Program = GO-9804, 
PI = Duncan;
\teff/\logg/[Fe/H]/\vt = 
5680~K/4.11/$-0.92$/0.90~\kmsec, \citealt{roederer12c}),
the similarly metal-poor dwarf
\object[HD94028]{HD~94028}
(GO-8197 and GO-14161, 
Duncan and Peterson;
6087~K/4.37/$-1.65$/1.10~\kmsec, \citealt{roederer18b}),
the more metal-poor dwarf 
\object[HD84937]{HD~84937} 
(GO-14161, 
Peterson;
6418~K/4.16/$-2.23$/1.50~\kmsec, \citealt{roederer18b}),
the more metal-poor subgiant
\object[HD160617]{HD~160617} 
(GO-8197, 
Duncan;
5950~K/3.90/$-1.77$/1.3~\kmsec, \citealt{roederer12b}),
the more metal-poor subgiant
\object[HD140283]{HD~140283} 
(GO-7348, 
Edvardsson;
5600~K/3.66/$-2.62$/1.15~\kmsec, \citealt{roederer12c}), and
the more metal-poor giant
\object[HD196944]{HD~196944}
(GO-14765, 
Roederer;
5170~K/1.60/$-2.41$/1.55~\kmsec, \citealt{placco15cemps}).
Absorption at 2090.768~\AA\ is clearly detected in
\mbox{HD~76932}, and a weak absorption line
is tentatively detected in \mbox{HD~196944}.

We check whether a fictitious 
Fe~\textsc{i} or Fe~\textsc{ii} line 
could consistently account for the absorption observed at this wavelength.
We consider four cases:\
a low-excitation weak Fe~\textsc{i} line 
 (E$_{\rm low}$/\loggf\ = 1.0~eV/$-3.7$),
a high-excitation strong Fe~\textsc{i} line (5.0~eV/$+0.1$),
a low-excitation weak Fe~\textsc{ii} line (1.0~eV/$-5.4$), and
a high-excitation strong Fe~\textsc{ii} line (5.0~eV/$-1.6$).
We set the \loggf\ values of these fictitious 
lines to reproduce the absorption observed in \hdtwo,
and we use those atomic data to 
generate synthetic spectra for each of the six other stars.
The Fe~\textsc{i} lines and the 
high-excitation Fe~\textsc{ii} line
fail to provide a satisfactory fit to all six stars.
A low-excitation Fe~\textsc{ii} line---not
unlike the Ga~\textsc{ii} line itself, when detectable---%
produces acceptable matches to the observed spectra.

We cautiously conclude that a Ga~\textsc{ii} line
is the most likely explanation for
absorption at 2090.769~\AA,
but an as yet unidentified low-excitation ion
from an Fe-group species cannot be excluded.
Under this assumption,
we also derive the Ga abundance in
\mbox{HD~76932} and
\mbox{HD~196944}, finding
\logeps{Ga} = 1.70 $\pm$~0.20 and 0.30 $\pm$~0.30, respectively,
assuming an \rpro\ isotope mix for \mbox{HD~76932} and an
\spro\ isotope mix for \mbox{HD~196944}.
If the absorption is not a Ga~\textsc{ii} line,
these values represent upper limits on the Ga abundance.
We urge further study of this line in additional stars.

\subsection{Germanium (Ge, $Z = 32$)}

We detect 5 Ge~\textsc{i} lines in our spectrum,
including the lines at 2691.341 and 3039.067~\AA\ that have 
been used in many previous studies.
These lines are illustrated in Figure~\ref{specplot3}.
\citet{li99} provide experimental \loggf\ values,
which are reliable to $\approx$10\% (0.05~dex), for these lines.

Ge has a moderately high FIP, 7.90~eV,
comparable to that of Fe, Co, Ni, and Cu.
Most Ge is found as Ge$^{+}$ in the atmosphere of \hdtwo.
Neutral Ge is a minority species,
and it could be susceptible to NLTE overionization.
No NLTE studies of Ge in late-type stars have been published.
The ions of Fe, Co, Ni, and Cu yield abundances higher by 
$\approx$0.15~dex than the neutral species in \hdtwo.
We report the LTE abundance of Ge in Tables~\ref{linetab}--\ref{finalabundtab},
but we caution that this abundance may underestimate the 
Ge abundance by a small amount.
We strongly encourage a study of
NLTE Ge line formation in cool stars.

\subsection{Arsenic (As, $Z = 33$)}
\label{arsenic}

We detect only one As~\textsc{i} line,
at 2288.115~\AA.~
It is moderately blended with a
Cd~\textsc{i} line at 2288.020~\AA\ and an
Fe~\textsc{i} line at 2288.045~\AA,
as illustrated in Figure~\ref{specplot2}.
We adjust the \loggf\ value of the Fe~\textsc{i} line to be $-4.0$,
using other STIS spectra where
absorption from the As~\textsc{i} and Cd~\textsc{i} lines
is minimized.
The As~\textsc{i} and Cd~\textsc{i} lines are sufficiently
separated in wavelength that they are both resolved in our spectrum.
The FIP of As is high, 9.79~eV,
so NLTE overionization is unlikely to impact the 
LTE As abundance substantially.

There is one stable isotope of As, $^{75}$As.
It has a nuclear spin of $I = 3/2$,
which produces HFS.~
The lower- and upper-level HFS $A$ constants are known for this line
\citep{bouazza87}.
We present the line component pattern for the As~\textsc{i} line
at 2288~\AA\ in Table~\ref{ashfstab}.

We adopt the theoretical transition probability
computed by \citet{holmgren75},
\loggf\ = $-0.06$.
This value is in good agreement
with the \loggf\ value ($-0.02$) 
computed using the theoretical upper-level radiative lifetime from
\citeauthor{holmgren75} and an experimental BF,
0.79 $\pm$~0.04, measured by \citet{berzins21}
and published after our analysis had been completed.
The \loggf\ uncertainty is dominated by the upper-level lifetime.
The NIST ASD suggests a conservative D grade,
and we adopt a \loggf\ uncertainty of 0.15~dex.
Our recommended As abundance is based on this one line.

\begin{deluxetable}{ccccccc}
\tablecaption{Hyperfine Structure Line Component Patterns for the 
As~\textsc{i} $\lambda$2288 Line
\label{ashfstab}}
\tabletypesize{\small}
\tablehead{
\colhead{Wavenumber} &
\colhead{$\lambda_{\rm air}$} &
\colhead{$F_{\rm upper}$} &
\colhead{$F_{\rm lower}$} &
\colhead{Component Position} &
\colhead{Component Position} &
\colhead{Strength} \\
\colhead{(cm$^{-1}$)} &
\colhead{(\AA)} &
\colhead{} &
\colhead{} &
\colhead{(cm$^{-1}$)} &
\colhead{(\AA)} &
\colhead{} 
}
\startdata
43690.625 & 2288.1149 & 3.0 & 4.0 & $-$0.045975 &    0.002408 & 0.375000 \\
43690.625 & 2288.1149 & 3.0 & 3.0 &    0.050825 & $-$0.002662 & 0.058333 \\
43690.625 & 2288.1149 & 3.0 & 2.0 &    0.123425 & $-$0.006464 & 0.004167 \\
\enddata
\tablecomments{%
The center-of-gravity wavenumber is from \citet{howard85}.  
The index of air from \citet{peck72} is
used to calculate the center-of-gravity air wavelength, $\lambda_{\rm air}$.  
The line component positions are given relative to those values.
The complete version of Table~\ref{ashfstab} is available in the online edition
of the journal in machine-readable format.
A short version is included here to demonstrate its form and content.
}
\end{deluxetable}

\subsection{Selenium (Se, $Z = 34$)}

Only one Se~\textsc{i} line, at 2074.784~\AA,
is both detected and unblended in our spectrum,
as illustrated in Figure~\ref{specplot1}.
We adopt the \loggf\ value from \citet{morton00},
with an uncertainty of 0.03~dex.
The HFS and IS of this Se~\textsc{i} line are 
expected to be small and negligible for our purposes,
as discussed in \citet{roederer12b}.
An additional Se~\textsc{i} line at 2039.842~\AA\
is detected, but it is too blended and the continuum placement
too uncertain to yield a reliable abundance.
The FIP of Se is high, 9.75~eV,
so, as in the case of As, NLTE overionization is 
unlikely to impact the LTE Se abundance.
Our recommended abundance is based on this one line.

\subsection{Strontium (Sr, $Z = 38$)}

We check our spectrum for the 10~lines of Sr~\textsc{ii} 
that are listed in the NIST ASD.~
Some are strong enough to be detectable, but they are
too blended with other features to be of use.
Our recommended Sr abundance in \hdtwo\ is based
on the optical Sr~\textsc{ii} lines from \citet{roederer18c}.

\subsection{Yttrium (Y, $Z = 39$)}

We detect and derive abundances from 
4 Y~\textsc{ii} lines in our spectrum.
One of them, at 2422.186~\AA,
is illustrated in Figure~\ref{specplot2}.
\citet{biemont11} presented \loggf\ values for these lines.
The weighted mean of these 4 Y~\textsc{ii} lines, 
\logeps{Y} = 1.06, is in excellent agreement with the
weighted mean abundance derived from the 40 optical Y~\textsc{ii} lines
studied by \citet{roederer18c},
\logeps{Y} = 1.04.
The FIP of Y is 6.22~eV, so most Y is found as Y$^{+}$
in the atmosphere of \hdtwo.
NLTE overionization is not expected to be an issue.
Our recommended Y abundance is based on the weighted average of these
44 Y~\textsc{ii} lines.

\subsection{Zirconium (Zr, $Z = 40$)}

We check the 55 lines of Zr~\textsc{ii} listed in \citet{ljung06},
and 22 of these lines yield abundances in our spectrum of \hdtwo.
Figure~\ref{specplot3} illustrates two of these lines.
The weighted mean of these 22 Zr~\textsc{ii} lines, 
\logeps{Zr} = 1.76, is in excellent agreement with the
weighted mean abundance derived from the 51 optical Zr~\textsc{ii} lines
studied by \citet{roederer18c},
\logeps{Zr} = 1.74.
The FIP of Zr is 6.63~eV, so Zr$^{+}$ is the dominant species of Zr
in the atmosphere of \hdtwo.
These Zr~\textsc{ii} lines likely form in LTE.~
Our recommended Zr abundance is based on the weighted average 
of these 73 Zr~\textsc{ii} lines.

\subsection{Niobium (Nb, $Z = 41$)}
\label{niobium}

We check the 112~lines of Nb~\textsc{ii} listed in \citet{nilsson08}
and \citet{nilsson10},
and 9 of these lines are sufficiently strong and unblended 
to yield abundances.
Figure~\ref{specplot3} illustrates two of these lines.
Nb has a low FIP, 6.76~eV, so Nb$^{+}$ dominates in the
atmosphere of \hdtwo,
and these Nb~\textsc{ii} lines likely form in LTE.~
\citet{roederer18c} derived an Nb abundance from a single
optical Nb~\textsc{ii} line, and 
our recommended Nb abundance is based on the weighted average 
of these 10 Nb~\textsc{ii} lines.

\subsection{Molybdenum (Mo, $Z = 42$)}

We check the 49~Mo~\textsc{ii} lines from the list of lines with 
experimental \loggf\ values from \citet{sikstrom01}.
Of these, 12 lines in our spectrum yield Mo abundances,
and Figure~\ref{specplot3} illustrates three of them.
The weighted mean Mo abundance, 
\logeps{Mo} = 1.36, is in reasonable agreement with the 
Mo abundance derived by \citet{roederer18c}
from 3 Mo~\textsc{i} lines in the optical spectrum,
\logeps{Mo} = 1.30.
The FIP of Mo is 7.09~eV, so most Mo is found as Mo$^{+}$
in the atmosphere of \hdtwo.
The agreement between these two values suggests that NLTE
effects are minimal, 
as previously found for a few other stars 
(e.g., \citealt{peterson11,roederer14d}).
Our recommended Mo abundance is the weighted average of the
abundances derived from 15 Mo~\textsc{i} and Mo~\textsc{ii} lines.

\subsection{Ruthenium (Ru, $Z = 44$)}

There are 12 Ru~\textsc{ii} lines from the study of \citet{johansson94}
that are covered by our spectrum, and 2 of them are
sufficiently strong and unblended to yield Ru abundances.
One of these lines, at 2281.720~\AA,
is illustrated in Figure~\ref{specplot2}.
These two lines yield \logeps{Ru} = 1.26,
which is in fair agreement
with the Ru abundance derived by \citet{roederer18c}
from 5 optical Ru~\textsc{i} lines, 
\logeps{Ru} = 1.34.
\citet{peterson11} found similarly good agreement in two stars.
The FIP of Ru is 7.36~eV, and most Ru is Ru$^{+}$ in the
atmosphere of \hdtwo.
Any NLTE overionization of Ru appears to be minimal.
Our recommended Ru abundance is the weighted average of
the abundances derived from 7 Ru~\textsc{i} and Ru~\textsc{ii} lines.

\subsection{Rhodium (Rh, $Z = 45$), Palladium (Pd, $Z = 46$), and
Silver (Ag, $Z = 47$)}
\label{silver}

We check our UV spectrum for a handful of the 
potentially strongest lines of 
Rh~\textsc{ii} \citep{quinet12,backstrom13},
Pd~\textsc{i} \citep{morton00},
Ag~\textsc{i} \citep{morton00}, and
Ag~\textsc{ii} (NIST ASD).
All are too weak or blended to be useful as abundance indicators.
Rh and Ag have low FIPs, 7.46 and 7.58~eV, respectively,
while the FIP of Pd is higher, 8.34~eV.
Rh and Ag are two of the few \rpro\ elements whose abundances in \hdtwo\
are derived from their minority states.
The abundances of Rh and Ag could be underestimated in 
LTE, but no NLTE calculations are available to 
confirm or refute this assertion.
We advise using these abundances with caution
until such calculations are available.
Our recommended Rh, Pd, and Ag abundances are based on the 
3 Rh~\textsc{i} lines,
3 Pd~\textsc{i} lines, and
1 Ag~\textsc{i} line
detected in the optical spectrum of \hdtwo\ by \citet{roederer18c}.

\subsection{Cadmium (Cd, $Z = 48$)}
\label{cadmium}

We detect the Cd~\textsc{i} line at 2288.020~\AA,
as shown in Figure~\ref{specplot2}.
It is blended with the As~\textsc{i} line at 2288.115~\AA,
as noted in Appendix~\ref{arsenic},
and both lines are sufficiently resolved
to yield independent abundance results.
The Cd~\textsc{ii} line at 2144.394~\AA, previously used by
\citet{roederer12b},
is too blended in our spectrum to yield a reliable abundance.

There are 5 naturally occurring isotopes of Cd that are 
accessible to the \rpro.
\citet{roederer10b} noted that the odd-$A$ Cd isotopes
comprise a small fraction of all Cd isotopes,
so their HFS can be neglected.
The IS are expected to be small.
We adopt the \loggf\ value for this line from \citet{xu04},
and its uncertainty is $\approx$0.05~dex.
Recent theoretical calculations
of the radiative lifetime of the $^{1}P_{1}$ upper level
\citep{yamaguchi19} and the oscillator strength \citep{shukla22}
are in agreement with the \citeauthor{xu04}\ 
experimental value.
The FIP of Cd is 8.99~eV, so a substantial fraction of
neutral Cd is present in the atmosphere of \hdtwo.
No NLTE investigation of Cd line formation is available at present.
Our recommended Cd abundance is based only
on the single Cd~\textsc{i} line.

The relatively low Cd abundance in \hdtwo\ (Section~\ref{rpropattern})
warrants additional scrutiny.
Neglecting to model an unidentified blending feature
would cause us to overestimate the Cd abundance, not underestimate it.
Similarly, neglecting to model multiple isotopes or HFS of 
$^{111}$Cd and $^{113}$Cd, which have only a small
nuclear spin, $I = 1/2$, would also cause us to 
overestimate the abundance.
Neutral Cd has a relatively simple spectrum, and its
partition function is very low in
the relevant temperature range
($\approx$1.01 at $\log\tau \sim$~0 in this model atmosphere),
so it is also unlikely that an incomplete partition function
is at fault.
The continuum is reasonably well identified in this spectral region
(Figure~\ref{specplot2}), 
and the abundances derived from other features in this wavelength range
appear normal (Figure~\ref{fegroupplot}).
We find no reason to discount the abundance derived from the
Cd~\textsc{i} line at 2288.020~\AA.~

\subsection{Indium (In, $Z = 49$)}

We detect absorption at the wavelength of the 
In~\textsc{ii} line at 2306.064~\AA,
as illustrated in Figure~\ref{specplot2}.
This line has not previously been examined in cool stars.
It connects the
$5s^{2}$ $^{1}S_{0}$ ground state to the
$5s5p$ $^{3}P_{1}$ excited state.

There are two stable isotopes of In, $^{113}$In and $^{115}$In.
$^{113}$In only comprises 4.28\% of In atoms in the solar system,
and it is blocked by $^{113}$Cd from the \rpro\ $\beta$-decay
chains, so we assume that all In is $^{115}$In in \hdtwo.
These isotopes have a large nuclear spin, $I = 9/2$,
and wide HFS, which is apparent in Figure~\ref{specplot2}.
The IS ($<$ 0.002 \AA; see below) is small compared to the 
wide-HFS components ($\approx$0.12~\AA), so the
$^{113}$In and $^{115}$In isotopes absorb at virtually the same wavelengths.
The derived abundance is thus 
virtually insensitive to the isotope fraction.
HFS $A$ and $B$ constants for the upper level 
of $^{115}$In
measured by \citet{konig20} agree with the earlier experimental work
by \citet{larkins93} and \citet{karlsson01}
and with the theoretical calculations by \citet{jonsson07}.
We adopt the \citeauthor{konig20}\ values,
which are effectively interchangeable with
others for the purposes of stellar abundance analyses.

We adopt the IS from \citet{wang07}.
The sign convention now in use yields a positive IS
when a heavier isotope is to the blue,
but this convention was not always followed 
in older literature.
The IS depends primarily on the upper and lower
configurations of a transition.
The mass shifts are small in the middle of the periodic table,
and if either configuration includes an $s$-electron
then the field shift of that configuration 
will often dominate the total IS.~
This results in the IS of both signs.
In subsequent discussions of elements with more than one isotope,
we resolve the sign ambiguity using a statement 
that the heavier or lighter isotopes are to the blue
for transitions of interest.
The lighter isotope is to the blue of the heavier isotope
for the In~\textsc{ii} line at 2306~\AA\
that connects to the ground $5s^{2}$ configuration.
This is clear from Figures~1 and 3 of \citeauthor{wang07},
and follows from first-order perturbation theory. 
We present the complete line component pattern 
in Table~\ref{inhfstab}.
The center-of-gravity wavenumber and air wavelength listed
in Table~\ref{inhfstab} are given for a solar system isotopic mixture.

\begin{deluxetable}{cccccccc}
\tablecaption{Hyperfine Structure and Isotope Shift 
Line Component Pattern for the In~\textsc{ii} $\lambda$2306 Line
\label{inhfstab}}
\tabletypesize{\small}
\tablehead{
\colhead{Wavenumber} &
\colhead{$\lambda_{\rm air}$} &
\colhead{$F_{\rm upper}$} &
\colhead{$F_{\rm lower}$} &
\colhead{Component Position} &
\colhead{Component Position} &
\colhead{Strength} &
\colhead{Isotope} \\
\colhead{(cm$^{-1}$)} &
\colhead{(\AA)} &
\colhead{} &
\colhead{} &
\colhead{(cm$^{-1}$)} &
\colhead{(\AA)} &
\colhead{} &
\colhead{}
}
\startdata
43350.582 & 2306.0645 & 5.5 & 4.5 &    1.060596 & $-$0.056419 & 0.40000 &  113 \\
43350.582 & 2306.0645 & 4.5 & 4.5 & $-$0.199477 &    0.010611 & 0.33333 &  113 \\
43350.582 & 2306.0645 & 3.5 & 4.5 & $-$1.258196 &    0.066931 & 0.26667 &  113 \\
43350.582 & 2306.0645 & 5.5 & 4.5 &    1.040281 & $-$0.055338 & 0.40000 &  115 \\
43350.582 & 2306.0645 & 4.5 & 4.5 & $-$0.223472 &    0.011888 & 0.33333 &  115 \\
43350.582 & 2306.0645 & 3.5 & 4.5 & $-$1.284827 &    0.068347 & 0.26667 &  115 \\
\enddata
\tablecomments{%
Energy levels from the NIST ASD and the index of air \citep{peck72}
are used to compute the
center-of-gravity wavenumbers and air wavelengths, $\lambda_{\rm air}$,
and component positions are given relative to those values.
Strengths are normalized to sum to 1 for each isotope.
Table~\ref{inhfstab} is available in the online edition
of the journal in machine-readable format.
}
\end{deluxetable}

The NIST ASD recommends the semiempirical \loggf\ value
from \citet{curtis00in}, $-$2.30,
which is assigned a grade of B+, (7\%, or 0.03~dex).
In has a low FIP, 5.79~eV.~
In$^{+}$ is the dominant ionization state in the atmosphere of \hdtwo,
so the line likely forms in LTE.~

There are two other nearby blends
that have a minimal impact on the In~\textsc{ii} line.
A high-excitation (E$_{\rm low}$\ = 6.09~eV)
Fe~\textsc{ii} line is detected at 2305.968~\AA,
and the NIST ASD grades the quality of its \loggf\ value as D+
($<$~40\%).
We adjust the strength of this line within this allowed range
to fit the observed line profile (\loggf\ = $-1.06$).
An Fe~\textsc{i} line is detected at 2306.172~\AA.~
The NIST ASD does not recommend a \loggf\ value for this line,
so we adjust its strength to match the observed spectrum
(\loggf\ = $-2.4$).

How likely is it that this absorption is due to In$^{+}$?
Confirmation of this identification using other In~\textsc{ii} lines
could be helpful, but no such lines are available.
The only other In~\textsc{ii} transition
with $\lambda >$ 1600~\AA\ and E$_{\rm low} < 5$~eV, at 
2364.686~\AA, is too blended with a strong Fe~\textsc{ii} line
at 2364.828~\AA\ to be of any use in confirming
the identification of the $\lambda$2306 line.
The predicted strengths of two other lines 
in the \citet{kurucz11} list with similar wavelengths,
Ni~\textsc{ii} $\lambda$2306.027~\AA\ and
Mn~\textsc{ii} $\lambda$2306.028~\AA,
are each more than 2~dex weaker than what would be required
to match the observed line in \hdtwo,
and their wavelengths are not as well matched to the
line center as the In~\textsc{ii} line is.
We conclude that there are no other lines in the
NIST ASD or \citet{kurucz11}
line lists that can plausibly account for this absorption.

We follow the same procedure as described in Appendix~\ref{gallium}
to check whether this absorption could be accounted for
by an unidentified line of an Fe-group element.
This wavelength is only covered by E230H spectra of
\mbox{HD~84937}, \mbox{HD~94028}, and \mbox{HD~140283},
so we supplement our search with the 
lower-resolution ($R$ = 30,000) STIS E230M spectrum 
of \mbox{HD~196944} (GO-12554, PI:\ Beers).
None of these spectra show absorption at this wavelength.
We consider 4 fictitious lines:
a low-excitation weak Fe~\textsc{i} line 
 (E$_{\rm low}$/\loggf\ = 1.0~eV/$-3.6$),
a high-excitation strong Fe~\textsc{i} line (5.0~eV/$+0.2$),
a low-excitation weak Fe~\textsc{ii} line (1.0~eV/$-5.3$), and
a high-excitation strong Fe~\textsc{ii} line (5.0~eV/$-1.5$).
We set the \loggf\ values of these fictitious 
lines to reproduce the absorption observed in \hdtwo,
and we generate synthetic spectra for each of the four other stars.
All of these fictitious lines 
overpredict the absorption in these four stars.
This result boosts our confidence that the absorption
detected at 2306.064~\AA\ is In~\textsc{ii}.
Our recommended In abundance is based on this one In~\textsc{ii} line.

\subsection{Tin (Sn, $Z = 50$)}
\label{tin}

We identify absorption at the wavelength of an Sn~\textsc{ii} line
at 2151.514~\AA.~ 
This line, illustrated in Figure~\ref{specplot1},
has not previously been examined in cool stars.
It connects the
$5s^{2}5p$ $^{2}P^{\rm o}_{1/2}$ ground state to the
$5s5p^{2}$ $^{4}P_{1/2}$ excited state.
The NIST ASD recommends \loggf\ = $-2.53$, 
with a grade of C+ (18\%, 0.09~dex),
based on the theoretical calculations of \citet{oliver10}.
There are 10 naturally occurring or stable isotopes of Sn,
and 6 of them are accessible to the \rpro.
We ignore any HFS or IS in our calculations.
The IS are small for this region of the periodic table.
The stable odd-$A$ Sn nuclei accessible to the \rpro\
($^{117}$Sn and $^{119}$Sn) that might exhibit HFS
comprise only $\approx$25\% of the \rpro\ abundance of Sn \citep{sneden08},
and they have only a small nuclear spin, $I = 1/2$.
Sn has a relatively low FIP, 7.34~eV, and most Sn is
singly ionized in the atmosphere of \hdtwo.
We derive \logeps{Sn} = 1.39 $\pm$~0.15.

How likely is it that this absorption is due to Sn~\textsc{ii}?
We perform the same test as described in Appendix~\ref{gallium},
using 4~stars with spectra covering this wavelength:\
\mbox{HD~76932}, \mbox{HD~84937}, \mbox{HD~140283}, and \mbox{HD~196944}.
We again consider 4~cases of fictitious lines:
a low-excitation weak Fe~\textsc{i} line 
 (E$_{\rm low}$/\loggf\ = 1.0~eV/$-3.5$),
a high-excitation strong Fe~\textsc{i} line (5.0~eV/$+0.3$),
a low-excitation weak Fe~\textsc{ii} line (1.0~eV/$-5.2$), and
a high-excitation strong Fe~\textsc{ii} line (5.0~eV/$-1.4$).
In addition, 
there is one other plausible identification for this absorption
given in the NIST ASD or \citet{kurucz11} lists,
an Fe~\textsc{ii} line at 2151.508~\AA.~
We perform the same test for this known line,
which has E$_{\rm low}$ = 7.48~eV.
We derive \loggf\ = $+0.8$ for this line by 
assuming that it alone accounts for the absorption at this wavelength
and fitting its strength to the observed line profile in \hdtwo.
Our test excludes all possibilities except the
fictitious low-excitation Fe~\textsc{ii} line.

There are three other Sn~\textsc{ii} lines that could potentially 
be strong enough to be detected in our spectrum
($\lambda\lambda$2150.845, 2266.016, and 2368.226~\AA).~
Unfortunately, all are too blended with stronger features.
\citet{peterson20} detected a stronger Sn~\textsc{ii} line at
$\lambda_{\rm vacuum}$ = 1899.898~\AA\
in the spectra of several other stars,
but that study did not examine the line at 2151.514~\AA.~
One star in the comparison sample, \mbox{HD~94028}, 
covers both lines.
\citet{oliver10} also calculated the transition probability for
the Sn~\textsc{ii} line at 1899~\AA, and 
the NIST ASD recommends \loggf\ = $-0.22$, with a B+ grade
(7\% uncertainty, 0.03~dex).
We synthesize both lines in \mbox{HD~94028}.
The Sn~\textsc{ii} line at 1899~\AA\ is easily detectable
and yields \logeps{Sn} = 0.9 $\pm$~0.2,
which is reasonable \citep{roederer16c}.
The Sn~\textsc{ii} line at 2151~\AA\ is much weaker
and not detected.
The upper limit we infer from this line,
\logeps{Sn} $< 1.2$, is compatible with the
abundance derived from the $\lambda$1899 line.

We also check several lines of the neutral, minority species
that are covered by our spectrum.
No Sn~\textsc{i} lines are detected, and we derive upper limits
from three lines that are not blended with other strong features.
The most constraining upper limit is derived from the Sn~\textsc{i} line
at 2199.346~\AA.~
This value, \logeps{Sn} $< 1.5$, is compatible with the abundance
derived from the Sn~\textsc{ii} line at 2151~\AA.~

We conclude that an Sn~\textsc{ii} line
is the most likely explanation for
absorption at 2151.514~\AA,
given current data.
We caution, however, that an unidentified 
low-excitation ion from an Fe-group element
could mimic the behavior observed in the spectra
available to us, 
so we urge further study of this line in additional stars.

\subsection{Antimony (Sb, $Z = 51$)}

We detect absorption at the wavelengths of two Sb~\textsc{i} lines,
at 2068.344 and 2175.818~\AA.~
The $\lambda$2175 line is illustrated in Figure~\ref{specplot1}.
We adopt the atomic transition probabilities from
\citet{hartman10},
which are reliable to $\approx$5\%.
Both lines are dominant branches ($\approx$90\%) from the upper levels.
The FIP of Sb is 8.61~eV.
Most Sb is ionized in the atmosphere of \hdtwo,
but a substantial reservoir of neutral Sb is likely to be present.

There are two stable isotopes of Sb,
$^{121}$Sb and $^{123}$Sb.
These isotopes have nonzero nuclear spin 
$I = 5/2$ and $I = 7/2$, respectively.
We adopt the ground-level HFS $A$ and $B$ values reported by 
\citet{fernando60}, which are
based on atomic beam magnetic resonance measurements, and
are of high accuracy and precision.
Two studies \citep{buchholz78,hassini88}
reported excited-level HFS constants of $^{121}$Sb 
for the resonance lines of interest,
and their results are in near agreement.
\citet{sobolewski16a,sobolewski16b} 
also reported HFS $A$ constants of $^{123}$Sb 
and a $B$ constant for one ($\lambda$2175) of the two lines.
We scale the HFS $B$ values for $^{121}$Sb 
found by \citeauthor{hassini88}\ 
to generate HFS $B$ values for both $^{123}$Sb lines.
Our scaled $B$ value for the upper level of the
$^{123}$Sb $\lambda$2175 line
matches the \citet{sobolewski16b} experimental value.
Our choice of HFS $B$ constants
has a negligible impact on the derived abundances.

The IS of 287~MHz for the 2175~\AA\ line 
was computed by \citet{gamrath18} using a sophisticated 
multiconfiguration Dirac-Hartree-Fock code.
The two Sb~\textsc{i} lines at 2068 and 2175~\AA\
in Table~\ref{sbhfstab} both connect a common upper 
$5p^{2}(^{3}P)6s$ configuration and common lower 
$5p^{3}$ ground configuration. 
Table~4 of \citeauthor{gamrath18}\ shows that the
field shift from the participation of an 
$s$ electron in the upper or lower configuration dominates the
normal and specific mass shifts 
for heavy elements, such as Sb.
The heavier isotope is to the blue
if the $s$ electron is in the upper, but not lower,
configuration.
Sb has a positive IS using the standard sign convention.
Although their mixing percentages are slightly different, 
we adopt the same IS value (287~MHz) for both lines.
We present the line component patterns for the
Sb~\textsc{i} lines at 2068 and 2175~\AA\ 
in Table~\ref{sbhfstab}.
The center-of-gravity wavenumber and air wavelength 
in Table~\ref{sbhfstab} are given for a solar system isotopic mixture.

\begin{deluxetable}{cccccccc}
\tablecaption{Hyperfine Structure and Isotope Shift 
Line Component Patterns for Sb~\textsc{i} Lines
\label{sbhfstab}}
\tabletypesize{\small}
\tablehead{
\colhead{Wavenumber} &
\colhead{$\lambda_{\rm air}$} &
\colhead{$F_{\rm upper}$} &
\colhead{$F_{\rm lower}$} &
\colhead{Component Position} &
\colhead{Component Position} &
\colhead{Strength} &
\colhead{Isotope} \\
\colhead{(cm$^{-1}$)} &
\colhead{(\AA)} &
\colhead{} &
\colhead{} &
\colhead{(cm$^{-1}$)} &
\colhead{(\AA)} &
\colhead{} &
\colhead{}
}
\startdata
48332.424 & 2068.3440 & 5.0 & 4.0 &    0.399814 & $-$0.017112 & 0.30555 &  121 \\
48332.424 & 2068.3440 & 4.0 & 4.0 &    0.102274 & $-$0.004377 & 0.06250 &  121 \\
48332.424 & 2068.3440 & 4.0 & 3.0 &    0.062277 & $-$0.002665 & 0.18750 &  121 \\
\enddata
\tablecomments{%
Energy levels from the NIST ASD and the index of air \citep{peck72}
are used to compute the
center-of-gravity wavenumbers and air wavelengths, $\lambda_{\rm air}$,
and component positions are given relative to those values.
Strengths are normalized to sum to 1 for each isotope.
The complete version of Table~\ref{sbhfstab} is available in the online edition
of the journal in machine-readable format.
A short version is included here to demonstrate its form and content.
}
\end{deluxetable}

The $\lambda$2068.344 line is found in the blue wing of a stronger
Cr~\textsc{ii} line at 2068.395~\AA,
and it is nearly coincident with an Fe~\textsc{ii} line
at 2068.320~\AA.~
This Sb~\textsc{i} line is too blended to be used as an
abundance indicator in \hdtwo.
We derive an upper limit on the Sb abundance by assuming
all of the absorption is due to Sb~\textsc{i} and Cr~\textsc{ii},
\logeps{Sb} $< 0.8$.

The $\lambda$2175.818 line is less blended and clearly detected,
with a depth $\approx$35\% of the continuum,
as illustrated in Figure~\ref{specplot1}.
There are no obvious blends at this wavelength,
and a nearby Fe~\textsc{ii} ($\lambda$2175.725)
line can easily be fit in our syntheses
with minimal impact on the derived Sb abundance.
An unidentified absorption line at 
2175.907~\AA\ is present.
We model it as
an Fe~\textsc{i} line with E$_{\rm low}$ = 1.0~eV and \loggf = $-2.95$, 
and it, too, has minimal impact on the derived Sb abundance.
We derive \logeps{Sb} = 0.37 $\pm$~0.20
from this line, which is compatible with the
upper limit derived from the $\lambda$2068 line.
Our recommended Sb abundance is based on the 
Sb~\textsc{i} line at 2175.818~\AA.~

\subsection{Tellurium (Te, $Z = 52$)}

We detect two Te~\textsc{ii} lines in our spectrum,
at 2259.034 and 2385.792~\AA.~
Other Te~\textsc{i} lines, 
at 2142.822 and 2383.277~\AA, are too blended to be of use
as abundance indicators in \hdtwo.
\citet{roederer12a} discussed the $\lambda$2385 line in detail,
and we adopt the \loggf\ for this line from that study.
The uncertainty in its \loggf\ value is $\approx$16\% (0.08~dex).
This line is in a region where the continuum is depressed $\approx$25\%
by strong Fe~\textsc{ii} lines at 2382.039 and 2388.629~\AA.~
We account for these lines in our syntheses.

The Te~\textsc{i} line at 2259.034~\AA\ has not previously been
studied in cool stars,
and it is illustrated in Figure~\ref{specplot2}.
It connects the
$5p^{4}$ $^{3}P_{2}$ ground level to the
$5p^{3}$($^{4}S^{\rm o}$)$6s$ $^{5}S^{\rm o}_{2}$ excited level.
This line is virtually unblended and unsaturated in \hdtwo,
so it is an ideal abundance indicator.
\citet{morton00} quotes a \loggf\ value for this line,
$-1.31$,
using an upper-level radiative lifetime measurement (accurate to 3\%)
from \citet{garpman71} and branching ratios from \citet{ubelis83}.
This line is a major ($\approx$92\%) branch.
The \loggf\ value should be reliable to within
a few hundredths of a dex.

The Te~\textsc{i} line at 2002.028~\AA\ is covered by our spectrum.
We calculate a \loggf\ value of $-1.60 \pm 0.17$ from the 
experimental transition probability obtained by \citet{ubelis91}.
This line is probably detected,
but the S/N in our spectrum is too low to be useful.
Furthermore, 
analysis of this line in other stars with STIS spectra
yields Te abundances considerably higher than
those from other Te~\textsc{i} lines,
suggesting the presence of an unidentified blend at this wavelength.
We discard this line from further consideration.

Four stable Te isotopes are accessible to the \rpro,
$^{125}$Te, $^{126}$Te, $^{128}$Te, and $^{130}$Te.
Only $^{125}$Te has a nonzero nuclear spin $I = 1/2$,
which is small.
This isotope comprises only $\approx$7\% of
the \rpro\ Te isotope mix \citep{sneden08},
so we ignore the HFS from this isotope.
We also ignore the small IS of these Te isotopes.
The FIP of Te is high, 9.01~eV, 
so a substantial fraction of neutral Te is present
in the atmosphere of \hdtwo.
We derive \logeps{Te} = 1.76 $\pm$~0.25
from the Te~\textsc{i} $\lambda$2385 line
and \logeps{Te} = 1.61 $\pm$~0.10
from the Te~\textsc{i} $\lambda$2259 line.
Our recommended Te abundance is based on a weighted average 
of these two values.

\subsection{Lanthanide Elements}
\label{ree}

Some of the lanthanide elements, including
gadolinium (Gd, $Z = 64$),
terbium (Tb, $Z = 65$),
dysprosium (Dy, $Z = 66$),
erbium (Er, $Z = 68$),
thulium (Tm, $Z = 69$),
and ytterbium (Yb, $Z = 70$)
are detectable in our spectrum of \hdtwo.
\citet{roederer18c} derived abundances
of these elements from the optical spectrum,
often using many more lines.
We derive abundances from
6~lines of Gd~\textsc{ii} in the UV spectrum
(compared with 38~lines in the optical spectrum),
1~line of Tb~\textsc{ii} (3~lines),
1~line of Dy~\textsc{ii} (32~lines),
3~lines of Er~\textsc{ii} (13~lines),
3~lines of Tm~\textsc{ii} (7~lines),
and
1~line of Yb~\textsc{ii} (1~line).
We use the same sources of \loggf\ values,
from the Wisconsin group's work,
so our results are on a consistent scale with
previous abundance derivations.
The uncertainties in these \loggf\ values are generally 
better than 5\% (0.02~dex).
The abundances derived from UV lines
and optical lines are in excellent agreement.
Our recommended abundances for these elements are
based on the weighted averages of these values.
Our recommendations for all other lanthanide elements,
plus Ba,
except Lu (Appendix~\ref{lutetium}),
are adopted from \citeauthor{roederer18c} 

The Ho abundance is slightly low
relative to the other lanthanide elements in \hdtwo\
(see Section~\ref{rpropattern}), so it deserves special mention.
We reexamine
all nine lines that were used by \citet{roederer18c} to derive
the Ho abundance,
recheck the atomic data used to generate our syntheses,
and confirm that our synthesis parameter files are correct.
We find no fault with the original analysis.
We also assess how much the original line fits
could be adjusted to maximize the abundance;
for example, by making different choices about the
strengths of blending features, while 
continuing to adequately reproduce the observed spectrum.
This approach could only increase the Ho abundance
by $\approx$0.04~dex, which is insufficient to 
explain the discrepancy.
The MOOG partition function for Ho~\textsc{ii} agrees with 
the NIST ASD value to within $\approx$5\% at $\log\tau \sim$~0.
Using the NIST partition functions would lead to a
small decrease in the Ho abundance, $\approx$0.03~dex or less,
exacerbating the discrepancy.
We find no reason to discount the Ho abundance
presented previously.

\subsection{Lutetium (Lu, $Z = 71$)}
\label{lutetium}

We check for 22 UV lines of Lu~\textsc{ii}, and 
7 of them are sufficiently strong and unblended to yield 
reliable abundances.
Figure~\ref{specplot3} illustrates the Lu~\textsc{ii} line at
2911.392~\AA.~
We adopt \loggf\ values for these lines from
\citet{quinet99}, \citet{lawler09}, and \citet{roederer10b}.
There are two stable isotopes of Lu,
$^{175}$Lu and $^{176}$Lu, and only
the majority isotope 
$^{175}$Lu (97.4\% in the solar system) is accessible to the \rpro.
It has nuclear spin $I = 7/2$,
which produces wide HFS patterns,
as is apparent in Figure~\ref{specplot3}.
We adopt the HFS patterns recently published by \citet{denhartog20}.
\citet{roederer18c} derived the Lu abundance in \hdtwo\
from two optical lines.
That study accounted for HFS and used a consistent set of \loggf\ values.
Lu has a low FIP, 5.43~eV, so
virtually all Lu atoms are ionized in the atmosphere of \hdtwo.
Our recommended Lu abundance 
reflects the weighted mean of the 7 UV lines and 2 optical lines.

\subsection{Hafnium (Hf, $Z = 72$)}

We check for 28 UV lines of Hf~\textsc{ii}, and 
15 of them yield reliable abundances.
Figure~\ref{specplot3} illustrates the
Hf~\textsc{ii} line at 2647.297~\AA.~
We adopt \loggf\ values from \citet{lawler07} and
\citet{denhartog21hf}.
Four stable Hf isotopes are accessible to the \rpro,
$^{177}$Hf, 
$^{178}$Hf, 
$^{179}$Hf, and
$^{180}$Hf, 
and the $^{177}$Hf and $^{179}$Hf isotopes have
large nonzero nuclear spins 
$I = 7/2$ and $9/2$, respectively.
The HFS and IS of the Hf~\textsc{ii} lines of interest have 
not been studied in the laboratory.
We observe no substantial broadening of the Hf~\textsc{ii} lines 
in our spectrum, and most of the Hf~\textsc{ii} lines
are on the weak part of the curve of growth.
Hf has a low FIP, 6.83~eV, and singly ionized Hf is the 
dominant species in the atmosphere of \hdtwo.
Our recommended Hf abundance is based on a weighted average of 
the 15~UV lines and 5~optical lines studied by \citet{roederer18c}.

\subsection{Tantalum (Ta, $Z = 73$)}

We have checked 20 potential Ta~\textsc{ii} lines
in our of spectrum of \hdtwo, but
we do not detect any of them.
Ta has a low FIP, 7.55~eV, so Ta$^{+}$ is the
dominant species in the atmosphere of \hdtwo.

\citet{siqueiramello13} reported detections of two Ta~\textsc{ii} lines,
at 2635.583 and 2832.702~\AA,
in a STIS E230M ($R = 30,000$) spectrum of the
metal-poor, \rpro-enhanced red giant
\object[BPS CS 31082-001]{CS~31082-001}.
Absorption is detected at the $\lambda$2635 line
in our spectrum of \hdtwo, but we cannot 
confidently attribute this absorption to Ta~\textsc{ii}.
Lines of Mn~\textsc{i} ($\lambda$2635.561) and
OH ($\lambda$2635.581) are potential absorbers.
We calculate 
\loggf\ = $-0.16 \pm 0.03$ for this Ta~\textsc{ii} line
from the transition probability published by \citet{quinet09}.
This value is 0.86~dex lower than the
\loggf\ value adopted by \citeauthor{siqueiramello13}\ from VALD,
and it predicts a substantially weaker line.
No absorption is detected at the $\lambda$2832 line in \hdtwo, 
despite the continuum being clear in this region.
\citeauthor{quinet09}\ did not cover the $\lambda$2832 line in their study,
and we are reluctant to derive an upper limit from this line
using the VALD \loggf\ value.
Our upper limit is instead derived from the Ta~\textsc{ii} line at
2752.486~\AA.~

\subsection{Tungsten (W, $Z = 74$)}

We check our spectrum 
for the 40 W~\textsc{ii} lines 
studied by \citet{kling00} and \citet{nilsson08w}.
Six of these lines yield acceptable abundances in \hdtwo,
and Figure~\ref{specplot1} illustrates two of them.
The FIP of W is 7.86~eV, so most W atoms are
singly ionized in the atmosphere of \hdtwo.

W has five stable isotopes, four of which are produced
by the \rpro:\
$^{182}$W, $^{183}$W, $^{184}$W, and $^{186}$W.~
Only $^{183}$W has nonzero nuclear spin $I = 1/2$.
No information on the HFS of this isotope is available.
The IS of W~\textsc{ii} lines have been studied by \citet{aufmuth95},
but none of the W~\textsc{ii} lines
in our spectrum have measured IS.~
Furthermore the assignment configurations and terms of levels 
in W$^{+}$ are also incomplete.
Low-lying even-parity levels are typically a mixture of 
$5d^{5}$, $5d^{4}6s$, and $5d^{3}6s^{2}$ configurations.
As noted previously, transitions involving an $s$-electron produce large 
field shifts.
The odd-parity levels studied by \citeauthor{aufmuth95}\
are primarily mixtures of the $5d^{4}6p$ and $5d^{3}6s6p$ configurations.
The upper levels of the relatively strong spectral lines in our study 
are also likely to include significant contributions 
from these configurations.
\citeauthor{aufmuth95}\ found negative field shifts as large as 
$-$0.155~cm$^{-1}$ for pure $5d^{3}6s^{2}$ to $5d^{4}6p$ transitions 
of $^{186}$W compared to $^{184}$W.~
\citeauthor{aufmuth95}\ used the standard sign convention,
and we use their results to estimate the IS for the W~\textsc{ii} lines.
We present the line component patterns for eight lines of W~\textsc{ii},
including several lines that may be useful abundance indicators in other stars,
in Table~\ref{wistab}.
These eight W~\textsc{ii} lines 
all connect to the ground $5d^{4}(^{6}S)6s$ configuration,
and have lighter isotopes to the blue of the heavier isotope.

\begin{deluxetable}{ccccc}
\tablecaption{Estimated Isotope Shift Line Component Patterns 
for W~\textsc{ii} Lines
\label{wistab}}
\tabletypesize{\small}
\tablehead{
\colhead{Wavenumber} &
\colhead{$\lambda_{\rm air}$} &
\colhead{Component Position} &
\colhead{Component Position} &
\colhead{Isotope} \\
\colhead{(cm$^{-1}$)} &
\colhead{(\AA)} &
\colhead{(cm$^{-1}$)} &
\colhead{(\AA)} &
\colhead{} 
}
\startdata
49245.361 & 2029.9948 &     0.301529 &   $-$0.012431 & 180 \\
49245.361 & 2029.9948 &     0.146529 &   $-$0.006041 & 182 \\
49245.361 & 2029.9948 &     0.069029 &   $-$0.002846 & 183 \\
\enddata
\tablecomments{%
Energy levels from the NIST ASD and the index of air \citep{peck72}
are used to compute the
center-of-gravity wavenumbers and air wavelengths, $\lambda_{\rm air}$,
and component positions are given relative to those values.
The complete version of Table~\ref{wistab} is available in the online edition
of the journal in machine-readable format.
A short version is included here to demonstrate its form and content.
}
\end{deluxetable}

Among the six W~\textsc{ii} lines we use to derive an abundance,
the IS are estimated for four of them
($\lambda\lambda$2088.204, 2094.751, 2118.875, and 2194.528).
We do not include the IS in our syntheses of the W~\textsc{ii} lines
at 2204.489 and 2658.032~\AA.~
We assess the abundance sensitivity to the IS
by deliberately excluding it
from the test syntheses of the lines for which an IS is estimated.
The inferred abundances change by $<$~0.01~dex,
so the impact of the unknown IS for the other two
weak lines is likely minimal.
We adopt an \rpro\ isotope mix in these syntheses.

These six transitions are all major decay branches from
upper levels, whose radiative lifetimes have been measured 
to better than $\approx$10\%.
The \citet{kling00} and \citet{nilsson08w} \loggf\ values
agree to within 0.03~dex for the one line of these six in common
($\lambda$2118).
The W~\textsc{ii} line at 2204.489~\AA\ was not included in
either study.
Roederer et al.\ (in preparation) estimated a \loggf\ value
for this line using a reverse abundance analysis
based on five other W~\textsc{ii} lines in the
STIS E230H ($R = 114,000$) spectrum of \object[HD196944]{HD~196944}.
We consider this \loggf\ value to be of lower quality than the
other ones, but it is likely reliable at the $\sim$~0.10--0.15~dex level.
Our recommended W abundance in \hdtwo\ is based on 
a weighted average of these six W~\textsc{ii} lines.

\subsection{Rhenium (Re, $Z = 75$)}

We detect two Re~\textsc{ii} resonance lines in our spectrum,
at 2214.277 and 2275.255~\AA,
as illustrated in Figures~\ref{specplot1} and \ref{specplot2}.
The FIP of Re is 7.83~eV, and Re$^{+}$ is the dominant ionization state
in the atmosphere of \hdtwo.

There are two stable isotopes of Re,
$^{185}$Re and $^{187}$Re, and
both are accessible to the \rpro.
Their nuclear spins are $I = 5/2$,
so they both exhibit HFS.~
We adopt the ground-level HFS $A$ and $B$ values 
reported by \citet{wahlgren97}, 
based on their measurements using a Fourier Transform Spectrometer (FTS).~
The IS of 118.6~mK for the Re~\textsc{ii} 
line at 2214~\AA\ and 117.8~mK for the Re~\textsc{ii} 
line at 2275~\AA\ were also extracted from the FTS data 
by \citeauthor{wahlgren97} 
Their Figure~1 shows that the lighter and rarer isotope
is to the blue,
and these two lines have negative IS,
using the standard convention.
The two Re~\textsc{ii} lines in Table~\ref{rehfstab}
both connect to the ground $5d^{5}$($^{6}S$)$6s$ configuration.
The common upper configuration and common lower configuration 
of the two Re lines yield very similar measured IS, as expected.
We present the line component patterns for these 
two Re~\textsc{ii} lines
in Table~\ref{rehfstab}.

\begin{deluxetable}{cccccccc}
\tablecaption{Hyperfine Structure and Isotope Shift 
Line Component Patterns for Re~\textsc{ii} Lines
\label{rehfstab}}
\tabletypesize{\small}
\tablehead{
\colhead{Wavenumber} &
\colhead{$\lambda_{\rm air}$} &
\colhead{$F_{\rm upper}$} &
\colhead{$F_{\rm lower}$} &
\colhead{Component Position} &
\colhead{Component Position} &
\colhead{Strength} &
\colhead{Isotope} \\
\colhead{(cm$^{-1}$)} &
\colhead{(\AA)} &
\colhead{} &
\colhead{} &
\colhead{(cm$^{-1}$)} &
\colhead{(\AA)} &
\colhead{} &
\colhead{}
}
\startdata
45147.402 & 2214.2770 & 5.5 & 5.5 & $-$1.195604 &    0.058646 & 0.25325 & 185 \\
45147.402 & 2214.2770 & 5.5 & 4.5 & $-$0.405968 &    0.019913 & 0.03247 & 185 \\
45147.402 & 2214.2770 & 4.5 & 5.5 & $-$1.017998 &    0.049934 & 0.03247 & 185 \\
\enddata
\tablecomments{%
Center-of-gravity wavenumbers and air wavelengths, $\lambda_{\rm air}$,
are given with component positions relative to those values.
Strengths are normalized to sum to 1 for each isotope.
The complete version of Table~\ref{rehfstab} is available in the online edition
of the journal in machine-readable format.
A short version is included here to demonstrate its form and content.
}
\end{deluxetable}

We adopt the \loggf\ value
for the Re~\textsc{ii} line at 2214~\AA, $-0.019$, 
from \citet{palmeri05re}.
Measurements by \citet{ortiz13} support this value.
The atomic transition probability for the Re~\textsc{ii} line 
at 2275~\AA\ was determined by both \citet{wahlgren97} 
and \citeauthor{palmeri05re} 
The radiative lifetime measurements of the upper $^{7}$P$_{2}$ level 
are in excellent agreement, $4.47 \pm  0.22$~ns and 
$4.5 \pm 0.3$~ns, respectively.
Both studies report a dominant BF for the
Re~\textsc{ii} line at 2275~\AA.~  
Unfortunately, the experimental BF
of $0.6 \pm 0.04$ from \citeauthor{wahlgren97}\ 
does not agree with the theoretical BF of 0.928 
from \citeauthor{palmeri05re} 
We adopt a simple average of 0.76, resulting in a 
\loggf\ value of $-0.180$.
The \loggf\ values for the Re~\textsc{ii} lines at 2214 and
2275~\AA\ are likely reliable to 
$\approx$7\% (0.03~dex)
and
$\approx$20\% (0.10~dex),
respectively.

Both lines are strong and are the dominant absorbers at their
respective wavelengths in our spectrum, as
shown in Figures~\ref{specplot1} and \ref{specplot2}.
Other blending features are present nearby.
The known lines cannot reproduce the observed line profile
in either case, without the presence of
strong Re~\textsc{ii} lines broadened by HFS.~
We adopt an \rpro\ isotope mix.
The abundances derived from the 
$\lambda$2214 and $\lambda$2275 lines,
\logeps{Re} = $0.10 \pm 0.15$ and $0.25 \pm 0.15$,
respectively,
are in good agreement.
Our recommended Re abundance is based on a
weighted average of these two lines.

\subsection{Osmium (Os, $Z = 76$)}

We check for 12~lines of Os~\textsc{i} and 
22~lines of Os~\textsc{ii} in our spectrum, and
we derive abundances from 
4 Os~\textsc{i} lines and 5 Os~\textsc{ii} lines.
Figure~\ref{specplot2} illustrates two Os~\textsc{ii} lines, and
Figure~\ref{specplot3} illustrates one Os~\textsc{i} line.
The FIP of Os is 8.44~eV, and both neutral and singly ionized
Os are present in \hdtwo.

There are four stable isotopes of Os that are accessible to the \rpro,
$^{188}$Os, 
$^{189}$Os, 
$^{190}$Os, and
$^{192}$Os.
Only $^{189}$Os has a nonzero nuclear spin $I = 3/2$,
and it only comprises $\approx$17\% of the 
predicted \rpro\ isotope mixture \citep{sneden08}.
Little is known about the HFS or IS for 
the Os~\textsc{i} and \textsc{ii} lines.
We assess the importance of HFS and IS as follows.
We mount an Os hollow cathode lamp 
in front of 25~$\mu$m and 10~$\mu$m pinhole entrance slits
to the 3~m focal length laboratory echelle spectrometer
at the University of Wisconsin \citep{wood12}.
This setup produces a resolving power of $R >$~250,000. 
The lines are not resolved, but broadening or hints of structure
are apparent for the Os~\textsc{ii} lines at 2067, 2227, and 2282~\AA.~
This preliminary result suggests that some HFS or IS may be
present at a level that may impact stellar abundance work.
Further investigation of this issue will be presented elsewhere.
We do not include any HFS or IS in our syntheses of
Os~\textsc{i} or \textsc{ii} lines,
which could, in principle, cause us to overestimate the
abundances derived from lines where HFS and IS are important.

We adopt \loggf\ values for both Os~\textsc{i} and 
Os~\textsc{ii} lines from \citet{quinet06}.
\citet{ivarsson04} also published \loggf\ values
for several Os~\textsc{ii} lines, but
these two sets of values agree only moderately well
for the 3 Os~\textsc{ii} lines in common, which are
also used to derive abundances.
This disagreement is mainly attributable to the
different BF values, which were calculated from theory
(\citeauthor{quinet06}) or measured experimentally
and corrected by calculations of residual branches
that fall outside the wavelength range observed (\citeauthor{ivarsson04}).
The mean difference in their \loggf\ values is
$+0.05$~dex, with the \citeauthor{ivarsson04} values being larger,
but the differences range from $-0.09$ to $+0.14$~dex,
with a standard deviation of 0.10~dex.
Formally, the \loggf\ errors are stated to be 
$\approx$6--10\% ($\approx$0.03--0.05~dex), but 
we adopt a conservative uncertainty of 0.10~dex
on these \loggf\ values.

The Os~\textsc{ii} line at 2067.230~\AA\ appears to be
blended with an unidentified species.
This line is broader than would be expected from Os~\textsc{ii}
alone, and it is shifted to the blue by $\approx$0.01~\AA\
relative to the expected center of the Os~\textsc{ii} line.
If we assume that Os~\textsc{ii} is the only 
line absorber at this wavelength,
it yields an abundance $\approx$0.6~dex higher than the mean 
of the other Os~\textsc{i} and Os~\textsc{ii} lines.
We treat this value as an upper limit on the Os abundance.
We do not recommend using this line as an abundance indicator
without further study of the blending feature.

The mean abundance derived from 4 Os~\textsc{i} lines,
\logeps{Os} = 1.19 $\pm$~0.14,
agrees with that derived from 5 Os~\textsc{ii} lines,
\logeps{Os} = 1.09 $\pm$~0.12.
These results agree with the Os abundance derived by \citet{roederer18c}
from two weak optical Os~\textsc{i} lines,
\logeps{Os} = 1.26 $\pm$~0.15.
We note that the UV lines yielding the highest abundances
are not uniformly the ones where hints of broadening
are detected in the laboratory echelle data.
This result suggests that the abundance uncertainties are
dominated by factors other than the neglect of HFS and IS
in our syntheses, likely unidentified minor blends and continuum placement.
Our recommended Os abundance is based on a weighted average
of these 11 lines.

\subsection{Iridium (Ir, $Z = 77$)}

We check for 53~lines of Ir~\textsc{i} and \textsc{ii}
in our spectrum.
Four Ir~\textsc{i} and two Ir~\textsc{ii} lines
are detected and useful as abundance indicators,
and two of these lines are illustrated
in Figures~\ref{specplot2} and \ref{specplot3}.
Several other lines are detected, but they are too
blended to be useful.
The FIP of Ir is high, 8.97~eV, 
and both neutral and singly ionized Ir are present.

There are two stable isotopes of Ir, $^{191}$Ir and $^{193}$Ir.
Both are accessible to the \rpro\ and have
nuclear spin $I = 3/2$.
The HFS and IS have been measured
by \citet{buttgenbach78}, \citet{burger84}, and \citet{gianfrani93}
for one of the Ir~\textsc{i} lines that is useful for abundance work,
$\lambda$2924.790.
This line connects an upper
$5d^{7}6s$($^{5}F$)$6p$ configuration 
to the ground $5d^{7}6s^{2}$ configuration.
The lighter and rarer isotope is to the blue of the heavier isotope,
which is a negative IS,
as shown in Figure~2 of \citeauthor{gianfrani93}.
We present the complete line component pattern for this line
in Table~\ref{irhfstab}.
This line provides the best fit to the
\hdtwo\ spectrum when its wavelength
is shifted by 0.013~\AA\ toward the blue.

\begin{deluxetable}{cccccccc}
\tablecaption{Hyperfine Structure and Isotope Shift
Line Component Pattern for the Ir~\textsc{i} $\lambda$2924 Line
\label{irhfstab}}
\tabletypesize{\small}
\tablehead{
\colhead{Wavenumber} &
\colhead{$\lambda_{\rm air}$} &
\colhead{$F_{\rm upper}$} &
\colhead{$F_{\rm lower}$} &
\colhead{Component Position} &
\colhead{Component Position} &
\colhead{Strength} &
\colhead{Isotope} \\
\colhead{(cm$^{-1}$)} &
\colhead{(\AA)} &
\colhead{} &
\colhead{} &
\colhead{(cm$^{-1}$)} &
\colhead{(\AA)} &
\colhead{} &
\colhead{}
}
\startdata
34180.48  & 2924.7905 & 7   & 6   &    0.096458 & $-$0.008256 & 0.31250 &  191 \\
34180.48  & 2924.7905 & 6   & 6   &    0.023650 & $-$0.002024 & 0.01231 &  191 \\
34180.48  & 2924.7905 & 6   & 5   &    0.045641 & $-$0.003907 & 0.25852 &  191 \\
\enddata
\tablecomments{%
Energy levels from the NIST ASD and the index of air \citep{peck72}
are used to compute the
center-of-gravity wavenumbers and air wavelengths, $\lambda_{\rm air}$,
and component positions are given relative to those values.
Strengths are normalized to sum to 1 for each isotope.
The complete version of Table~\ref{irhfstab} is available in the online edition
of the journal in machine-readable format.
A short version is included here to demonstrate its form and content.
}
\end{deluxetable}

The \loggf\ values for the Ir~\textsc{i}
lines are adopted from the NIST ASD,
which combined upper-level radiative lifetime measurements
from \citet{gough83} with BFs from \citet{xu07}.
With the exception of the Ir~\textsc{i} $\lambda$2481 line,
whose \loggf\ uncertainty is graded by NIST as having D accuracy
($<$50\%, 0.30~dex), the other \loggf\ values are
reliable to better than 7\% (0.03~dex).
The \loggf\ values for the Ir~\textsc{ii} lines are 
adopted from \citet{ivarsson04}, who 
estimated uncertainties of $\approx$7\% (0.03~dex).

\citet{roederer18c} derived the Ir abundance from a single
optical Ir~\textsc{i} line.
We noticed an error in the treatment of the Ir isotope mix
in our synthesis of this line, at 3800.124~\AA, and 
we revise its abundance to
\logeps{Ir} = 1.32.
This value is 0.22~dex lower than that presented in \citeauthor{roederer18c} 

The abundance derived from the two Ir~\textsc{i} lines
with HFS and IS,
\logeps{Ir} = 1.31,
is comparable to that from the
three Ir~\textsc{i} lines without HFS and IS,
\logeps{Ir} = 1.24.
This result suggests that the impact of
the HFS and IS in the line lists
is small for this set of lines in this star.
Both values are much lower than the abundance derived
from the two Ir~\textsc{ii} lines,
\logeps{Ir} = 1.58,
which is not derived using HFS and IS.~
Our recommended Ir abundance
reflects a weighted average of the 
five Ir~\textsc{i} lines.

\subsection{Platinum (Pt, $Z = 78$)}

We check our spectrum of \hdtwo\ for 35 Pt~\textsc{i} and
8 Pt~\textsc{ii} lines among the stronger lines listed in
\citet{denhartog05} and \citet{quinet08}.
We derive abundances from 8 Pt~\textsc{i} lines
(Figures~\ref{specplot2} and \ref{specplot3})
and
1 Pt~\textsc{ii} line (Figure~\ref{specplot1}).
This marks the first 
detection of Pt~\textsc{ii} in a metal-poor star.
\citeauthor{denhartog05}\ quote transition probability 
uncertainties $\approx$5--7\% (0.02--0.03~dex) for 
the lines we use.
\citeauthor{quinet08}\ estimate that their \loggf\ values are 
reliable to $\approx$25\% (0.12~dex).
The high FIP of Pt, 8.96~eV, ensures that a
substantial fraction of both neutral and singly ionized Pt
are present in the atmosphere of \hdtwo.

There are four stable isotopes of Pt that are accessible to 
the \rpro:\
$^{194}$Pt, $^{195}$Pt, $^{196}$Pt, and $^{198}$Pt.
The $^{195}$Pt isotope has nuclear spin $I = 1/2$
and thus exhibits HFS.~
Previous laboratory studies have 
measured the HFS constants and IS for only a limited selection of the
levels required to compute the 
line component patterns for the lines detected in \hdtwo.
We include HFS and IS for two of the lines 
($\lambda\lambda$2646.881 and 2705.895)
from \citet{denhartog05}.
We compute the pattern
for one more line ($\lambda$2274.381)
using the HFS $A$ constants from 
\citet{labelle89} and \citet{basar96} and
the IS from \citeauthor{labelle89}\ and \citet{kronfeldt95}.
This line connects 
an upper $5d^{8}6s$($^{2}F$)$6p$ configuration
to a lower $5d^{9}6s$ configuration.
The lighter isotope is to the blue of the heavier isotope,
which is a negative IS.~
Table~\ref{pthfstab} presents the
complete line component pattern for this line.
Our recommended Pt abundance is based 
on a weighted average of all nine Pt~\textsc{i} and Pt~\textsc{ii} lines.

\begin{deluxetable}{cccccccc}
\tablecaption{Hyperfine Structure and Isotope Shift
Line Component Pattern for the Pt~\textsc{i} $\lambda$2274 Line
\label{pthfstab}}
\tabletypesize{\small}
\tablehead{
\colhead{Wavenumber} &
\colhead{$\lambda_{\rm air}$} &
\colhead{$F_{\rm upper}$} &
\colhead{$F_{\rm lower}$} &
\colhead{Component Position} &
\colhead{Component Position} &
\colhead{Strength} &
\colhead{Isotope} \\
\colhead{(cm$^{-1}$)} &
\colhead{(\AA)} &
\colhead{} &
\colhead{} &
\colhead{(cm$^{-1}$)} &
\colhead{(\AA)} &
\colhead{} &
\colhead{}
}
\startdata
43954.430 & 2274.3808 & 3.0 & 2.0 &    0.065542 & $-$0.003392 & 1.00000 &  190 \\
43954.430 & 2274.3808 & 3.0 & 2.0 &    0.041584 & $-$0.002152 & 1.00000 &  192 \\
43954.430 & 2274.3808 & 3.0 & 2.0 &    0.015028 & $-$0.000778 & 1.00000 &  194 \\
\enddata
\tablecomments{%
Energy levels from the NIST ASD and the index of air \citep{peck72}
are used to compute the
center-of-gravity wavenumbers and air wavelengths, $\lambda_{\rm air}$,
and component positions are given relative to those values.
Strengths are normalized to sum to 1 for each isotope.
The complete version of Table~\ref{pthfstab} is available in the online edition
of the journal in machine-readable format.
A short version is included here to demonstrate its form and content.
}
\end{deluxetable}

\subsection{Gold (Au, $Z = 79$)}

We detect two Au~\textsc{i} resonance lines
at 2427.950 and 2675.950~\AA.~
The $\lambda$2427 line is blended,
and only the $\lambda$2675 line is useful as an abundance indicator.
We also check for weaker Au~\textsc{i} lines at 
2126.630, 2352.580, 2376.240,
2641.480, 2748.250, and 3122.780~\AA\
\citep{fivet06,zhang18}.
None are detected.
Our spectrum is 
relatively unblended around $\lambda$2376 and $\lambda$3122,
and these lines might detectable with higher-S/N spectra.
The FIP of Au is high, 9.23~eV, and 
neutral Au is common in the atmosphere of \hdtwo.
Syntheses of a few of the strongest Au~\textsc{ii} lines
listed in \citet{fivet06} reveal that they
are too blended to be of use as abundance indicators in \hdtwo.

Au has only one stable isotope, $^{197}$Au,
which has nuclear spin $I = 3/2$.
\citet{demidov21} calculate HFS $A$ constants for the 
upper and lower levels of the $\lambda$2675 line,
and their values agree with earlier experimental work by 
\citet{dahmen67} and \citet{passler94}.
We use the \citeauthor{demidov21}\ HFS $A$ constants to
compute the complete line component pattern for this line,
which is given in Table~\ref{auhfstab}.

\begin{deluxetable}{ccccccc}
\tablecaption{Hyperfine Structure 
Line Component Pattern for the Au~\textsc{i} $\lambda$2675 Line
\label{auhfstab}}
\tabletypesize{\small}
\tablehead{
\colhead{Wavenumber} &
\colhead{$\lambda_{\rm air}$} &
\colhead{$F_{\rm upper}$} &
\colhead{$F_{\rm lower}$} &
\colhead{Component Position} &
\colhead{Component Position} &
\colhead{Strength} \\
\colhead{(cm$^{-1}$)} &
\colhead{(\AA)} &
\colhead{} &
\colhead{} &
\colhead{(cm$^{-1}$)} &
\colhead{(\AA)} &
\colhead{}
}
\startdata
37358.991 & 2675.9366 & 2.0 & 2.0 & $-$0.070073 &    0.005019 & 0.31250  \\
37358.991 & 2675.9366 & 2.0 & 1.0 &    0.137403 & $-$0.009842 & 0.31250  \\
37358.991 & 2675.9366 & 1.0 & 2.0 & $-$0.090688 &    0.006496 & 0.31250  \\
37358.991 & 2675.9366 & 1.0 & 1.0 &    0.116789 & $-$0.008365 & 0.06250  \\
\enddata
\tablecomments{%
Energy levels from the NIST ASD and the index of air \citep{peck72}
are used to compute the
center-of-gravity wavenumbers and air wavelengths, $\lambda_{\rm air}$,
and component positions are given relative to those values.
Strengths are normalized to sum to 1.
Table~\ref{auhfstab} is available in the online edition
of the journal in machine-readable format.
}
\end{deluxetable}

The observed center-of-gravity wavelength listed in the NIST ASD, 
2675.950~\AA,
is found to be a better match to the Au~\textsc{i} line
in our spectrum
than the wavelengths calculated directly from the Au~\textsc{i} 
energy levels,
2675.9366~\AA\ \citep{ehrhardt71}.
This discrepancy is larger than the stated uncertainty
in the energy levels measured by \citeauthor{ehrhardt71}
and the typical uncertainty present in the NIST ASD.~
Its cause is not immediately clear.
Several other Au~\textsc{i} lines
listed in the NIST ASD exhibit similar discrepancies
between their observed and calculated wavelengths.
The energies and wavelengths listed in Table~\ref{auhfstab}
are based on the calculated center-of-gravity wavelength, 
and we shift the wavelengths by $+0.022$~\AA\ in our syntheses.

The NIST ASD quotes a \loggf\ value for the 
$\lambda$2675 line
from \citet{hannaford81}, with a grade of
A+ (2\%, 0.01~dex).
Measurements by \citet{zhang18} of the 
radiative lifetime of the upper level
agree with the \citeauthor{hannaford81}\ value to 
within $\approx$10\%.
This agreement is worse than the stated mutual uncertainties
of the two measurements,
and translates into a 0.05~dex difference in the \loggf\ values.
For consistency with previous Au abundance derivations,
we continue to adopt the \citeauthor{hannaford81}\ \loggf\ value,
but we caution that the 0.01~dex uncertainty recommended by NIST
may be slightly optimistic.

The Au~\textsc{i} line at 2675~\AA,
illustrated in Figure~\ref{specplot3},
is the only one that has previously been used to
derive Au abundances in metal-poor stars.
Most of the absorption at this wavelength is Au~\textsc{i}.
The most significant blend is 
Nb~\textsc{ii}, at 2675.942~\AA.~
We are confident that this blend is synthesized correctly
because we know the Nb abundance well
(derived from 10 lines; Appendix~\ref{niobium}), 
NLTE effects are minimal (Appendix~\ref{niobium}), and
the \loggf\ value for this line is known experimentally
(38\% or 0.20~dex; \citealt{nilsson08}).
We adjust the strength of this blend within its known uncertainty,
illustrated by the gray band in Figure~\ref{specplot3},
which affects the derived Au abundance by about $\pm$~0.07~dex.
We derive \logeps{Au} = 0.53 $\pm$~0.22 from this line.
The $\lambda$2376 and $\lambda$3122 lines yield upper limits
\logeps{Au} $< 1.2$ and $< 1.1$,
which are compatible with this abundance.
Our recommended Au abundance is based on this one Au~\textsc{i} line.

\subsection{Lead (Pb, $Z = 82$)}

\citet{roederer20} reported the detection of the
Pb~\textsc{ii} line at 2203.534~\AA\
in \hdtwo,
as illustrated in Figure~\ref{specplot1}.
That study derived \logeps{Pb} = 1.14 $\pm$0.16.
The FIP of Pb is 7.42~eV, and
singly ionized Pb is the dominant ionization state
in the atmosphere of \hdtwo.
\citet{mashonkina12} showed that the ground state
of Pb$^{+}$ is formed in LTE, and we assume
that the low-excitation level that
gives rise to the $\lambda$2203 line is also 
formed in LTE.~
We adopt the \loggf\ from \citet{quinet07}
(uncertainty $\approx$14\%, or 0.07~dex),
the HFS and IS patterns from \citeauthor{roederer20}, and
an \rpro\ isotopic mix from \citet{sneden08} in our syntheses.

\citet{peterson21} noted the presence of 
absorption near this wavelength in two
stars more metal-poor than \hdtwo.
Pb absorption would not be expected in those stars,
and \citeauthor{peterson21}\ postulated that
this absorption could be assigned to an unidentified
Fe~\textsc{i} line with wavelength = 2203.526~\AA,
E$_{\rm low}$ = 2.18~eV, and \loggf\ = $-$2.06.
This postulated line is offset slightly to the blue
of the Pb~\textsc{ii} line, and it would decrease the
derived Pb abundance by $\approx$0.2~dex.
Until additional confirmation of this potential assignment is available,
we recommend the Pb abundance for \hdtwo\
as described in \citet{roederer20}.

\subsection{Bismuth (Bi, $Z = 83$)}

We check for the 13 Bi~\textsc{i} lines that are listed in the
NIST ASD, but we detect none of them.
The Bi~\textsc{i} line at 2230.609~\AA\
provides the best upper limit on the Bi abundance.  
It lies in a region between several weak lines of Fe-group species,
and the line profile can be reasonably well fit with no Bi present.
We derive an upper limit by minimizing the absorption from these
blends, while simultaneously providing a reasonable fit
to the observed line profile.

\subsection{Thorium (Th, $Z = 90$) and Uranium (U, $Z = 92$)}

There are no Th~\textsc{ii} or U~\textsc{ii} lines
known in this region of the UV spectrum.
\citet{roederer18c} derived a Th abundance from 5 optical lines
\citep{nilsson02th}
and a U upper limit from 2 optical lines \citep{nilsson02u}.
We perform a new check for the strongest U~\textsc{ii} lines
listed in the study by \citet{gamrath18u},
and none are detected or useful for further constraining the U abundance.
Our recommended Th and U abundances are taken from
\citeauthor{roederer18c} 

\section{Radial Velocity Measurements of HD~222925}
\label{rvappendix}

The radial velocity, $V_{r}$, of \hdtwo\
has been measured several times over the last decade.
All $V_{r}$ measurements known to us are listed in
Table~\ref{rvtab}.
We have made several new measurements during the
most recent observing season,
using the Magellan Echellette (MagE) Spectrograph
\citep{marshall08} and MIKE.~
The MagE observations were collected using the 1\farcs0 slit,
yielding $R \sim$~4,700,
and the MIKE observations were collected using the 0\farcs7 slit,
yielding $R \sim$~35,000.
We measure $V_{r}$ by cross-correlating the 
order containing the Mg~\textsc{i} ``b'' triplet 
against a stellar template, as described in \citet{roederer14c}.
We calculate heliocentric corrections using the
IRAF ``rvcorrect'' task.
The dates of some measurements are unspecified
by the original references, but the 
time baseline of the $V_{r}$ measurements from
high-resolution spectroscopy spans at least 7~yr.
No evidence for $V_{r}$ variations is found.

\begin{deluxetable}{cccc}
\tablecaption{Radial Velocity Measurements of HD~222925
\label{rvtab}}
\tabletypesize{\small}
\tablehead{
\colhead{Date} &
\colhead{$V_{r}$} &
\colhead{Unc.} &
\colhead{Reference} \\
\colhead{} &
\colhead{(\kmsec)} &
\colhead{(\kmsec)} &
\colhead{} 
}
\startdata
(unspecified) & $-$34    & 7    & \citet{beers14} \\
(unspecified) & $-$38.64 & 0.36 & \citet{navarrete15} \\
(unspecified) & $-$37.93 & 0.28 & Gaia DR2 \citep{katz19} \\
2017/09/28    & $-$38.9  & 0.6  & \citet{roederer18c} \\
2021/07/04    & $-$38.7  & 10   & new (MagE) \\
2021/07/05    & $-$25.2  & 12   & new (MagE) \\
2021/11/24    & $-$37.6  & 0.6  & new (MIKE) \\
2021/11/25    & $-$38.6  & 0.6  & new (MIKE) \\
2021/12/05    & $-$38.1  & 0.6  & new (MIKE) \\
2021/12/06    & $-$38.1  & 0.6  & new (MIKE) \\
\enddata
\end{deluxetable}

\bibliographystyle{aasjournal}
\bibliography{ms.bbl}

\end{document}